\documentclass[fleqn,usenatbib]{mnras}
\DeclareRobustCommand{\VAN}[3]{#2}
\let\VANthebibliography\thebibliography
\def\thebibliography{\DeclareRobustCommand{\VAN}[3]{##3}\VANthebibliography}
\usepackage{newtxtext,newtxmath}
\usepackage{stfloats}
\usepackage[flushleft]{threeparttable}
\usepackage{subfigure}
\usepackage[T1]{fontenc}
\usepackage{setspace}
\usepackage{longtable}
\usepackage{threeparttable}
\usepackage{soul}
\usepackage{supertabular}
\usepackage{graphicx}	
\usepackage{amsmath}	

\usepackage{amssymb}	
\usepackage[justification=centering]{caption}
\usepackage{float}
\usepackage{supertabular}
\usepackage{makecell}
\usepackage{natbib}
\usepackage{adjustbox}

\title[Periodic X-ray Sources in M31 Bulge]{A {\it Chandra} Search for Periodic X-ray Sources in the Bulge of M31}

\author[Zhang \& Bao \& Li]{
Jiachang Zhang$^{1,2}$\thanks{E-mail: zhangjiachang@smail.nju.edu.cn}
Tong Bao$^{1,2}$\thanks{E-mail: baotong@smail.nju.edu.cn}
Zhiyuan Li$^{1,2}$\thanks{E-mail: lizy@nju.edu.cn}
\\
$^{1}$School of Astronomy and Space Science, Nanjing University, Nanjing 210046, China\\
$^{2}$Key Laboratory of Modern Astronomy and Astrophysics (Nanjing University), Ministry of Education, Nanjing 210046, China
}

\date{}

\begin{document}
\label{firstpage}
\pagerange{\pageref{firstpage}--\pageref{lastpage}}
\maketitle
\begin{abstract}
We present a systematic search for periodic X-ray sources in the bulge of M31, using $\sim$2 Ms of archival {\it Chandra} observations spanning a temporal baseline of 16 years. Utilizing the Gregory-Loredo algorithm that is designed for photon-counting, phase-folded light curves, we detect seven periodic X-ray sources, among which four are newly discovered. 
Three of these sources are novae, the identified periods of which
range between 1.3–2.0 hour and is most likely the orbital period. The
other four sources are low-mass X-ray binaries, the identified periods of which range between 0.13--19.3 hour and are also
likely orbital due to a clear eclipsing/dipping behavior in the light
curve. 
We address implications on the X-ray binary population of the M31 bulge. 
Our study demonstrates the potential of using archival X-ray observations to systematically identify periodic
X-ray sources in external galaxies, which would provide valuable
information about the underlying exotic stellar populations.
\end{abstract}

\begin{keywords}
galaxies: bulges – galaxies: individual (M31) - X-rays: binaries - (stars:) novae, cataclysmic variables
\end{keywords}

\section{Introduction} \label{sec:intro}
The Andromeda galaxy (= M31, at a distance of 785 kpc; \citealp{2005MNRAS.356..979M}) is a massive spiral with a substantial stellar bulge. Thanks to its proximity to the Milky Way, M31 provides a unique opportunity for studying galaxies similar in physical properties with our Galaxy.
In particular, the X-ray emissions of M31 have been observed since the era of the {\it Einstein} satellite \citep{1979BAAS...11..609V} and have subsequently been studied with improved spatial resolution, spectral resolution and sensitivity through the utilization of many X-ray missions, especially {\it Chandra} \citep{2002ApJ...577..738K,2002ApJ...578..114K} and {\it XMM-Newton} \citep{2001A&A...365L.195S,2001A&A...378..800O,2005A&A...434..483P}.

M31 holds hundreds of bright X-ray point sources, mostly accretion-powered X-ray binaries (XRBs) involving a stellar-mass black hole (BH), a neutron star (NS) or a white dwarf (WD), which can be studied in detail thanks to low Galactic foreground absorption. 
These sources have been the subject of systematic investigations over the past decades, e.g., by \citet{2002ApJ...577..738K,2005A&A...434..483P,2006A&A...447...71V,2013A&A...555A..65H,2014ApJ...780...83B}, among others.
In the bulge of M31, where the stellar population is predominantly old, low-mass X-ray binaries (LMXBs) make up the majority, if not all, of the XRBs \citep{2013A&A...555A..65H}. 
Another important component is cataclysmic variables (CVs), which experience mass transfer from a main sequence star or a red giant star to a WD \citep{1976ARA&A..14..119R}. The thermonuclear runaway in the accreted matter triggers classical nova events. The classical novae in the central region of M31 have been monitored by {\it XMM-Newton}, {\it Chandra} and {\it Swift} from 2006 to 2011 \citep{2010A&A...523A..89H,2011A&A...533A..52H,2014A&A...563A...2H}. 
Additionally, ultraluminous X-ray sources (ULXs), super-soft sources (SSSs, \citealp{1997ARA&A..35...69K}), and quasi-soft sources (QSSs) are known to exist among the X-ray populations of M31. These diverse stellar populations serve as a natural and promising laboratory to advance our understanding of the underlying physical processes related to accretion and binary evolution.

Many X-ray sources in M31 have exhibited strong flux variability, a characteristic of accretion-powered systems. 
A particularly interesting feature is periodic variability, which arise due to either binary orbital motion or the spin of the accretor.  
Over the years, a number of periodic X-ray signals have been identified in M31 by either dedicated searches or serendipitous discoveries. Table~\ref{tab:known} summarizes the previously known X-ray periodic sources in M31. 
These include four novae with periodic signals, three SSSs with pulsation periods, four LMXBs with orbital or super-orbital periods, among which one also shows a pulsation period. 
These periodic variations provide valuable insights into the dynamics and characteristics of the X-ray binary populations in M31. 
Yet, compared to the fraction of the Galactic XRBs with known orbital periods (145 in 348 for LMXBs, \citealp{2023A&A...675A.199A}; 112 in 169 for high-mass X-ray binaries [HMXBs], \citealp{2023A&A...677A.134N}), 
far less is currently known for the M31 XRB populations.

\begin{table*}
    \centering
    \begin{threeparttable}
\caption{Known periodic X-ray sources in M31.}
    \begin{tabular}{cccc}
    \hline\hline
        Class& Name          & Period & Reference\\
            \hline
        Nova& M31N 2006-04a & 1.6 hr & {\cite{2010A&A...523A..89H}}\\
            & M31N 2011-11e & 1.3 hr & {\cite{2014A&A...563A...2H}}\\
            & M31N 2013-01b & 1.28 hr& {\cite{2018ApJ...866..125M}}\\
            & M31N 2007-12b & 4.9 hr & {\cite{2011A&A...531A..22P}}\\
        SSS & XMMU J004319.4+411758 & 865 s & {\cite{2001A&A...378..800O}} \\
            & XMMU J004252.5+411540 & 217 s & {\cite{2008ApJ...676.1218T}}\\
            & M31N 2007-12b         & 1110 s& {\cite{2011A&A...531A..22P}}\\
        LMXB& XMMU J004314.1+410724 & 2.78 hr, 5.65 d\tnote{a} & {\cite{2002ApJ...581L..27T,2015ApJ...801...65B}}\\
            & XMMU J004308.6+411247 & 1.78 hr & {\cite{2004A&A...419.1045M}}\\
            & 3XMM J004232.1+411314 & 3 s\tnote{b}, 4.15 hr & {\cite{2017ApJ...851L..27M,2018ApJ...861L..26R}}\\
            & 3XMM J004301.4+413017 & 1.2 s\tnote{c}, 1.27 d & {\cite{2016MNRAS.457L...5E}}\\
            \hline
    \end{tabular}
    \label{tab:known}
    \begin{tablenotes}
      \small
      \item
      Notes: 
      \tnote{a} Super orbital period. Source located in the globular cluster Bo 158. \tnote{b} Pulsation period of the neutron star. \tnote{c} Pulsation period of the accreting pulsar. Source located in the globular cluster GLC 377.
\end{tablenotes}
    \end{threeparttable}
\end{table*}

In this work, we conduct a systematic search for periodic X-ray signals from a large number of X-ray sources in the bulge of M31, based on extensive {\it Chandra} monitoring observations taken over the past two decades. 
These observations provide a rare but promising dataset for probing periodic X-ray signals from an extragalactic population of XRBs. 
In Section \ref{sec:xdata}, we describe the preparation of the X-ray data and the construction of a list of X-ray sources in the M31 bulge. In Section \ref{sec:period}, we detail the period searching method, procedures and results. 
The X-ray spectra of the periodic X-ray sources are analyzed in Section \ref{sec:spectra}, followed by an examination of the physical nature of the periodic sources based on their temporal and spectral properties in Section~\ref{sec:class}. 
In Section~\ref{sec:discusstion}, we address the implications and limitations of our results.
A brief summary of our study is given in Section~\ref{sec:sum}.

\section{X-ray Data Preparation}\label{sec:xdata}

The bulge of M31 has been extensively observed by {\it Chandra} using its Advanced CCD Imaging Spectrometer (ACIS) and High Resolution Camera (HRC). 
 We require that the aim-point of a selected observation was  $\lesssim 1\arcmin$ from the center of M31,
which helps to minimize variations in the point-spread function (PSF) across different observations at any given position, ensuring consistent and accurate photometry for the X-ray point sources. 
This results in a total of 116 ACIS observations taken between October 13, 1999 and November 28, 2015 with a total exposure time of $\sim 996.9$ ks, and 64 HRC observations taken between November 30, 1999 and June 1, 2012 with a total exposure of $\sim 1032.4$ ks. These observations together span a temporal baseline of over 16 yrs, but with a rather irregular cadence. 

The information of the ACIS and HRC observations utilized in this work is given in Table~\ref{tabacis} and Table \ref{tabhrc}. 

We downloaded and uniformly reprocessed the raw data for each observation,  following the standard procedure\footnote{http://cxc.harvard.edu/ciao} and using CIAO v.4.14 and calibration data files CALDB v.4.9.8. The ACIS and HRC observations were treated separately. The CIAO tool ${\it acis\_process\_event}$ and ${\it hrc\_process\_events}$ were used to reprocess the level 1 event files to create level 2 events,  correcting charge transfer inefficiency and filtering cosmic-ray afterglows. 
For the ACIS observations, we included only data from the I0, I1, I2 and I3 chips if the aim-point was placed on the I3 chip, and only data from the S2 and S3 chips if the aim-point was placed on the S3 chip. 
The analysis of the light curve for each observation revealed that the instrumental background remained quiescent for the majority of the time intervals. Hence all science exposures were preserved for the subsequent timing analysis, ensuring an uninterrupted light curve within each observation. Additionally, the CIAO tool $\it{axbary}$ was employed to correct the photon arrival time in each registered event to the Solar system barycenter (i.e. Temps Dynamique Barycentrique time).

\begin{figure*}
\centering
\includegraphics[width=1\textwidth,trim=0cm 0cm 1cm 0cm, clip]{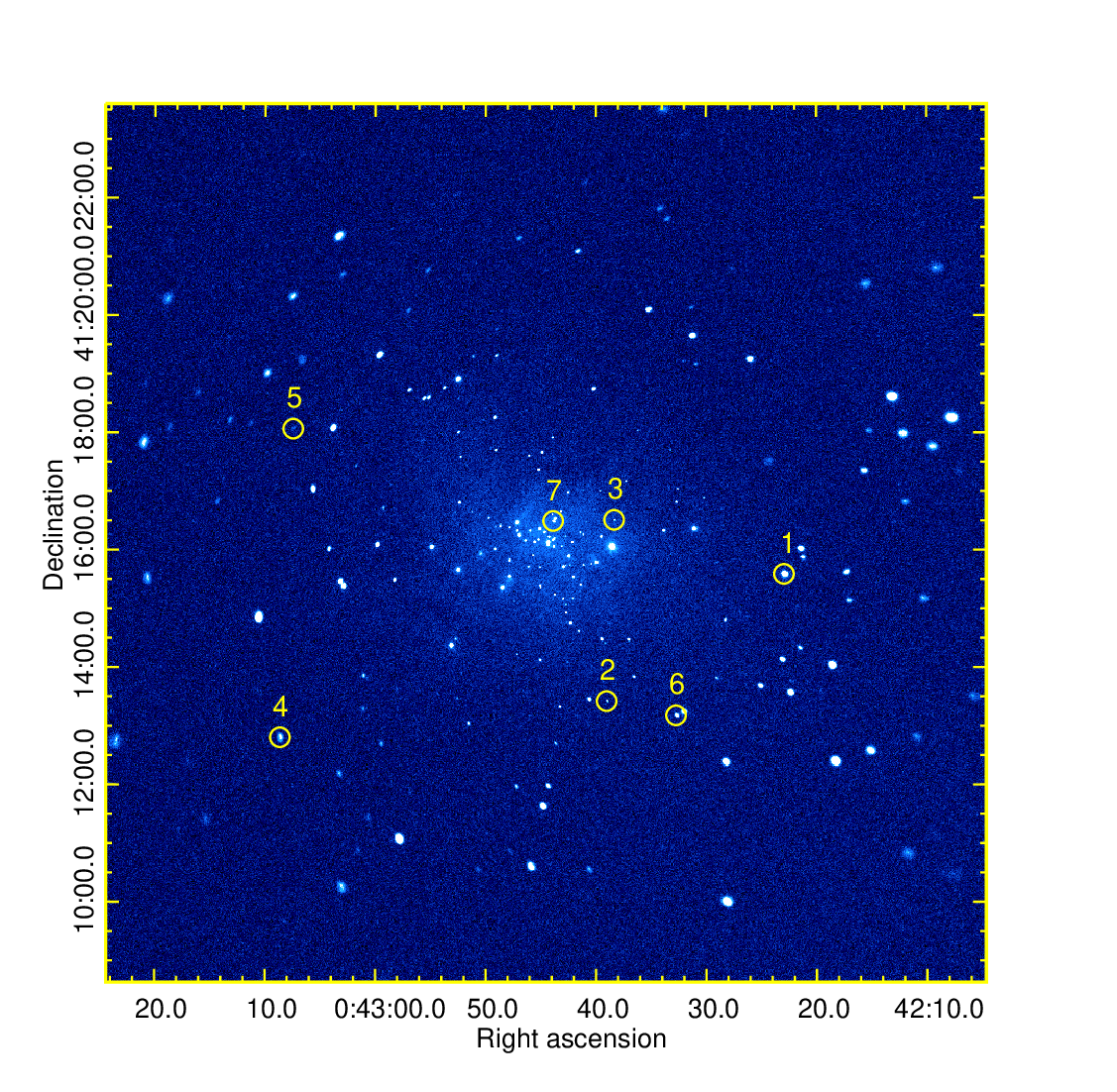}
\caption{Counts image of the M31 bulge, combining 64 {\it Chandra} HRC observations. The positions of the seven periodic sources are indicated by yellow circles, with source numbering as in Table \ref{tab:result}. }
\label{fig:ACISFoV}
\end{figure*}

To align the astrometry among individual ACIS or HRC observations, we performed centroid matching for commonly detected point sources using the CIAO tool \emph{reproject\_aspect}. The observation with the longest exposure time (i.e., ObsID 14196 for ACIS and ObsID 1912 for HRC) was selected as the reference frame for this alignment process. The stacked counts map over the energy range of 0.5–8 keV for ACIS and the PI range of 50--300 for HRC was then created respectively. The stacked HRC counts map of the inner $15\arcmin \times 15\arcmin$ ($\sim$ 3.4 kpc $\times$ 3.4 kpc) of M31 is shown in Figure~\ref{fig:ACISFoV}. 
Exposure and PSF maps were generated, with the PSF maps being computed according to a specified enclosed count radius (ECR). A fiducial source spectrum was employed for both the exposure and PSF maps. This spectrum is characterized by an absorbed power-law model with a photon index of 1.7 and a hydrogen column density consistent with the Galactic foreground value ($N\rm_H=7.0\times 10^{20}~cm^{-2}$), representative of the bulge X-ray sources \citep{2013A&A...555A..65H}.

Point source detection was conducted following the procedures outlined in \cite{2018ApJS..235...26Z}.  Briefly, the CIAO tool {\it wavdetect} was applied to the stacked energy = 0.5--8 keV counts map of ACIS and stacked PI = 50--300 counts map of HRC. 
We restricted the region of interest within a projected distance of 500$\arcsec$ from the M31 center, which not only ensures an optimal PSF, but also minimizes the fractional contamination by cosmic X-ray background (CXB) sources and potential HMXBs related to the star-forming disk of M31.
The algorithm was fed with the stacked exposure map and the 50\% ECR PSF map, setting a false-positive probability threshold of $10^{-6}$. The centroid of each source provided by {\it wavdetect} was further refined using a maximum-likelihood method that iterates over the recorded positions of the individual counts detected within the 90\% ECR.
Our detection resulted in a raw list of 456 ACIS sources and  403 HRC sources, which form the basis of the period searching (Section~\ref{sec:period}).
 A cross-matching of ACIS-detected and HRC-detected sources suggests a total of 647 positionally independent sources.
\cite{2006A&A...447...71V} published an X-ray source catalog based on a large fraction of the ACIS observations used here, whereas \cite{2013A&A...555A..65H} also published a source catalog using the same set of 64 HRC observations.
We find that almost all sources within the same region of interest in these two catalogs are recovered in our source lists.
We defer a detailed description and analysis of the updated source catalog of the M31 bulge to future work.

For each source in each ObsID, we adopted a default 90\% ECR to extract the source events, again focusing on the energy range of 0.5--8 keV for ACIS and PI range of 50--300 for HRC. Source events from all ACIS or all HRC observations together form the time series of a given source to be fed to the period searching algorithm (Section~\ref{subsec:GL}). Crowding of X-ray sources is not a general concern for the M31 bulge. 
For a handful of sources which have a close neighbor source, we have also tested an  extraction region of 75\% ECR, but found no significant difference in the result.
Local background counts were extracted from a concentric annulus with inner-to-outer radii of 2--4 times the 90\% ECR, masking pixels falling with the 90\% ECR of neighboring sources, if any.

We examined possible flux variability by quantifying the net photon flux in the individual observations. The total source counts, background counts, source area, background area, and local exposure were fed to the CIAO tool {\it aprates} to calculate the net photon flux and associated error, which accounts for the Poisson statistics in the low-count regime. If the 3$\sigma$ lower limit of the photon flux hits zero in a certain observation, the source is considered non-detected and we derive a 3$\sigma$ upper limit of the source photon flux for this particular observation.

\section{Period searching procedure}\label{sec:period}
\subsection{Method}
\label{subsec:GL}
We apply the Gregory-Loredo (GL) algorithm \citep{1992ApJ...398..146G} to detect potential periodic signals, which is a Bayesian-based, phase-folding method.
The GL algorithm is effective in identifying periodic signals from X-ray data that is often characterized by a moderate number of photon events and/or an irregular observing cadence, as is the case of M31 here.
This algorithm has been successfully employed to detect periodic X-ray sources in a Galactic bulge field \citep{2020MNRAS.498.3513B}, the {\it Chandra} Deep Field South \citep{2022MNRAS.509.3504B}, as well as a number of Galactic globular clusters \citep{2023MNRAS.521.4257B,2024MNRAS.527.7173B}, all based on deep {\it Chandra} observations. 
Readers are referred to these papers for details of the GL algorithm and specific implementation for {\it Chandra} data, as well as potential caveats.
A concise description of the working principle of the GL algorithm is given in Appendix~\ref{appendix}.
Main steps of the period searching procedure are as follows.

By design, the GL algorithm folds the time series of a given source at trial frequencies (periods).
The chosen resolution and frequency range is a compromise of efficiency and computational resource. 
We primarily explore period ranges of (100, 500), (500, 1000), (1000, 5000), (5000, 10000) and (10000, 20000) seconds, with a frequency resolution of $10^{-5}$, $10^{-6}$, $10^{-7}$, $10^{-8}$ and $10^{-9}$ Hz, respectively. 
Given the timespan of $\sim 5\times10^8$ sec between the first and last {\it Chandra} observations, the chosen frequency resolutions are optimal for an efficient search of periodic signals. 
Since the GL algorithm only determines the most probable period, once a tentative period is identified, a second search is performed excluding a narrow interval around the identified period, which ensures that a possible second period within the same period searching range will not be missed.

In the second step, we rerun the algorithm on the time series from each individual observation. 
If a signal of roughly the same period were detected in more than one observations, we rerun the GL algorithm  on the combined time series from those observations to refine the period. We do not attempt to combine the ACIS and HRC observations due to their substantially different instrumental response.

Although we are also interested in signals of 
lower frequencies (longer periods), they are hard to identify due to potentially significant flux variations among observations that may cause numerous spurious signals.
Therefore, we search for periods between 20000--40000 sec using only a series of closely-packed long observations (ObsID 13825,13826,13827,13828,14195,15267,14196 for ACIS) and (ObsID 5925,6177,5926,6202,5927,5928 for HRC) at a frequency resolution of  $10^{-9}$ Hz. 

We note in passing that the ACIS and HRC data in principle allow for probing shorter periods ($< 100$ sec), e.g., due to spin modulation of neutron stars. 
For ACIS, a natural lower limit is set by the CCD frame time, which is typically 3.2 sec; For HRC, the actual limit inversely depends on the rate of registered counts, which could be also as low as $\sim$20 ms for the present case. 
However, essentially all the X-ray sources in M31 have a moderate or low count rate, which introduces a substantial Poisson noise in the phase-folded light curve, challenging the identification of a true periodic signal.
Nevertheless, this has been attempted by us, which 
resulted in several periods on the order of a few tens of sec.
However, all such short periodic signals were only found in a single observation and had a relatively low value ($\sim0.9$) of  probability threshold ($P_{\rm GL}$). Given the large number of single-epoch light curves fed to the GL algorithm, we cannot rule out the possibility that these short-period signals are simply false detections due to noise. Moreover, these periods are not only too low to be associated with an orbital modulation, but also too short-lived to be related to a spin modulation, which ought to be persistent. Therefore, we do not consider them genuine periodic signals.

We adopt a probability threshold $P_{\rm GL}$ of 90\% for selecting the tentative periods returned by the GL algorithm. 
Some of these tentative periodic signals, however, could be spurious due to one of the following reasons. Therefore we apply further filtering to the tentative periodic signals: 
\begin{enumerate}
 	\item Eliminate periodic signals caused by the dither pattern of the ACIS/HRC detectors. For observations with ACIS, the nominal periods are 1000 s in yaw and 707 s in pitch, while for observations with the HRC, dithering follows nominal periods of 1087 s in yaw and 768 s in pitch\footnote{https://cxc.cfa.harvard.edu/ciao/why/dither.html}. 
    When part of the source region falls onto a bad pixel or goes outside the CCD edge, a spurious signal modulated by the dithering could occur.
    Any signals detected at the two dither periods or their harmonics to within 1 per cent are considered spurious and excluded.

	\item Exclude fake signals produced by strong, aperiodic source variability. 
    For sources exhibiting strong (up to a factor of 10 variation in the net photon flux) long-term (i.e. inter-observation) variations, we repeat the period search in two subsets of the light curve: one covering only the high state (defined as the observation[s] with the highest photon flux) and the other excluding the high state. If a periodic signal arising from the full time series could not be reproduced in either subset, we exclude it as a false detection. 
    Moreover, for each tentative periodic source, we inspect its light curve to identify short-term (i.e. intra-observation) flares, and subsequently repeat the period search after discarding any observation(s) in which significant short-term flares are present. 
    Only when a periodic signal survives this procedure it is considered a genuine signal.
\end{enumerate}

In Section~\ref{subsec:rednoise}, we further address the possibility of false detection of periodic signals due to an intrinsic red noise of the X-ray sources. 

\subsection{Resultant periodic signals}

\begin{table*}
\caption{Detected periodic sources.}\label{tab:result}
\begin{adjustbox}{width=1.0\textwidth}
\begin{threeparttable}
    \small 
    \centering
\begin{tabular}{ccccccccccccc}
\hline
\hline
ID	&	R.A.	&	Decl.  & ObsID	& \multicolumn{4}{c}{ACIS} &\multicolumn{4}{c}{HRC} & Notes
\\
    &          &     &           & Period (s) &$P_{\rm GL}$& C &$\rm C_B$    & Period (s) &$P_{\rm GL}$& C &$\rm C_B$ 
\\ 
(1) & (2) & (3)  &(4)& (5) & (6) & (7) & (8) & (9) & (10) & (11)  &(12)&(13)
\\
\hline
1 & 00 42 22.953  & +41 15 35.52 &\makecell{11808,12113,12114,13179,\\13180,11838,11839,12164}	&461.68  &1.0000&  1291 & 1.8 &463.24 & 1.0000 &10353 & 44.1	 & LMXB, 464 s\tnote{a}    \\
2 & 00 42 39.030  & +41 13 25.50   &13179,13180	& /&/& /& /& 4685.82 &0.9999 & 358	& 4.2 &Nova\tnote{b}    \\
3 & 00 42 38.350  & +41 16 30.87    &13278,13279,13280	& /&/ &/ & /&  4933.40&1.0000 &335	& 3.2		& Nova, 1.3 h\tnote{c}\\
4 & 00 43 08.636  & +41 12 48.42   & All Observations  & 6432.94 &1.0000&4979 & 85.2 & 6431.44 &0.9894 &3096 & 747.8 &LMXB, 1.78 h\tnote{d}	   \\
5 & 00 43 07.453  & +41 18 03.98   &5927, 5928		&/&/ & /& /&  7183.91&0.9997 & 185	& 18.2    &Nova\tnote{e}\\
6 & 00 42 32.746 & +41 13 10.94    &All Observations	& 22835.09 & 1.0000&4795 & 65.2 &22834.53 & 1.0000& 2486 & 191.5  &LMXB\\
7 & 00 42 43.877 & +41 16 29.72   &All Observations	&69417.50& 1.0000 &6263 & 75.2 &69411.24 & 1.0000& 2258 & 59.9 &LMXB \\
\hline
\end{tabular}
\begin{tablenotes}
      \small
      \item
      Notes: 
(1) Source sequence number assigned in the order of increasing period. (2) and (3) Right Ascension and Declination (J2000) of the source centroid.(4) The ID of observations based on which the periodic signal is detected.
(5)-(8) The period and the $P_{\rm GL}$ determined by the GL algorithm, the number of total counts in the 0.5-8 keV band and the number of estimated background counts of ACIS observations.  (9)-(12) The period and the $P_{\rm GL}$ determined by the GL algorithm, the number of total counts in the PI range of 50--300 and the number of estimated background counts of HRC observations. (13) Tentative source classification and periods reported in previous work (if any). \tnote{a}3XMM J004222.9+411535 \citep{2021A&A...650A.167D}; \tnote{b}M31N 2011-01b;  \tnote{c}M31N 2011-11e \citep{2014A&A...563A...2H}; \tnote{d}XMMU J004308.6+411247 \citep{2004A&A...419.1045M}; \tnote{e}M31N 2004-11b.
\end{tablenotes}
\end{threeparttable} 
\end{adjustbox}
\end{table*}
The above procedures result in a total of seven periodic signals/sources in the M31 bulge, among which four are newly discovered. 
Table \ref{tab:result} lists these periodic sources in the order of increasing period, giving information on the source ID, position, period, the GL probability and the relevant ObsID, as well as tentative classification (see details in Section \ref{sec:class}). 
The detected periods range from 463.24 sec to 69417.50 sec.
It is noteworthy that the longest period, found for Seq.7, actually exceeds our primary period searching range. The GL algorithm originally reported a period half of this value ($\sim$ 34707 sec). After further examination of multiple, adjacent observations, we determine the true period, which is $\sim$ 69417.50 sec, from the eclipsing behavior of this source (see details in Section \ref{subsec:rednoise}).

\begin{figure*}
    \centering
\setlength{\abovecaptionskip}{0.cm} 
\setlength{\abovecaptionskip}{0.cm} 
\subfigtopskip=0.cm
\subfigbottomskip=0.cm
\subfigure{\includegraphics[width=0.34\linewidth]{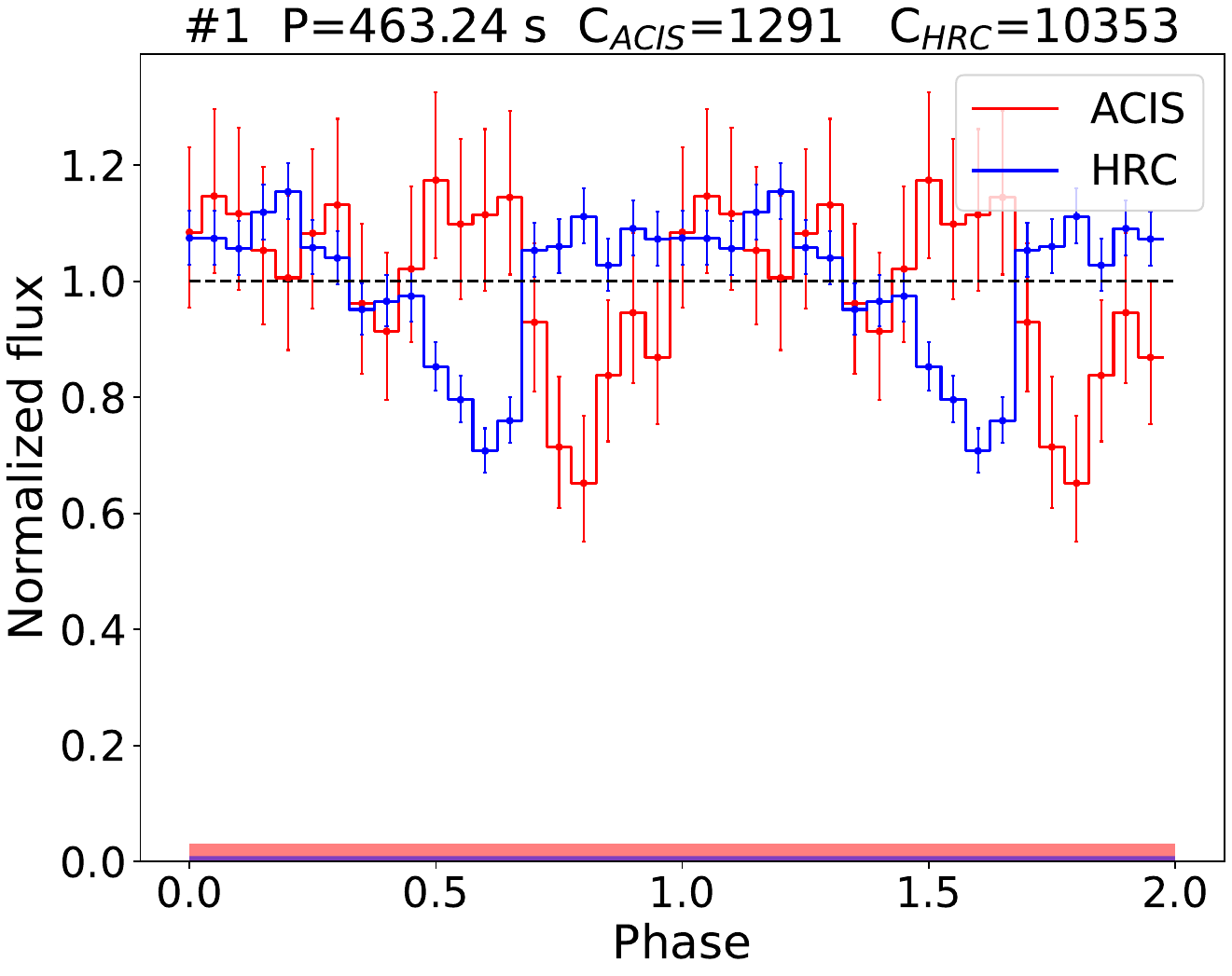} \includegraphics[width=0.345\linewidth]{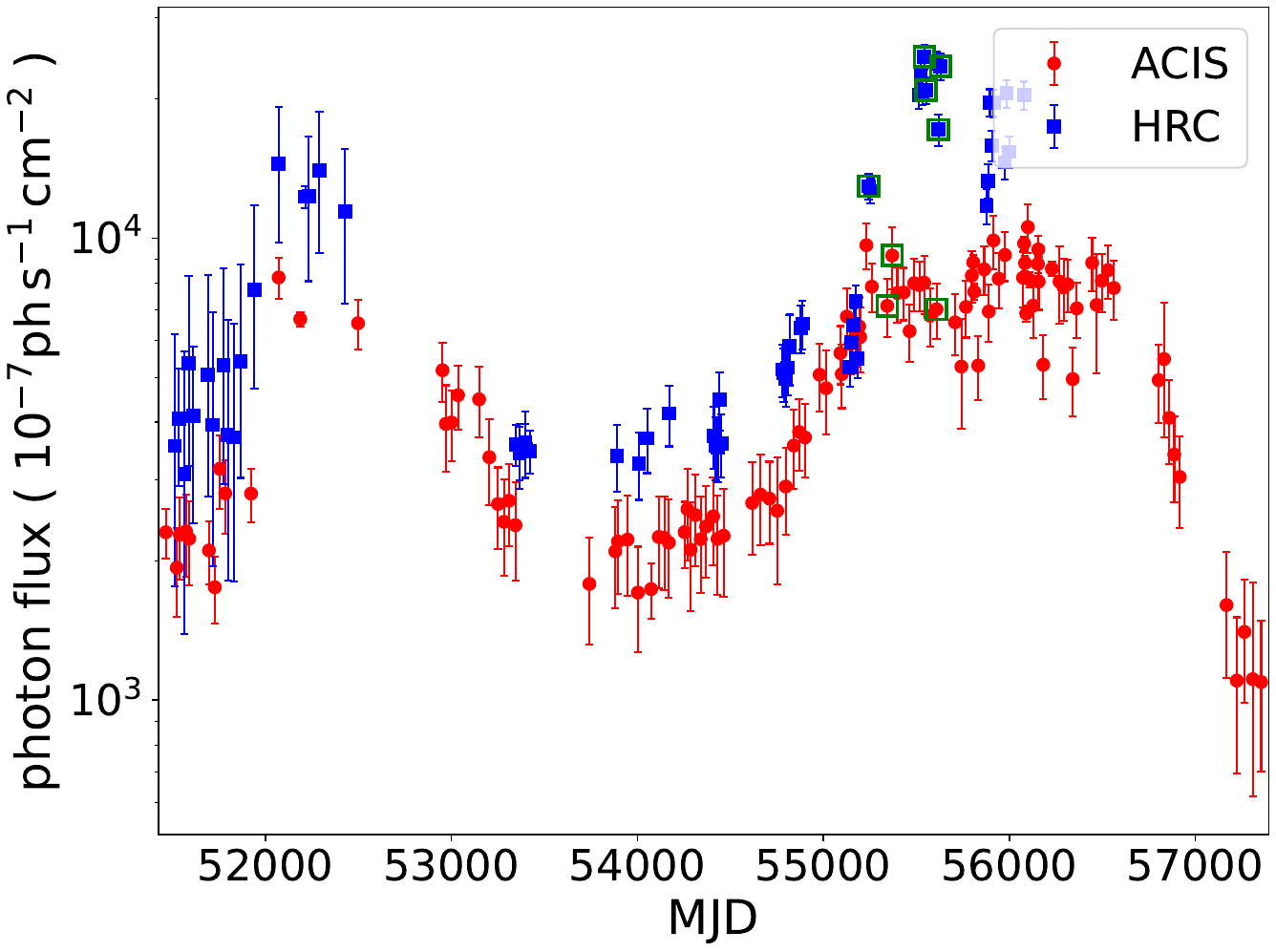}\includegraphics[width=0.35\linewidth,trim=0cm 0.3cm 3cm 2cm, clip]{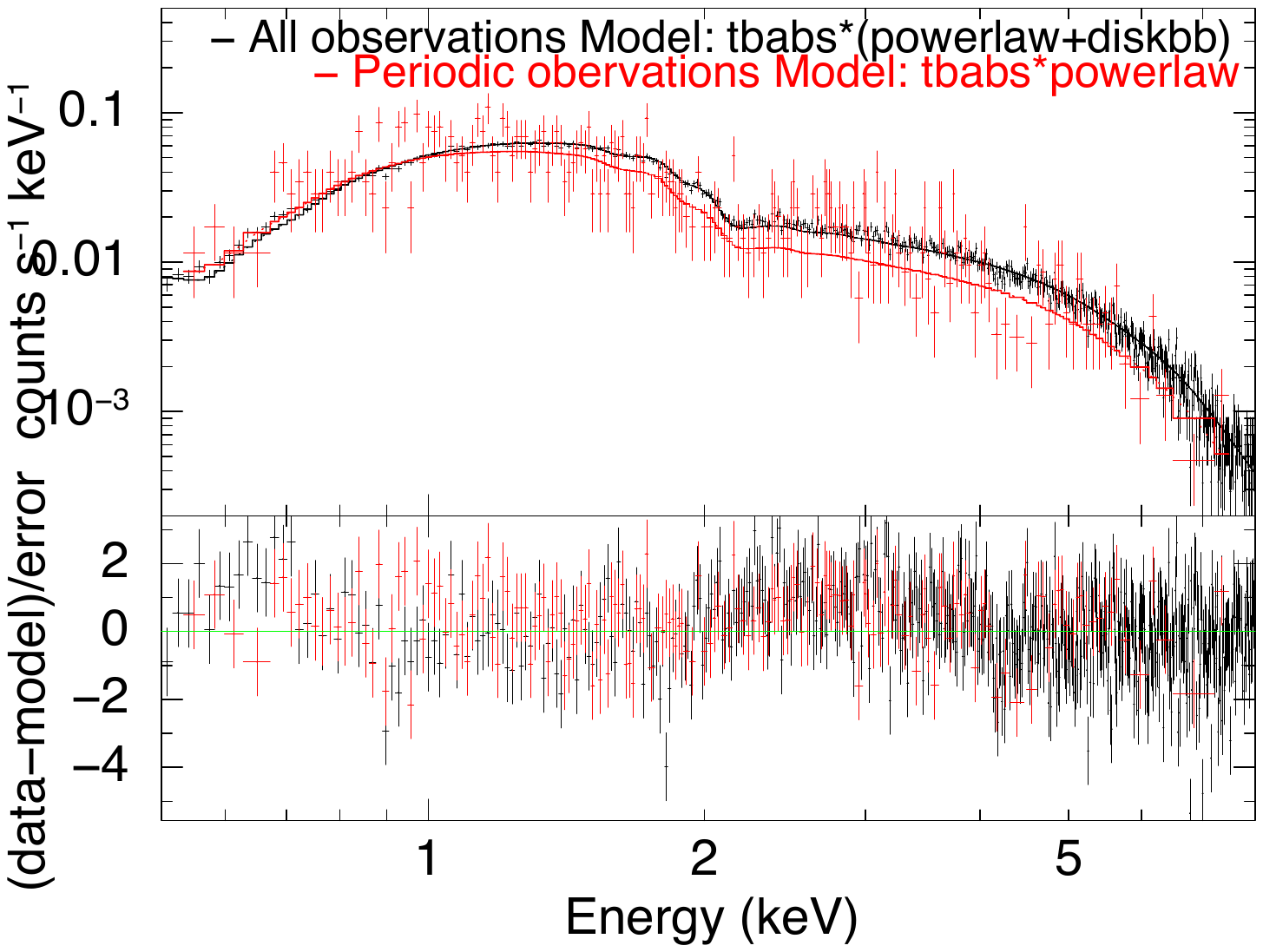}}
\subfigure{\includegraphics[width=0.34\linewidth]{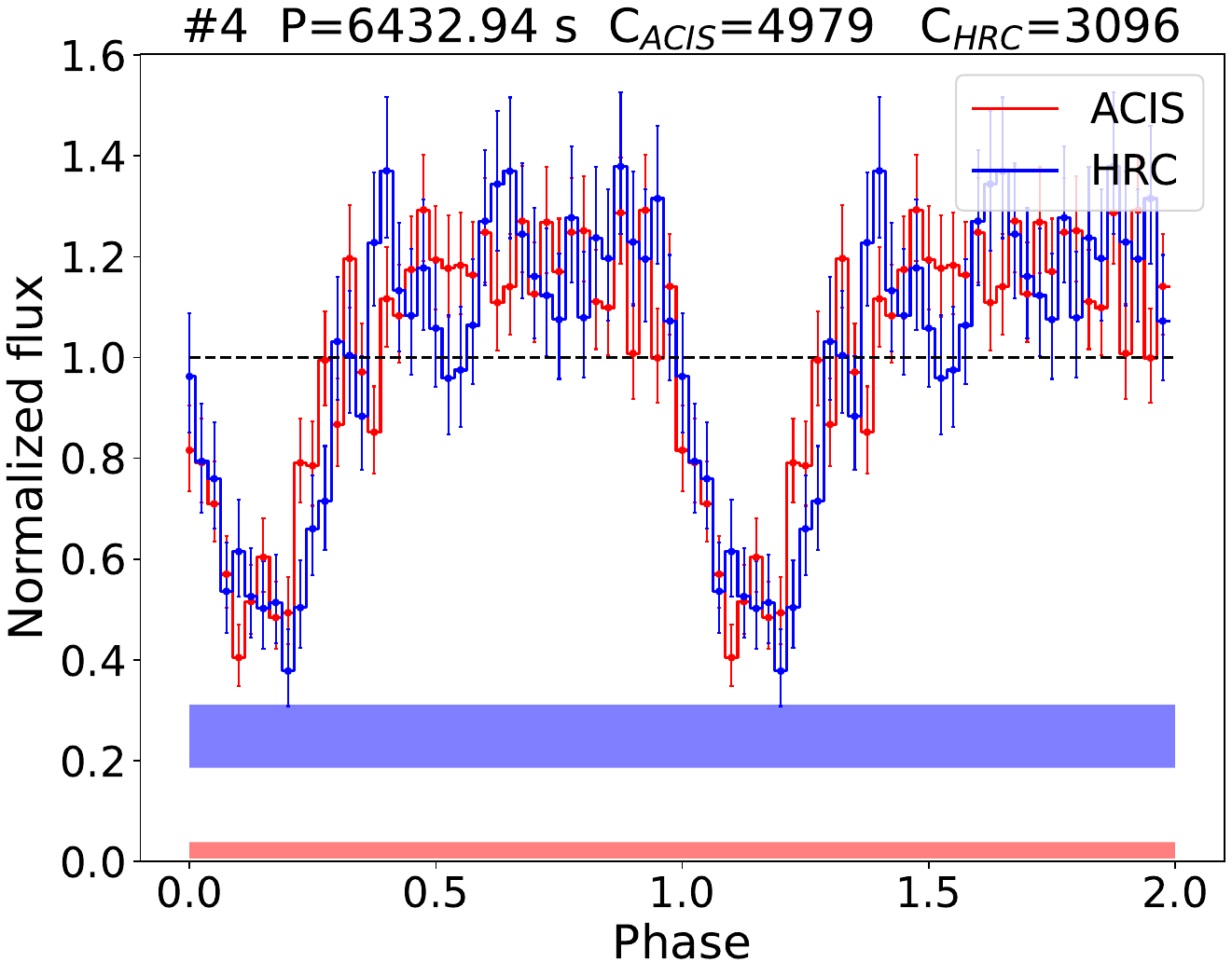}\includegraphics[width=0.345\linewidth]{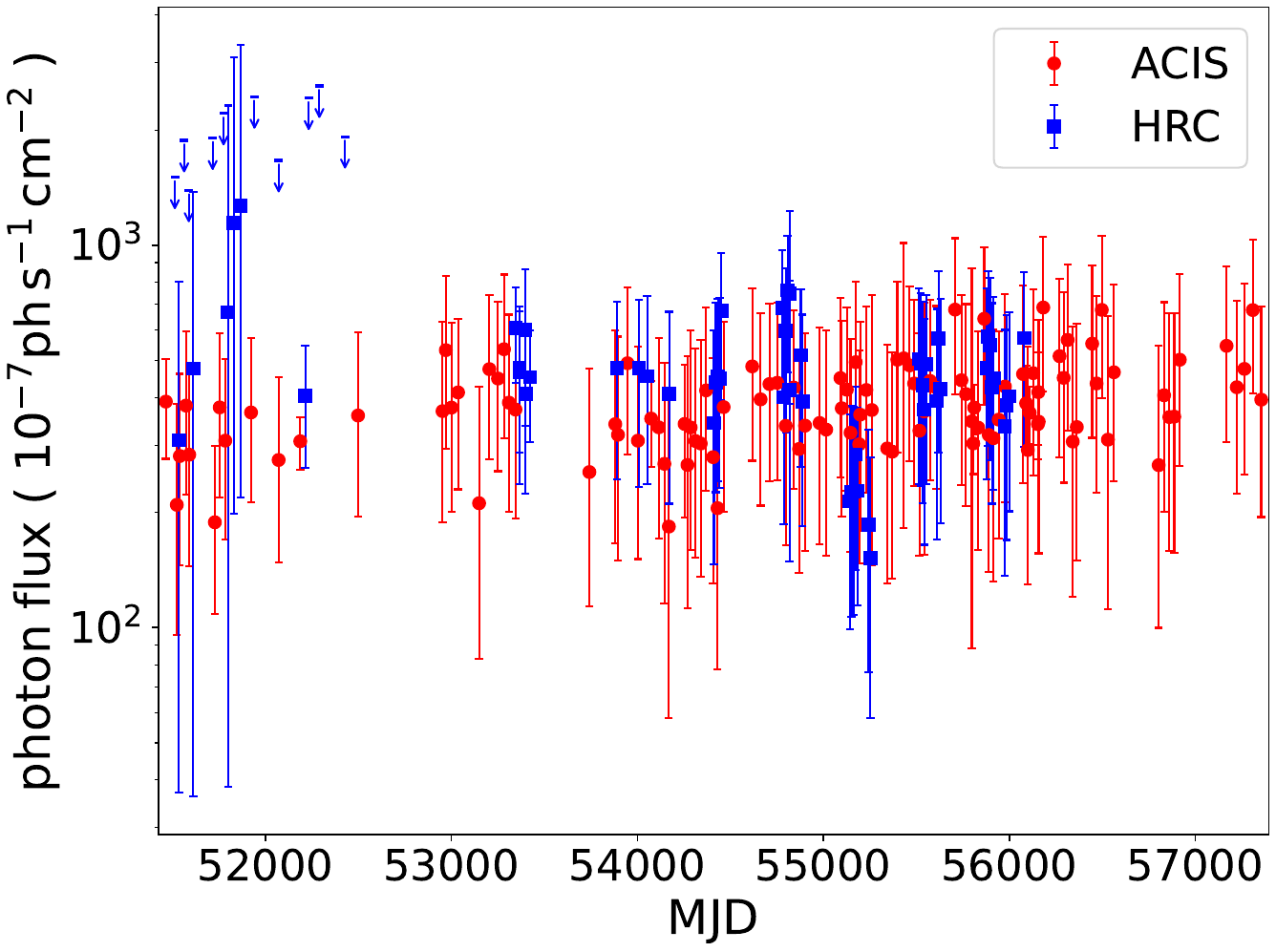}\includegraphics[width=0.35\linewidth,trim=0cm 0.3cm 3cm 2cm, clip]{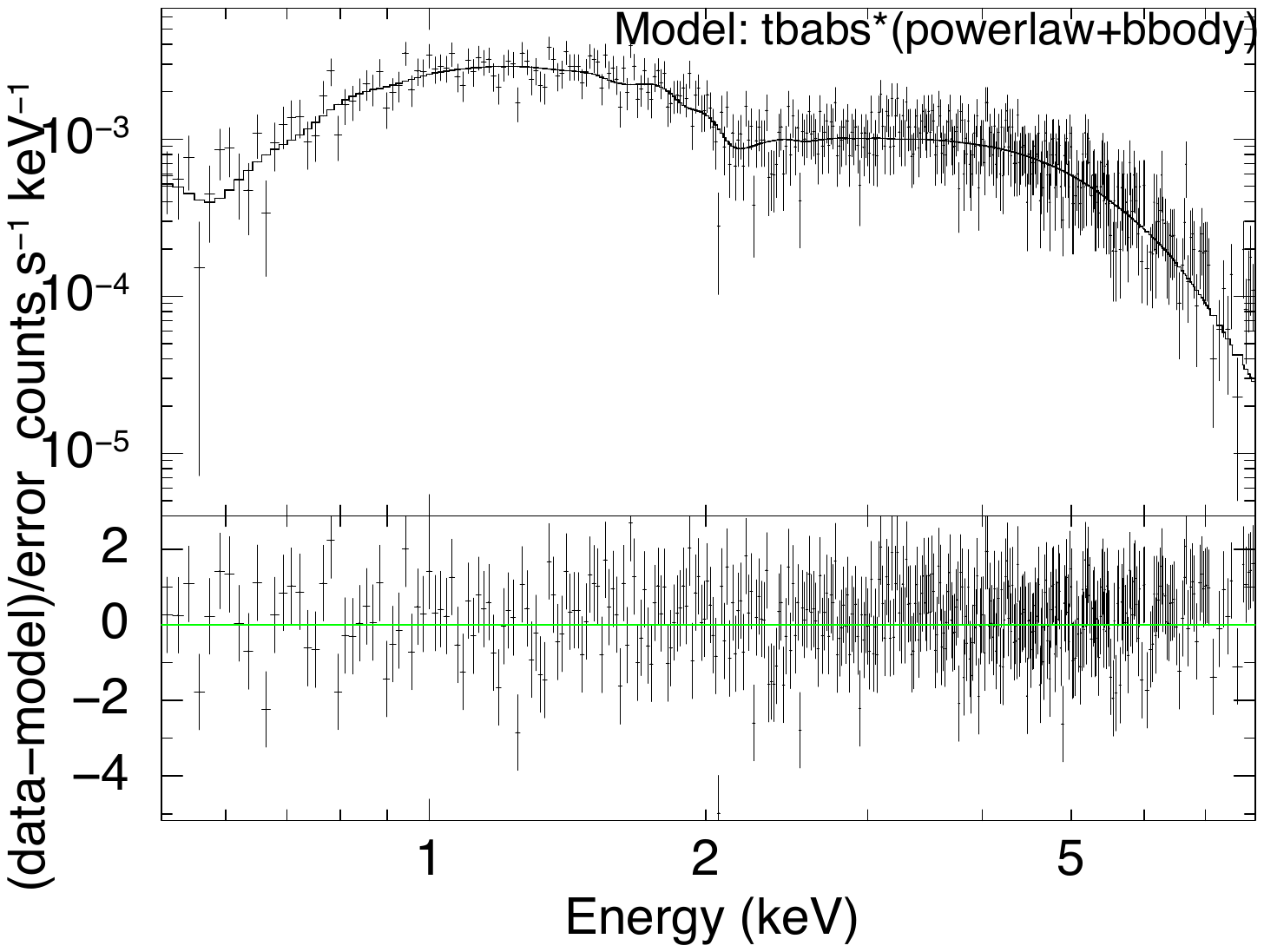}}
\subfigure{\includegraphics[width=0.34\linewidth]{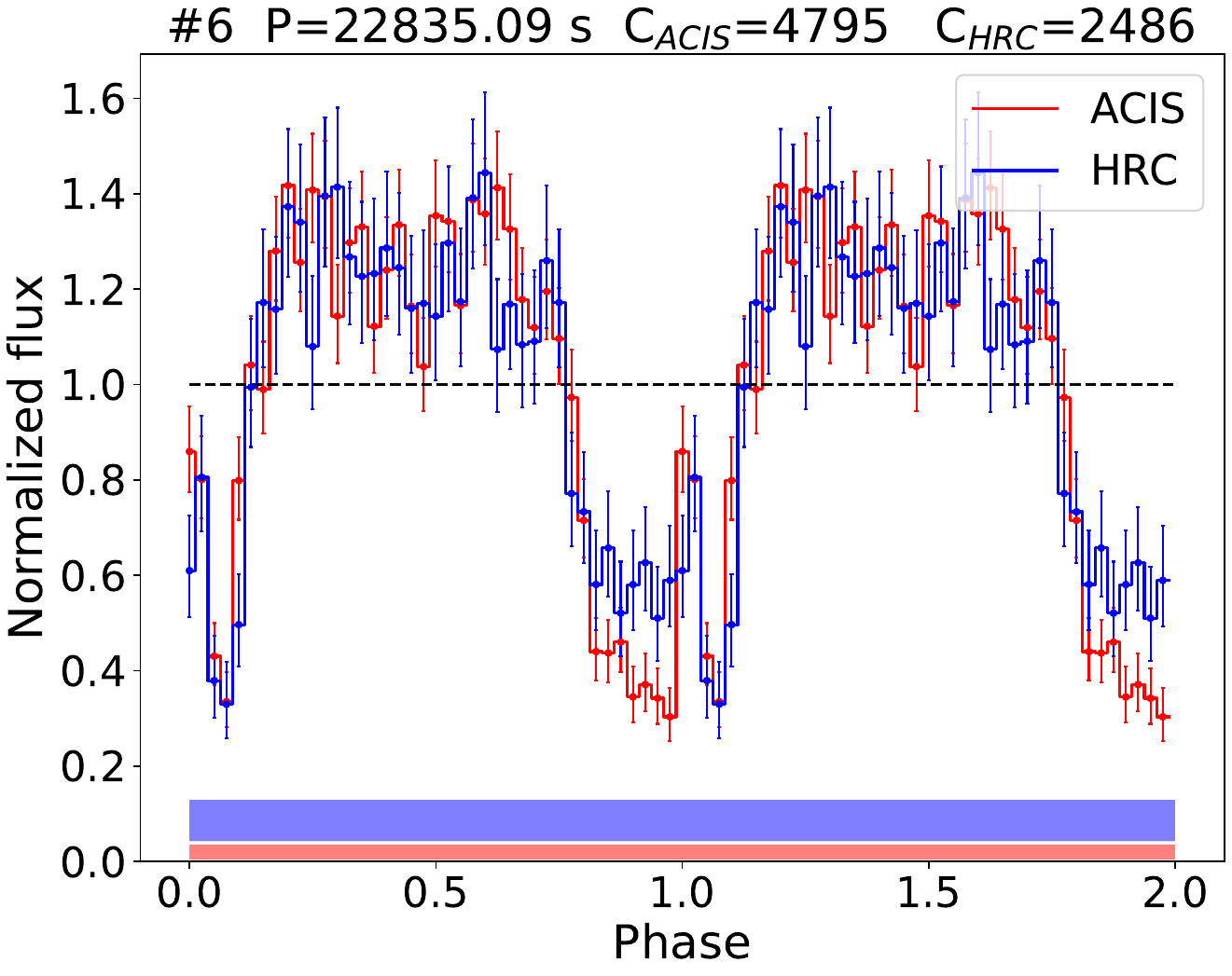}\includegraphics[width=0.345\linewidth]{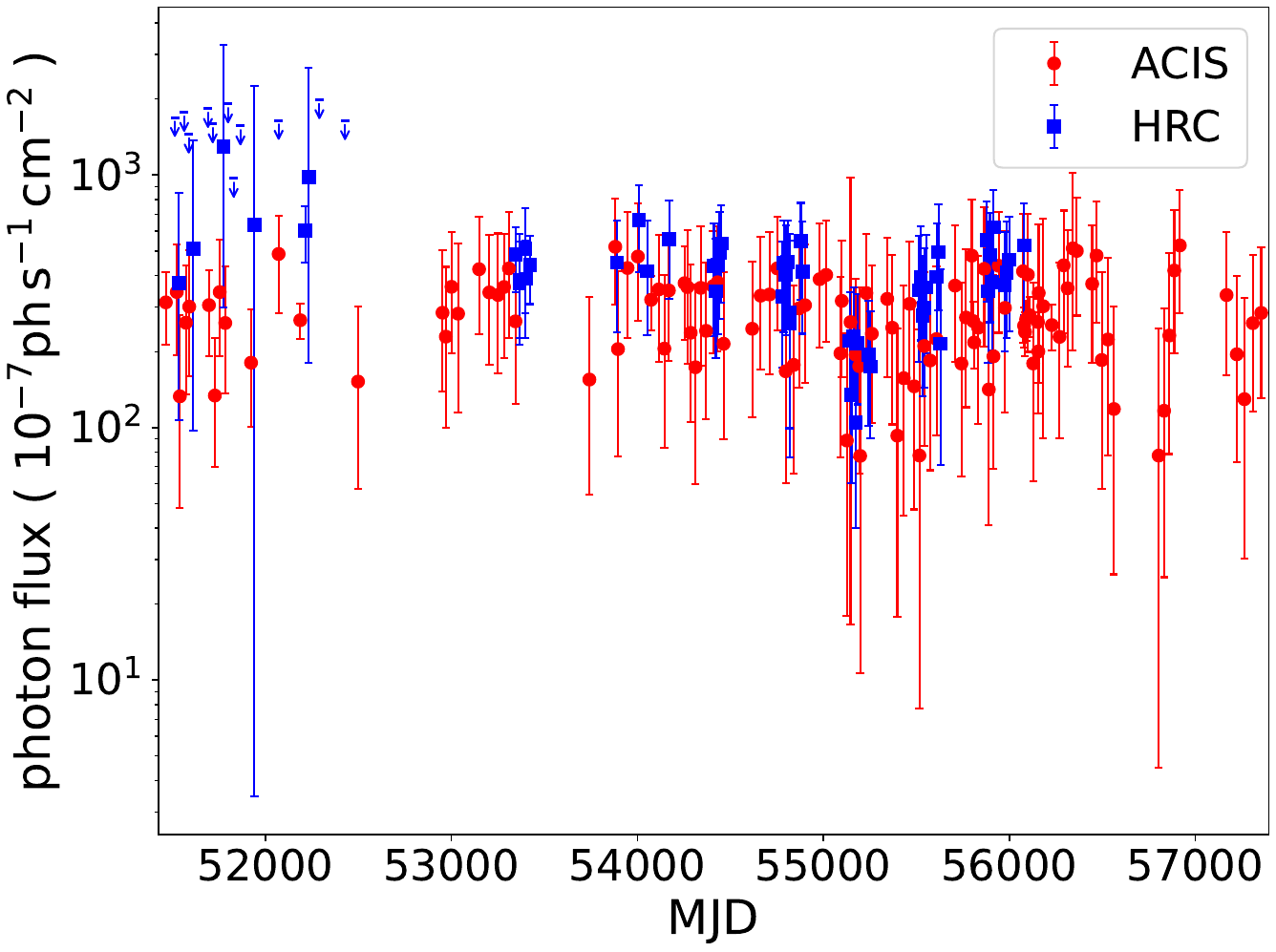}\includegraphics[width=0.35\linewidth,trim=0cm 0.3cm 3cm 2cm, clip]{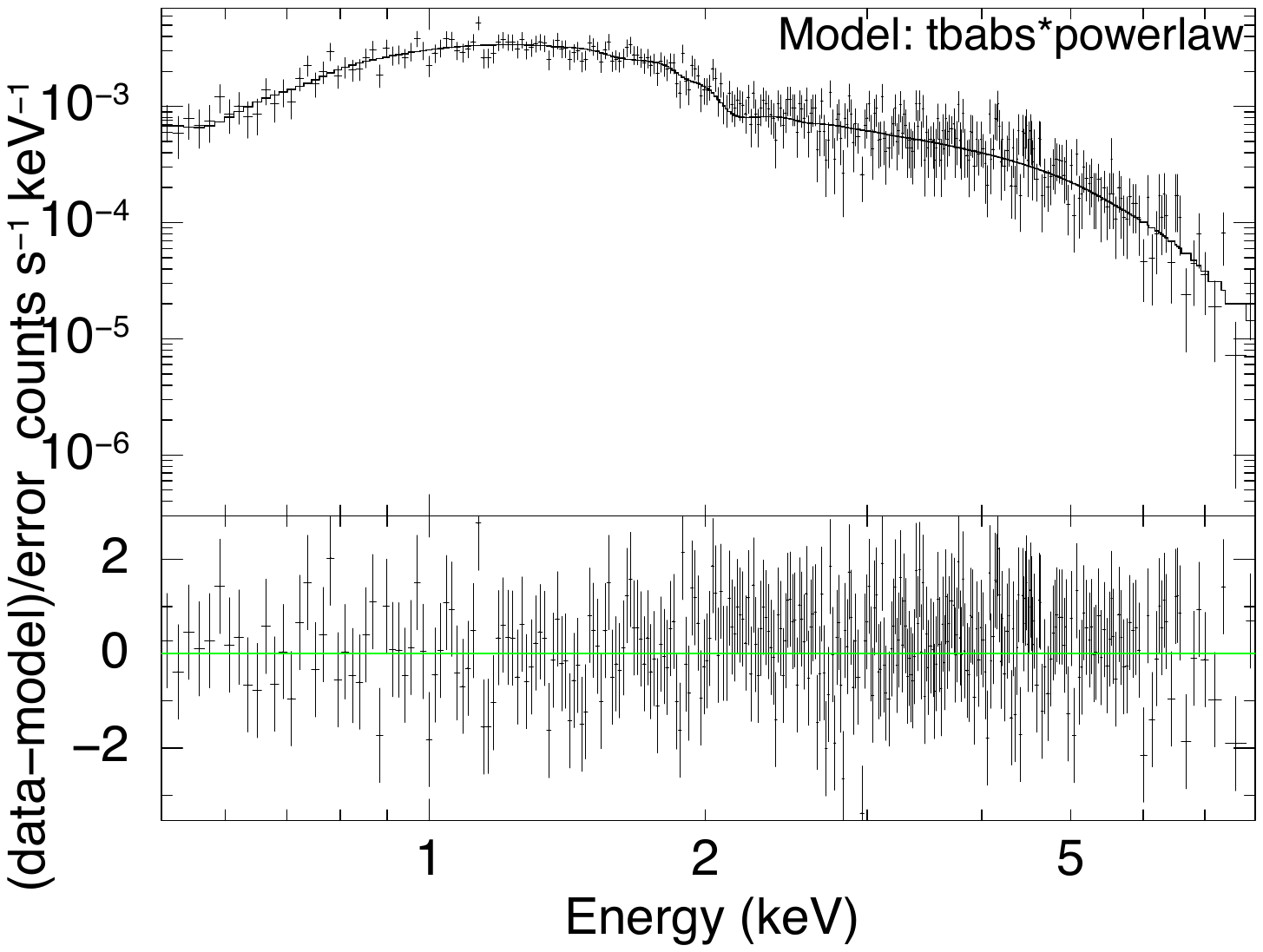}}
\subfigure{\includegraphics[width=0.34\linewidth]{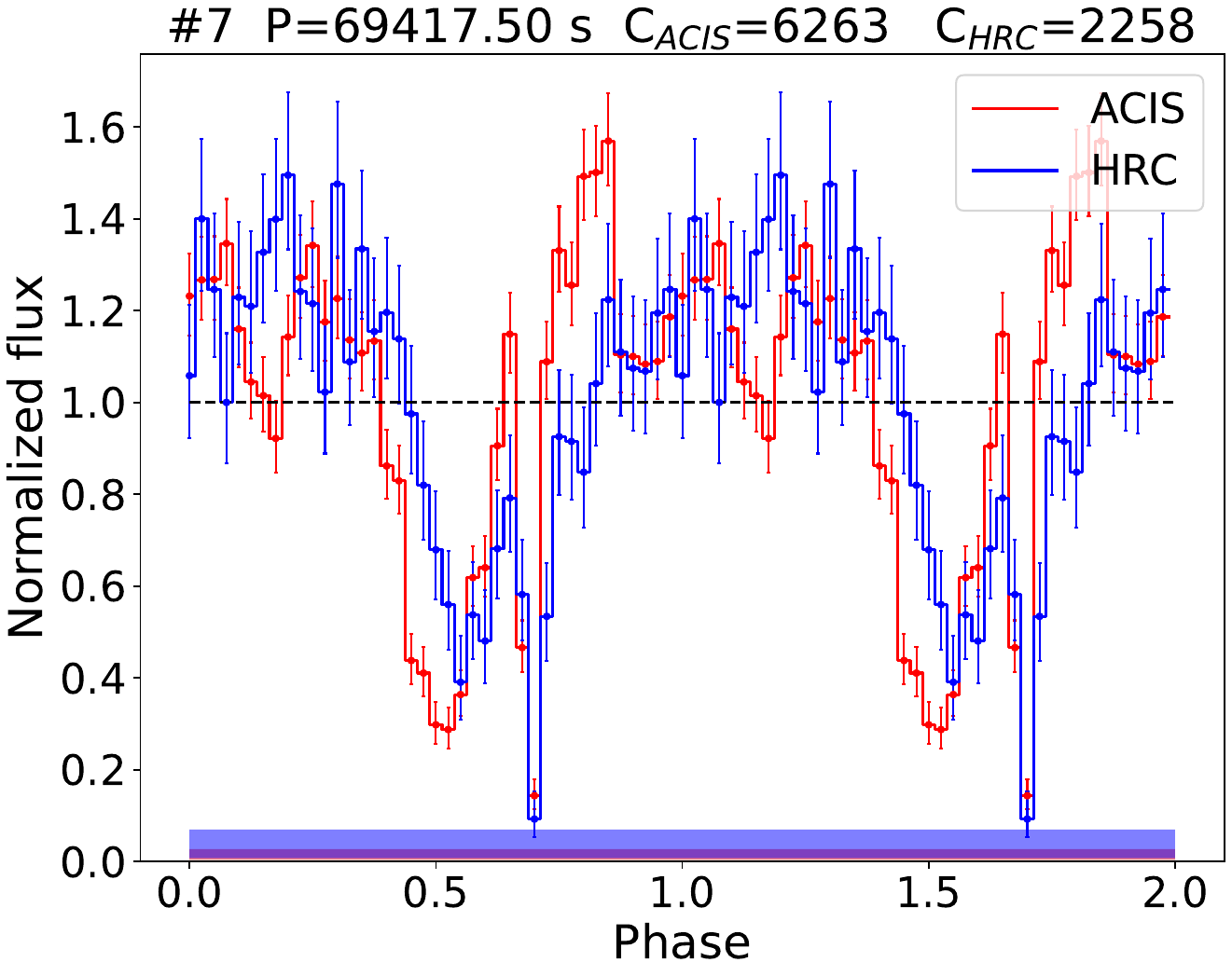}\includegraphics[width=0.345\linewidth]{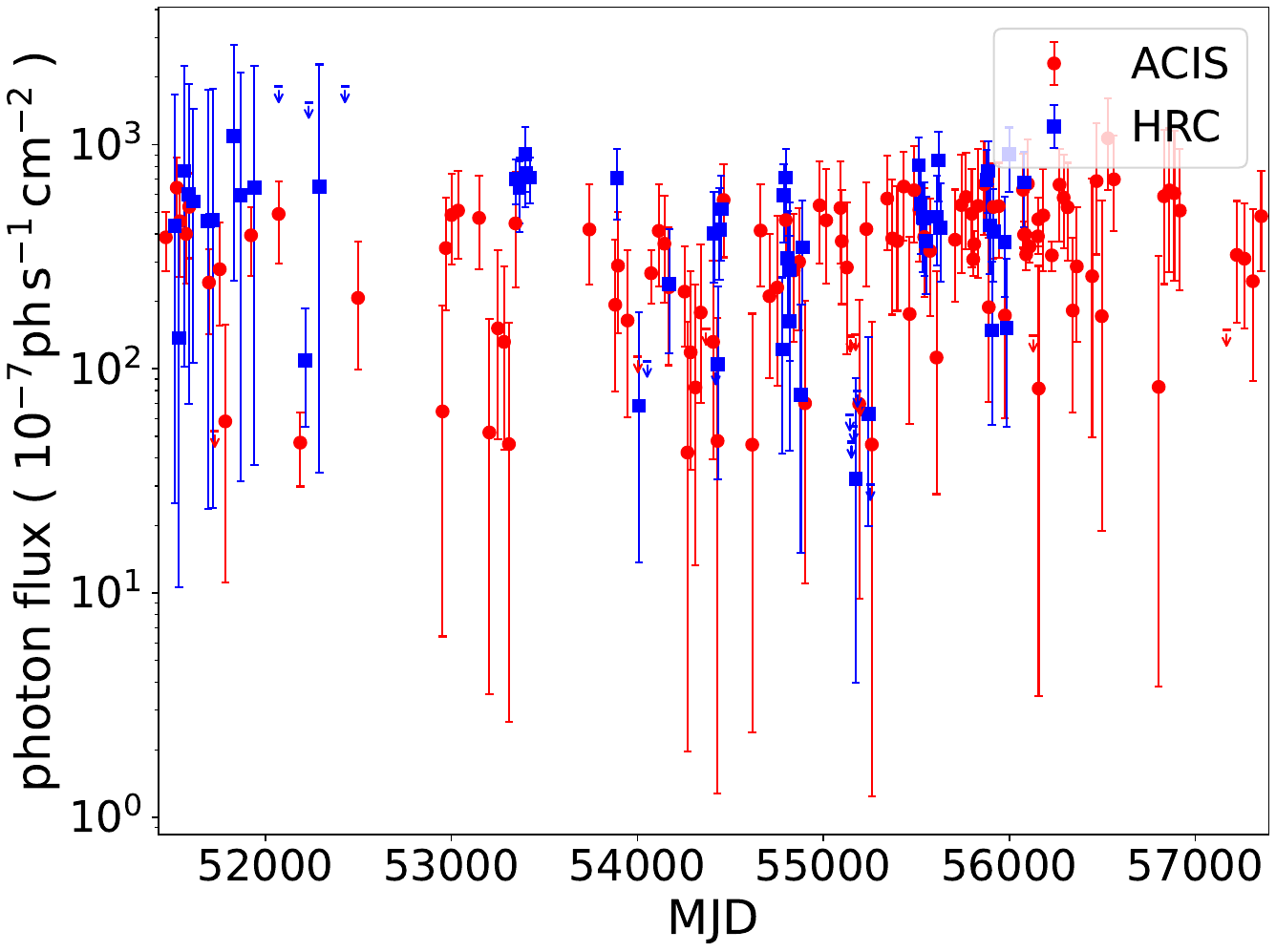}\includegraphics[width=0.35\linewidth,trim=0cm 0.3cm 3cm 2cm, clip]{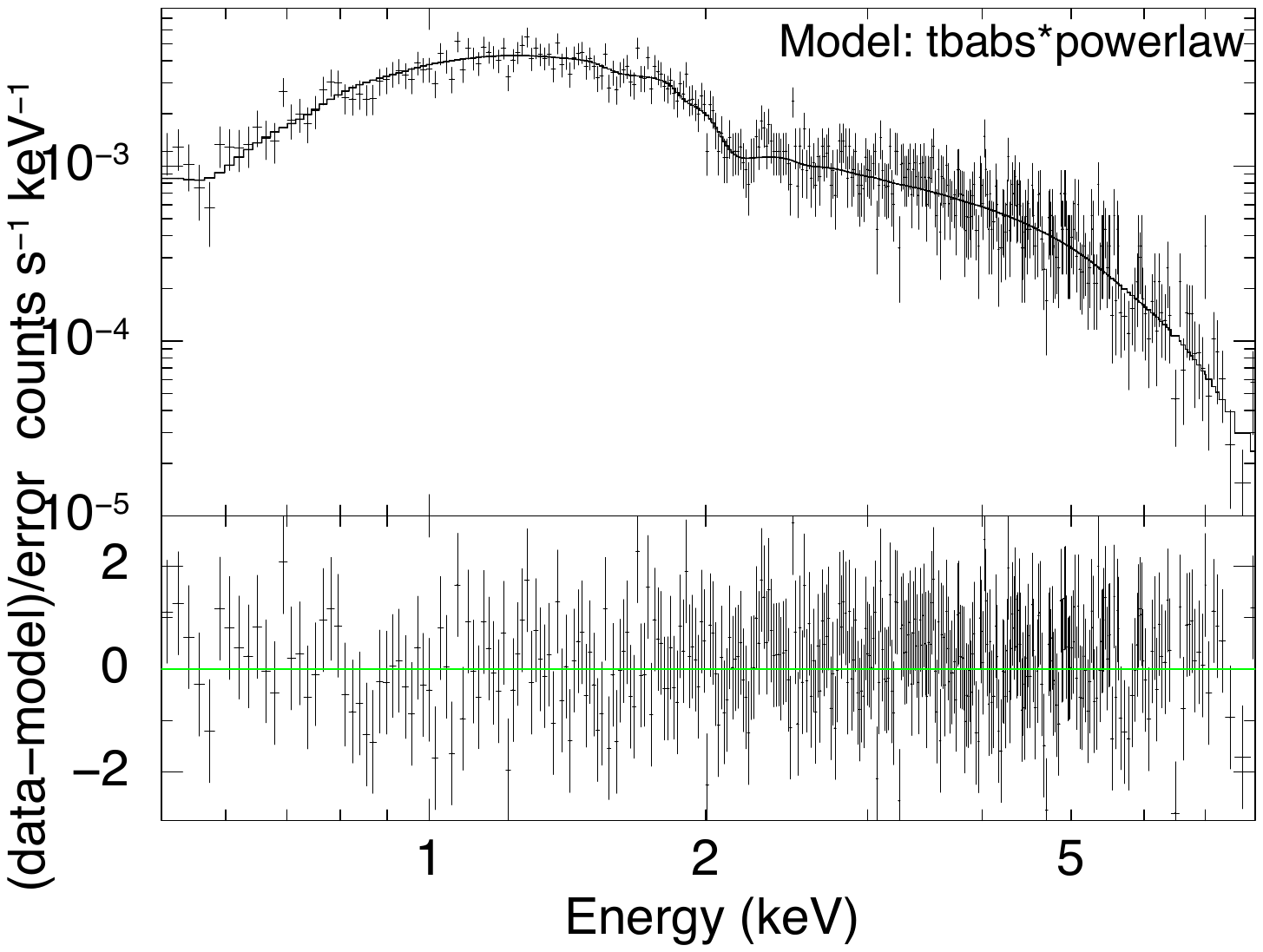}}
\caption{The phased-folded light curve (left panels), the background-subtracted inter-observation light curve (middle panels) and the cumulative ACIS spectrum with the best-fit model (right panels) of Seq.1, Seq.4, Seq.6 and Seq.7.  For each source, the ACIS light curves are shown in red, while the HRC light curves are in blue.  
The estimated local background in the phase-folded light curve is marked by a strip. 
Arrows in the inter-observation light curves represent 3$\sigma$ upper limit. 
The periodic signal of Seq.1 is only detected in five HRC and three ACIS observations, which are marked with a square. For this source, a spectrum combining the three ACIS observation is also fitted (shown in red).
The fitted spectral models are denoted.
}
\label{fig:pf}
\end{figure*}

\begin{figure*}
\centering
\subfigure{\includegraphics[width=0.33\linewidth]{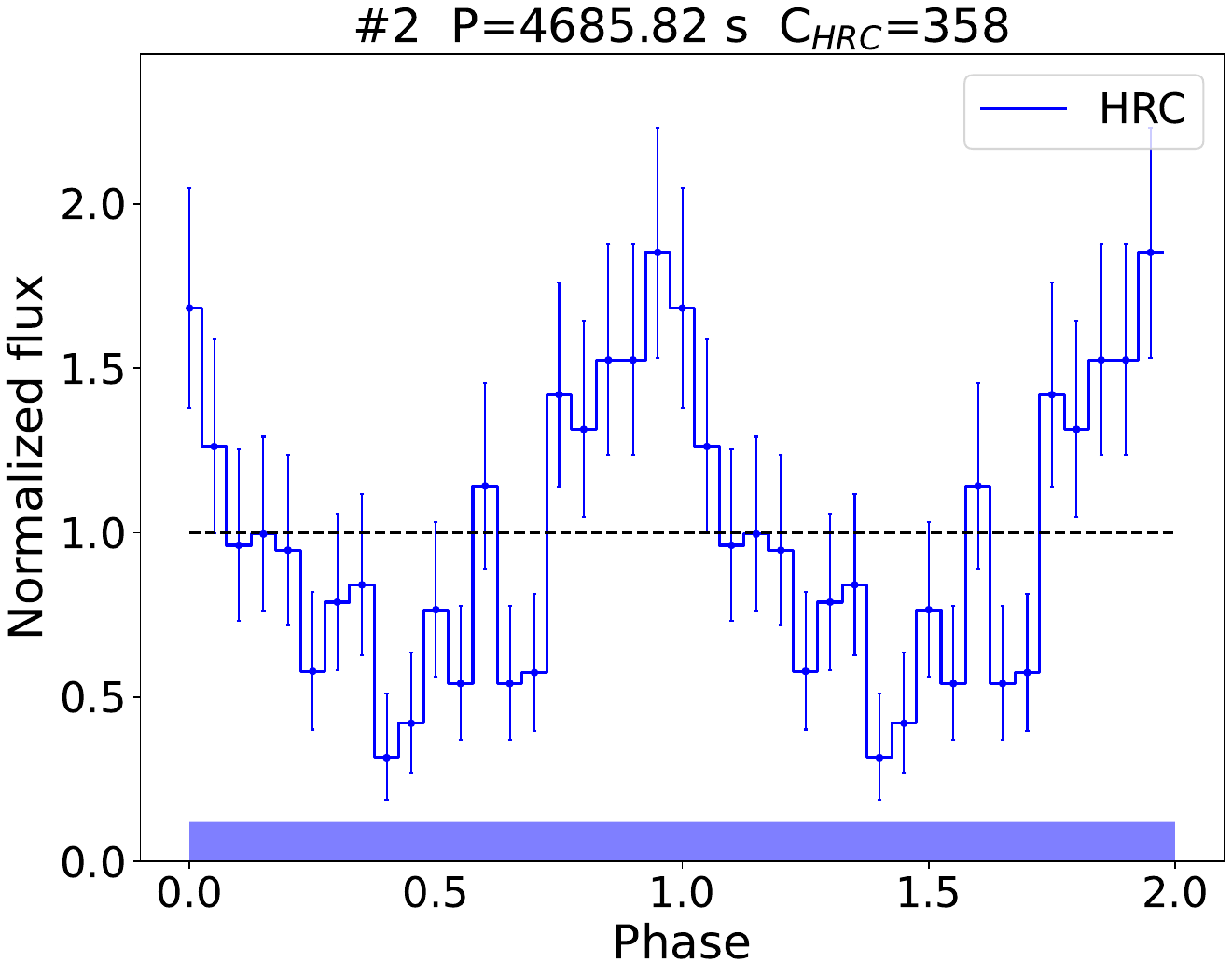}\includegraphics[width=0.33\linewidth]{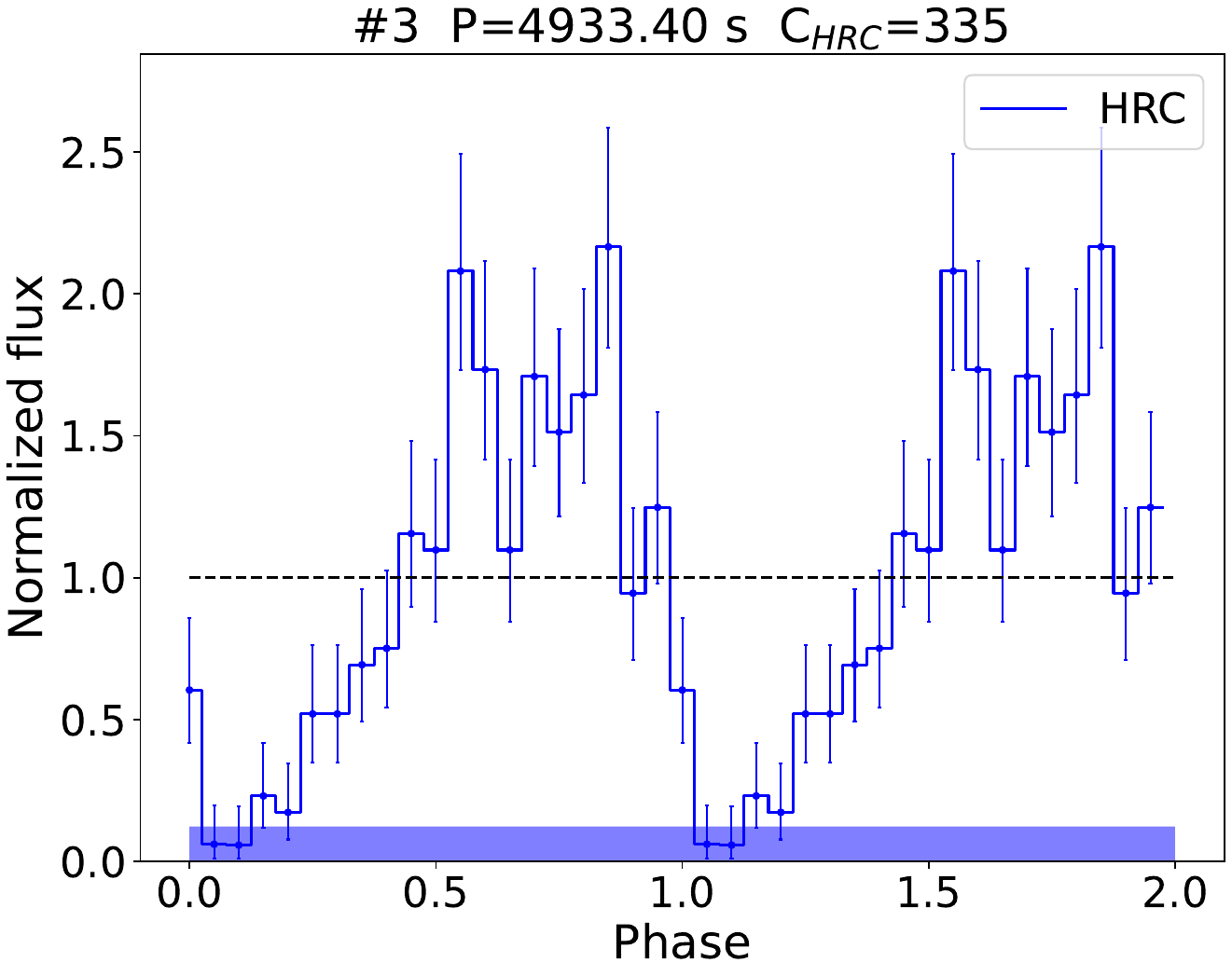}\includegraphics[width=0.33\linewidth]{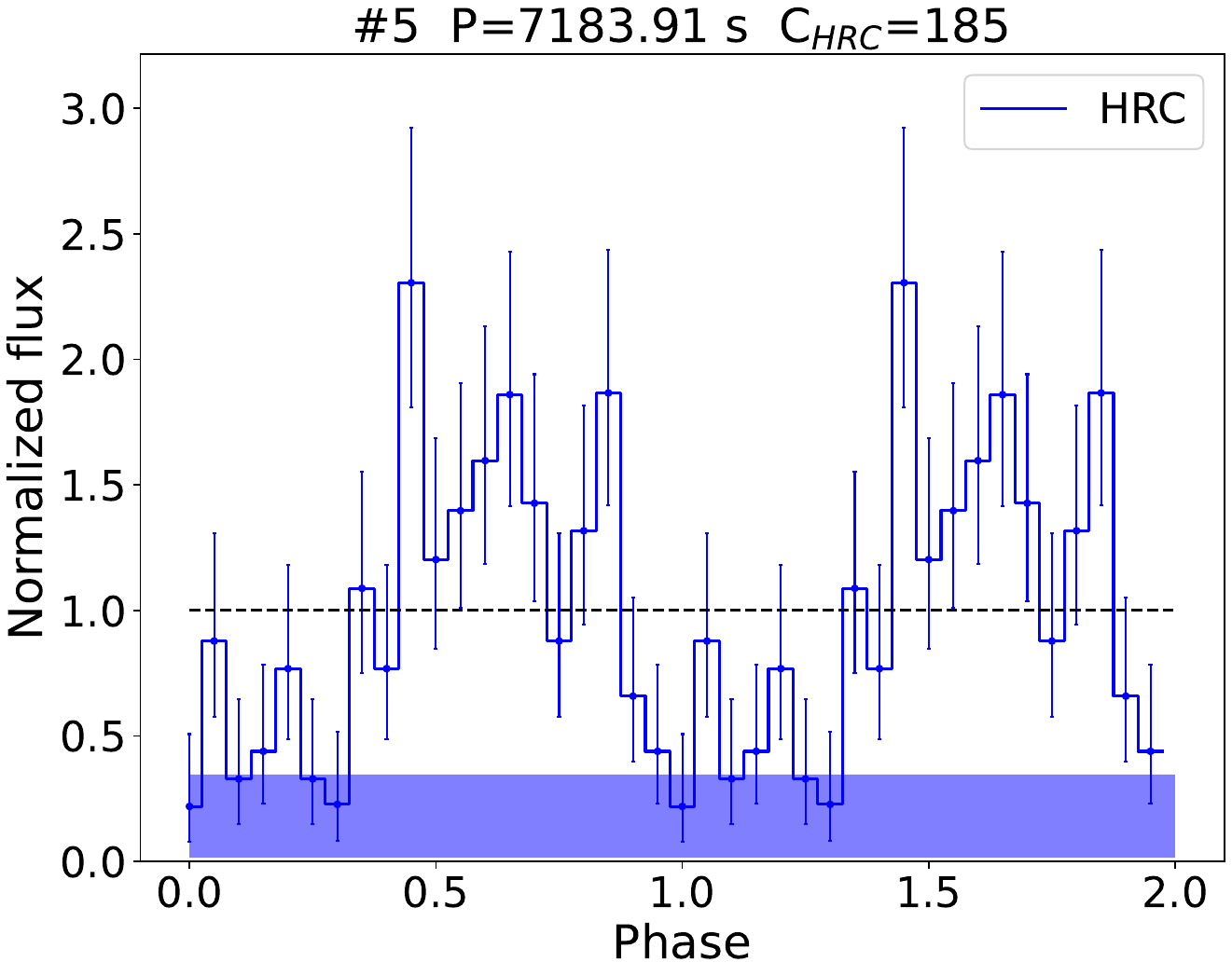}}
\captionsetup{justification=raggedright}
    \caption{Phased-folded light curves for Seq.2, Seq.3 and Seq.5, whose periodic signals are only detected in a few HRC observations during their outburst.}
    \label{fig:pf2}
\end{figure*}

Among the seven sources, three (Seq.4, Seq.6 and Seq.7 in Table~\ref{tab:result}) have a persistent periodic signal (i.e., detected over all ACIS and all HRC observations).
Seq.1, while itself being a persistent source, has its periodic signal detected in only 5 HRC observations and 3 ACIS observations (marked in the top middle panel of Figure~\ref{fig:pf}). 
For these four sources, the periods derived from the ACIS and HRC observations differ by $\sim$0.01\%--0.3\%, which can be understood as the typical uncertainty of periodic signals determined by the GL algorithm \citep{2023MNRAS.521.4257B}. 
The phase-folded light curve, inter-observation light curve and cumulative ACIS spectrum (Section~\ref{sec:spectra}) of these four sources are shown in Figure~\ref{fig:pf}. 
To phase-fold the light curve, we adopt either the ACIS-detected or HRC-detected period, whichever has the more source counts. 
A phase consistency between the ACIS and HRC light curves is clearly evident in Seq.4, Seq.6 and Seq.7. For Seq.1, due to a $\sim$0.3\% difference between the ACIS- and HRC-derived periods, a significant phase shift is seen. 
The remaining three sources (Seq.2, Seq.3 and Seq.5) have their periodic signal detected in only 2--5 observations (all are HRC) when each source was caught in an outburst. 
The phase-folded light curves of these three sources are shown in Figure~\ref{fig:pf2}.
We defer more detailed discussions of the periodic signals to Section~\ref{sec:class}.

\subsection{The effect of red noise}
\label{subsec:rednoise}
Accretion-powered systems, such as CVs and LMXBs, are known to exhibit aperiodic variability across a wide range of time scales. The so-called {\it red noise}, which is a significant component of the aperiodic variation, can potentially introduce false periodic signals, particularly at lower frequencies \citep{1989IBVS.3383....1W}. 
It is therefore instructive to estimate the possibility of false alarms among the reported signals by the GL algorithm.   

\begin{figure}
\hspace{-0.5cm}\includegraphics[width=0.5\textwidth,trim=0cm 0cm 0cm 0cm, clip]{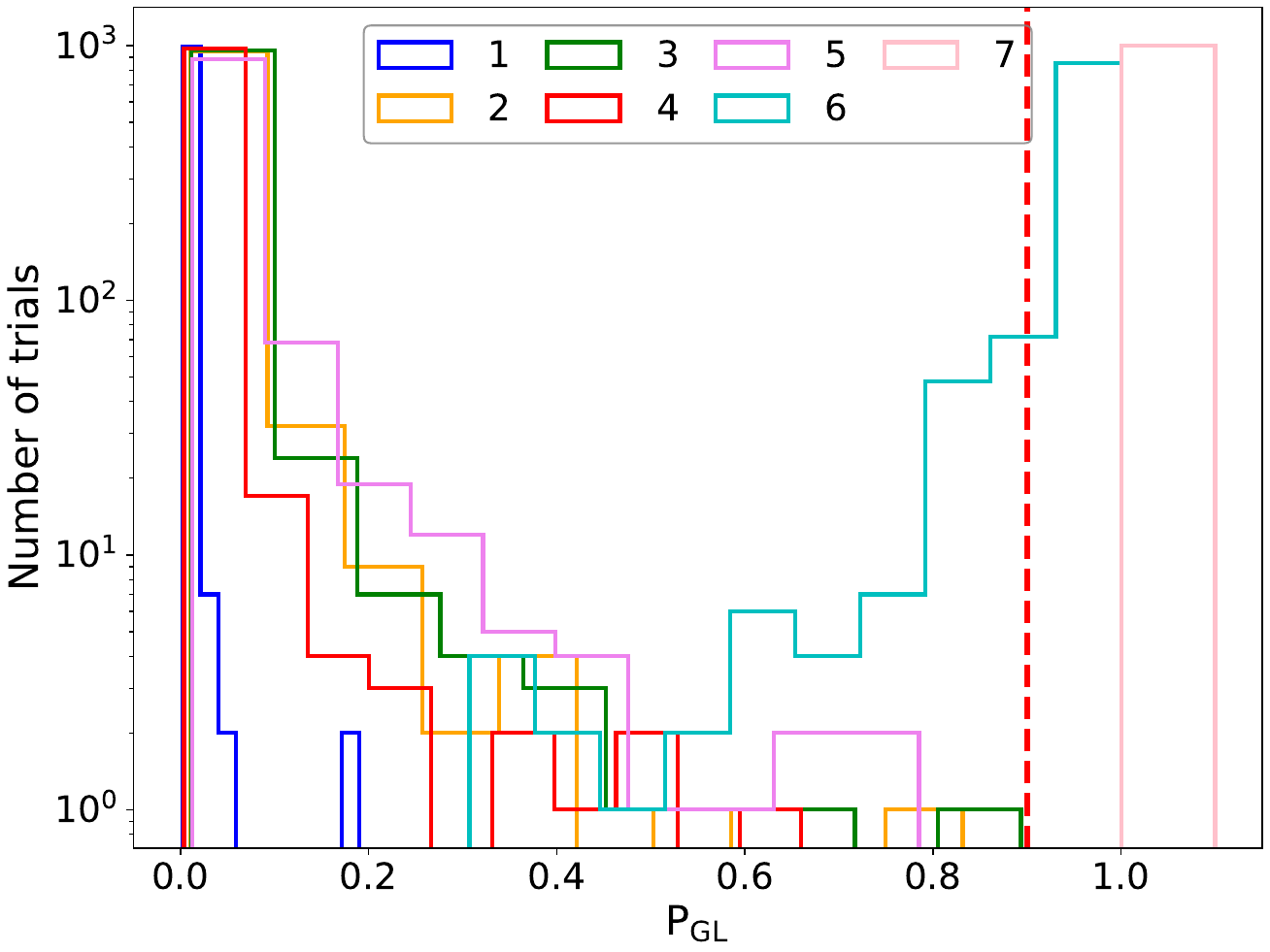}
\caption{The distribution of $P_{\rm GL}$ from 1000 simulated light curves based on the modelled red noise of each periodic source (coloured histograms).
The red dashed line marks $P_{\rm GL}=0.90$, the adopted threshold for a periodic signal to be identified.}
\label{fig:glhist}
\end{figure}

Following \citet{2023MNRAS.521.4257B},
we adopt a power-law model to describe the source power spectrum in order to account for the presence of red noise:
\begin{equation}
P(\nu)=N \nu^{-\alpha}+C_{\rm p}.
\label{eqn:LMXBps}
\end{equation}
Here $N$ is the normalization factor, $\alpha$ is the spectral index, and $C_{\rm P}$ represents the Poisson noise dictated by the mean photon flux of the source.
To mitigate the potential effect of interrupted observations in the Fourier analysis, we utilize the longest ACIS or HRC observation for each periodic source to characterize their power spectrum. 
This approach ensures a continuous and extended observation period, avoiding the influence of interruptions on the analysis.
The power spectrum of a given source is fitted with Eq.~\ref{eqn:LMXBps} using the Markov Chain Monte Carlo approach (with the python \emph{emcee} package, \citealp{2013PASP..125..306F}) to determine the best-fit parameters and errors. 
Next, following the procedure proposed by \citet{1995A&A...300..707T}, we simulate 1000 time series using the best-fit power spectrum model of each source, which are fed to the GL algorithm. 
The histogram of the resultant $P_{\rm GL}$ for each source is shown in Figure~\ref{fig:glhist}.
The fraction of false signals (i.e., those with $P_{\rm GL}$ above the default threshold of 0.9) represents the false alarm probability (FAP) for the source.

\begin{figure*}
\hspace{-0.5cm}
\centering
\noindent\makebox[\textwidth][c]{
\subfigure{\includegraphics[width=0.458\paperwidth]{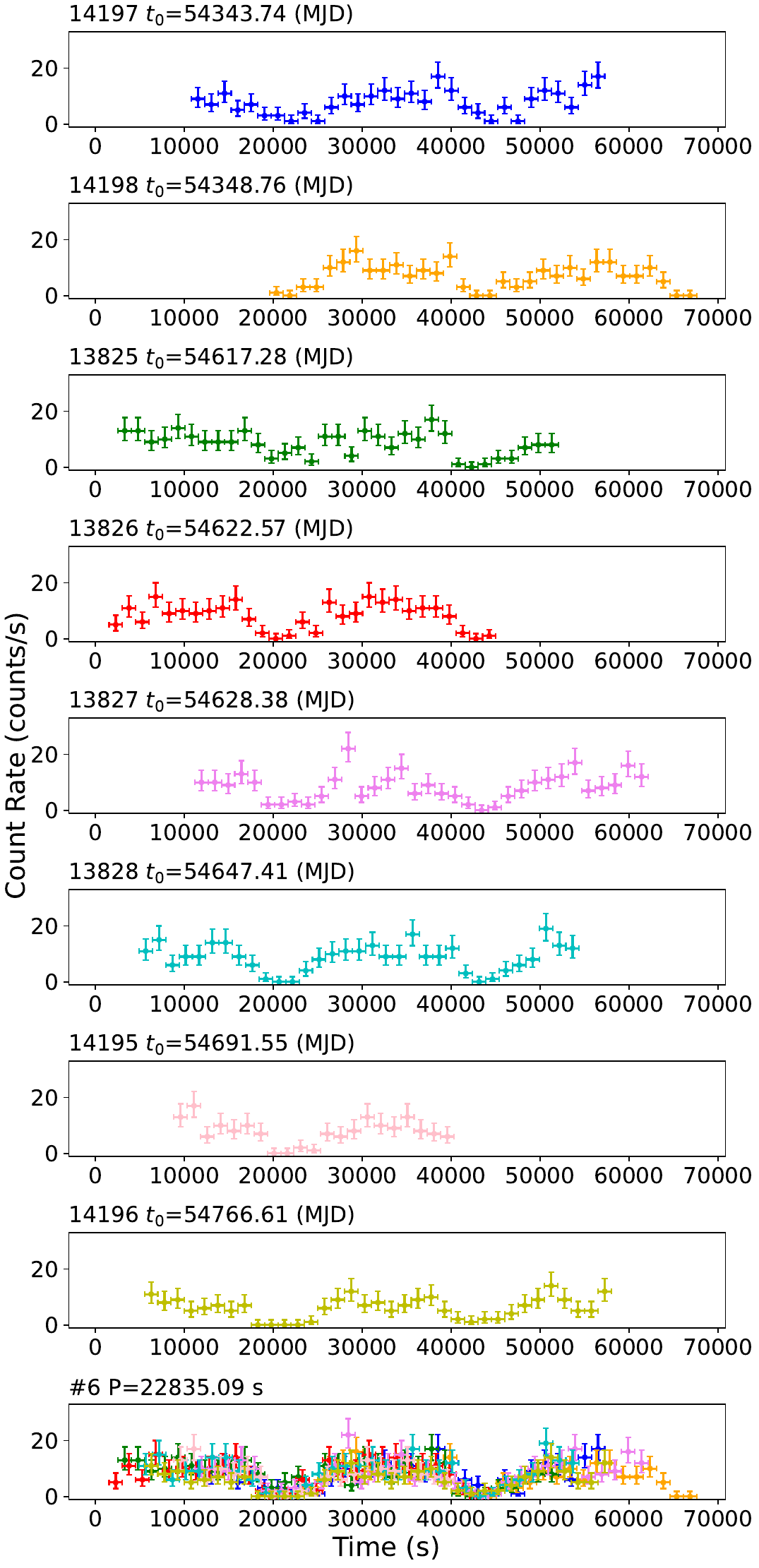}}
\subfigure{\includegraphics[width=0.450\paperwidth,trim=0cm 0cm 0cm 0.1cm, clip]{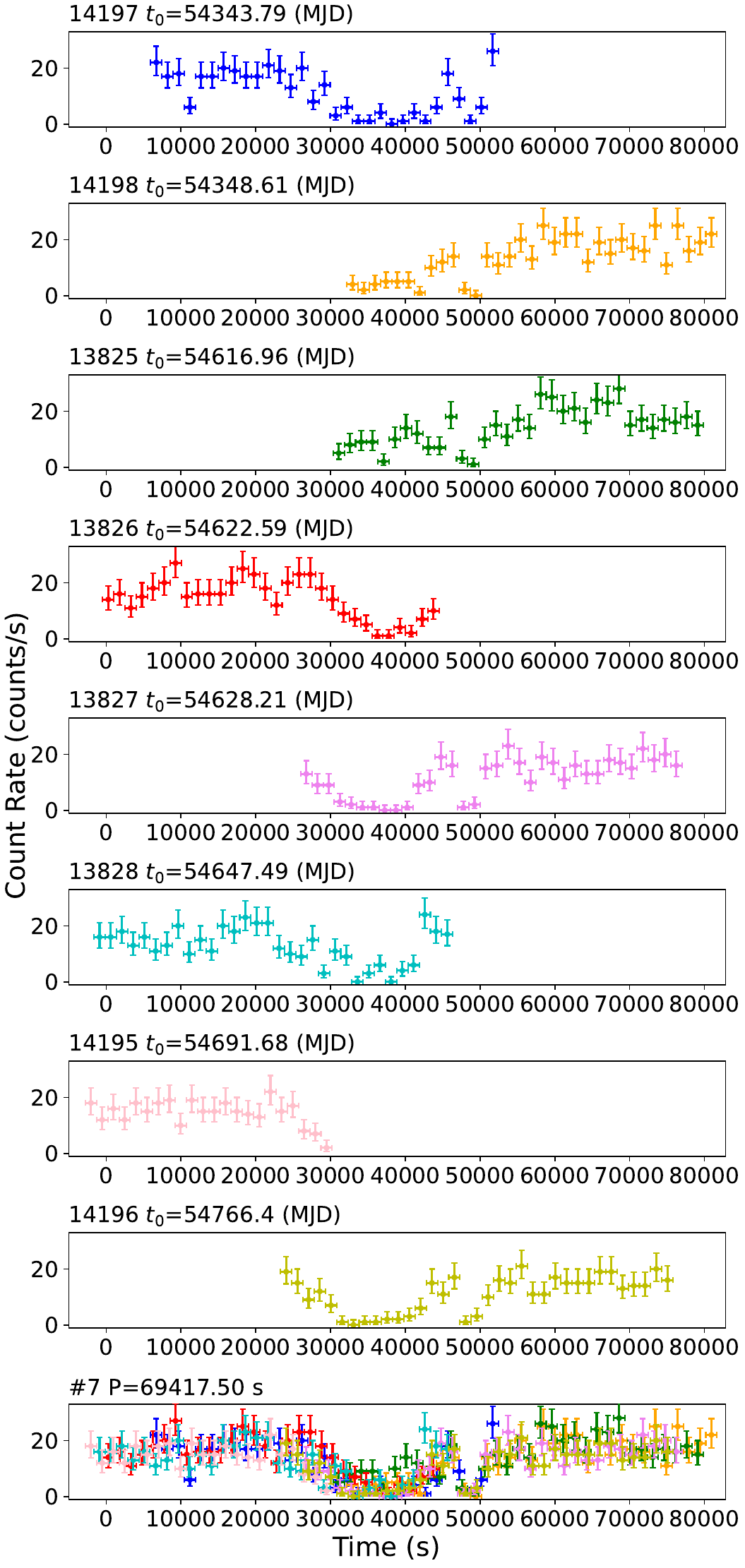}}}
    \caption{Phase-aligned light curves of 8 long observations (ObsID 14197, 14198, 13825, 13826, 13827, 13828, 14195, 14196) for Seq.6 (left panels) and Seq.7 (right panels). The top eight rows  show the light curves of each individual observation, for which the ObsID and start time of the light light curve are labeled. The botton row overlays the eight light curves, highlighting the eclipsing behavior in both sources. 
   Two eclipses are evident in Seq.7.
    }
    \label{fig:manylc}
 \end{figure*}

The majority of detected periodic sources have a FAP $< 1\%$, which can be taken as strong evidence for an intrinsic periodicity. 
The two sources with the longest reported periods, Seq.6 and Seq.7, have their FAP much greater than 1\%, which suggests a substantial false alarm probability due to red noise. However, this does not necessarily exclude their possibility of being true periodic variations, since the GL algorithm can still uncover intrinsic periodic signals in the presence of substantial red noise.
To confirm that Seq.6 and Seq.7 are genuine periodic signals, we select for both sources eight adjacent and long observations (ObsID 14197, 14198, 13825, 13826, 13827, 13828, 14195, 14196) and align the light curves from each individual observation according to the phase determined by the identified period. 
The aligned light curves are shown in Figure~\ref{fig:manylc}, which clearly reveal an eclipsing behavior in both sources. 
Furthermore, the consistency in their eclipse phases across these observations serves as compelling evidence for the accuracy of the identified period.
Therefore, we conclude that these two periodic signals are genuine, despite their large FAP inferred from the simulated red noise-based light curves.

\section{X-ray spectral analysis}\label{sec:spectra}
It is instructive to investigate the X-ray spectral properties of the periodic sources, which may provide insights into their physical nature.
For this purpose, we focus on the ACIS data, but discard the HRC data due to its poor energy resolution. 
However, three periodic sources, Seq.2, Seq.3 and Seq.5, are previously known to be novae (see Section~\ref{sec:class} for details). Their detection by {\it Chandra} occurs solely during the outburst phase, which were all through HRC observations (Table~\ref{tab:result}), and all three sources fall below the detection limit in the ACIS observations, precluding the availability of a meaningful spectrum.
Hence, we estimate their peak  0.5--8 keV luminosity during outburst from their net photon flux, by assuming an absorbed blackbody  spectrum with a column density of $N_{\rm H}=1.9 \times 10^{21}\ \rm cm^{-2}$ and a black-body temperature of $kT=40\rm~eV$, taken from the spectral model of Seq.2 determined by XMM-Newton observations \citep{2014A&A...563A...2H}.

For Seq.1, Seq.4,Seq.6 and Seq.7, we examine the source spectrum combining all available ACIS observations. For Seq.1, we further examine the spectrum combining only the three ACIS observations over which the periodic signal was detected.
We extract the source and background spectra for each periodic source by utilizing the CIAO tool \emph{specextract} from the aperture defined in Section~\ref{sec:xdata}.
For each source, we then generate the combined spectra by CIAO tool {\it combine\_spectra}, with the corresponding ARFs and RMFs weighted by the effective exposure. 
The spectra are analyzed using XSPEC v12.12.1, after adaptively binned over 0.5--8 keV such that a minimum of 20 counts and a S/N greater than 2 per bin are achieved.

To fit the spectra, we typically employ an absorbed power-law model, and when necessary, introduce an additional soft component, either {\it diskbb} or {\it bbody} in XSPEC.  The best-fit model parameters are given in Table \ref{tab:spec}.
The unabsorbed 0.5--8 keV luminosity is estimated according to the best-fit model and an assumed distance of 785 kpc for M31 \citep{2005MNRAS.356..979M}.

\begin{table*}
\renewcommand\arraystretch{1.5}
\caption{Spectral fit results.}\label{tab:spec}
\begin{threeparttable}
\begin{tabular}{lccccccccc}
\hline
\hline
ID	& Model	&$\chi ^2$/dof & $L_{\rm X}$\tnote{1}	& $N_{\rm H}$ 	& $\Gamma$	& $kT$
\\
	&		& 			&$10^{36}\ \rm{erg\ s^{-1}}$	&$10^{20}\ \rm{cm^{-2}}$	& 		& keV  \\ 
(1) & (2) & (3) & (4) & (5) & (6) \\
\hline
1   &	tbabs*(powerlaw+diskbb)	&609.95/502	&	$109_{-1}^{+2}$&$8_{-1}^{+2}$	&$1.5_{-0.1}^{+0.2}$	&$1.4_{-0.2}^{+0.3}$				\\
1$^{*}$  &tbabs*powerlaw           &192.25/181 &   $108_{-8}^{+14}$    &$7(<14)$       &$1.8_{-0.1}^{+0.2}$   &\\
          
$2^{\dag}$ &tbabs*bbody	&\/			&	$7_{-1}^{+1}$ 	&$19.0$	    &\/								&$0.04$\\
$3^{\dag}$ &tbabs*bbody	&			&	$3.4_{-0.9}^{+0.9}$	&$19.0$     &\/								&$0.04$\\
4  &	tbabs*(powerlaw+bbody)	&401.96/399	&	$15_{-3}^{+5}$ &$47_{-14}^{+14}$	&$3.9_{-0.8}^{+0.8}$	&$1.5_{-0.1}^{+0.1}$				\\
$5^{\dag}$ &tbabs*bbody	&\/			&	$1.5_{-0.4}^{+0.4}$	&$19.0$     &\/					&$0.04$\\
6  &	tbabs*powerlaw			&279.63/320	&	$5.9_{-0.2}^{+0.2}$ 	&$17_{-3}^{+3}$	&$1.85_{-0.08}^{+0.08}$	&				\\
7	&tbabs*powerlaw	&285.68/359	&$7.1_{-0.2}^{+0.2}$		&$20_{-3}^{+3}$		&$1.72_{-0.07}^{+0.07}$\\
\hline
\end{tabular}
\begin{tablenotes}    
\small \item
Notes: 
(1) Source sequence number.
(2) The best-fit model in XSPEC.
(3) $\chi^2$ and degree of freedom of the best-fit model
(4) 0.5--8 keV unabsorbed luminosity. 
*: Spectrum combining only three ACIS observations over which the periodic signal was detected. $\dag$: Peak luminosity during the outburst, which is converted from the observed photon flux in the HRC observations. 
(5) Line-of-sight absorption column density.
(6) Photon index of the power-law model.
(7) Temperature of diskbb or bbody model. Quoted errors are at the 90\% confidence level.\\
\end{tablenotes}         
\end{threeparttable}
\end{table*}

\section{Classifying the periodic X-ray sources}
\label{sec:class}

Based on the timing and spectral properties of the periodic sources, as well as information from the literature, we attempt to classify them in order below. 
We note in passing that, while a non-negligible fraction of all X-ray sources detected in the direction of the M31 bulge arise from the CXB ($\sim$0.1 per arcmin$^{-2}$; \citealp{2007A&A...468...49V}), our recent study based on the {\it Chandra} Deep Field South \citep{2022MNRAS.509.3504B} suggests that the CXB has a negligible contribution to the periodic signals found here. 

{\bf Seq.1:} This source is the most luminous one among the seven periodic sources, naturally making it an active LMXB. 
It has been suggested that this source is a BH-XRB candidate \citep{2013ApJ...770..148B}, 
based on an argument that its X-ray spectrum at the high hard state is distinctively softer than that of NS-XRBs. 
We detected its periodic signal, which exhibits an eclipsing or dipping behavior during its high state between 2010--2011. 
The presence of eclipsing/dipping signatures strongly suggests that the observed signal originates from the binary orbital motion.
That this signal is only present in a small fraction of observations is somewhat puzzling, especially in the scope of an orbital modulation. 
Nevertheless, this signal should be genuine, given the fact that it is detected in both HRC and ASIC observations, disfavoring an instrumental artifact.
We speculate that this eclipse/dipping is due to some absorbing structure in the accretion disk, e.g., an accretion disk corona  \citep[ADC,][]{1982ApJ...257..318W,2014RAA....14.1367C}. In this scenario, the central source of the XRB system evaporates material from the surface of its accretion disk.
If the evaporated material does not escape from the system, it will corotate above and below the disk as a corona and can partly absorb or scatter the X-rays. This may happen even when the viewing angle of the binary system is relatively small. Such an ADC is likely a transient structure depending on the level of the central illuminating source and could be consistent with what is observed in Seq.1.
However, we find no significant evidence for an enhanced absorption in the X-ray spectrum of Seq.1 at the epochs when the periodic signal was detected (Figure~\ref{fig:pf} and Table~\ref{tab:spec}).

Regardless of the nature of the transient dipping, the short period of $\sim$463 sec, if truly an orbital period, would make Seq.1 a candidate of the shortest-period ultra-compact X-ray binary (UCXB) observed so far.
Such an exceptionally tight orbit also indicates that Seq.1 has the potential of becoming an extragalactic gravitational wave source for future detectors. According to \cite{2023ApJ...953..153H},  at least 0.8 merging BH-WD binaries and $\sim$0.1--0.4 merging NS-WD binaries are expected to be detected in M31 during a 4-yr LISA mission. The gravitational wave of a NS-WD or BH-WD system with $ P_{\rm orb}=463.24$ sec is about $10^3$ times below the LISA sensitivity\citep{2021arXiv210801167B}. Next generation gravitational wave detectors are eagerly awaited to detect extragalactic UCXBs. 

{\bf Seq.2, Seq.3 and Seq.5:} All three sources have been previously confirmed as novae due to their intense outbursts.
During the nova outburst, the X-ray luminosity may rise to $\gtrsim10^{36}\rm~erg~s^{-1}$ (Table~\ref{tab:spec}), temporarily offering an X-ray light curve sufficient for probing the periodicity.  

Seq.2, also known as M31N 2011-01b, displayed an optical counterpart identified as PNV J00423907+4113258 \citep{2014A&A...563A...2H}. In ObsIDs 13179 and 13180, when the nova burst occurred, a periodic signal with a period of $\sim$4686 sec is detected. 
Seq.3 (M31N 2011-11e) with an optical counterpart PNV J00423831+4116313 exhibited a periodic variation of $P\sim1.3\pm 0.1 \ \rm h$ during its outburst \citep{2014A&A...563A...2H}. By utilizing the same HRC observations in 2012 (ObsID 13278,13279,13280), we detect a nearly identical period of 4933.40 sec.
Seq.5 was identified as the nova M31N 2004-11b \citep{2010A&A...523A..89H}. We detect a periodic signal at $\sim$ 7184 sec during an outburst in 2005 caught by ObsID 5927 and 5928. 
Judging from the phase-folded light curves (Figure~\ref{fig:pf2}), an eclipsing/dipping behavior is seen in at least two sources, Seq.3 and Seq.5, suggesting an orbital modulation.
Notably, these periods fall below the so-called {\it orbital period gap (2--3 h)} of CVs \citep{2011ApJS..194...28K}. This might be partly due to a selection effect given the relatively short exposures of the individual {\it Chandra} observations.

{\bf Seq.4:}
This source has a period of $\sim 6432.94$ sec, with an eclipsing behavior strongly suggesting an orbital modulation of a binary system. An orbital period of 6420 sec was previously discovered 
by {\it XMM-Newton} observations in 2000--2002 and {\it Chandra} observations in 2001 \citep{2004A&A...419.1045M}. Here we detect this period consistently in all ACIS and HRC observations from 1999 to 2015. 
\cite{2018ApJ...862...28L} argued that this source is an HMXB, based on the tentative identification of an optical counterpart exhibiting a B-type star multi-band spectral energy distribution.
However, given the short orbital period identified here, we argue that this source is more likely an LMXB. In this regard, the previously suggested optical counterpart was probably a mismatch.
The X-ray spectrum of this source is best-fitted with a power-law plus a blackbody, consistent with an LMXB. 

{\bf Seq.6:}
This source exhibits a period of $\sim$ 22835 sec in both the ACIS and HRC observations. 
The eclipsing behavior observed in its light curve strongly supports the notion that the period is from the binary orbital modulation. 
In both the ACIS and HRC phase-folded light curves, there appears a small spike (at phase $\sim1.0$) in the otherwise broad eclipse (Figure~\ref{fig:pf}), which starts around phase $\sim$ 0.75 and ends at phase $\sim$ 0.2. 
However, such a feature is not clearly evident in the unfolded light curves shown in Figure~\ref{fig:manylc}, which casts doubt to its reality. 
The cumulative X-ray spectrum is well described by an absorbed power-law model, with $\Gamma \approx 1.85$ and $L_{\rm X}\approx 6\times 10^{36}\rm~erg~s^{-1}$.
The combination of its periodicity and spectral characteristics is well consistent with an LMXB. 

{\bf Seq.7:}
This source exhibits a highly distinctive light curve characterized by two eclipses, a wider one (phase between $\sim$ 0.4--0.6) followed by a narrow one (centering at phase $\sim$ 0.7), consistently observed in the ACIS and HRC observations. 
This dual-eclipse feature is most likely real, because it is directly seen in the unfolded light curves (Figure~\ref{fig:manylc}), and the narrower eclipse is almost a full eclipse. 
The cumulative X-ray spectrum can be well described by a simple power-law, with $\Gamma \approx 1.72$ and $L_{\rm X} \approx 7\times 10^{36}\rm~erg~s^{-1}$.
The projected position of this source is closest to the center of M31, raising the possibility that it is an LMXB formed in dynamical encounters in the crowed nuclear region (a projected radius $\lesssim 1\arcmin$; \citealp{2007MNRAS.380.1685V}).

To our knowledge, a dual-eclipse feature is rarely seen in LMXBs. 
GRS 1747-312, an LMXB residing in the Galactic globular cluster Terzan\,6, is known to exhibit eclipses/dips at different orbital phases \citep{2024MNRAS.529..245P}, but these peculiar features are not persistent, unlike the case of Seq.7, in which the dual-eclipse remains highly stable throughout the {\it Chandra} observations spanning 16 years.
We propose two possible explanations for the behavior of Seq.7. 
The first explanation involves absorption by an ADC and a nearly edge-on viewing angle of the orbit. Two eclipses will appear in the light curve:a narrow eclipse by the donor star and a wide eclipse by the ADC.
A couple of examples of this kind have been discovered in Galactic XRBs (e.g., 4U2129+47, 4U1822-37,  \citealp{1982ApJ...257..318W}).
The second explanation involves a three-body system, which was also suggested for the case of GRS 1747-312 \citep{2024MNRAS.529..245P}.
In this scenario, the XRB system consists of three objects: a compact object (NS or BH), a donor star, and
a third object (a normal star or a degenerate star), likely sharing the same orbit of the donor but with a quasi-stable phase difference of 1/6 orbit (similar to what is observed in Seq.7).  The donor star transfers mass to the compact
object, producing the bulk of the X-rays, which is obscured by the donor at a nearly edge-on viewing angle. 
The presence of the third object can introduce additional obscuration, either by itself or by a mini accretion disk around it. A more quantitative test of this picture is beyond the scope of the present study.

\section{Discussion}\label{sec:discusstion}

\subsection{Comparison with previous works}\label{subsec:compare}
We have detected seven periodic signals from seven X-ray sources in the M31 bulge, among which four are newly discovered. 
On the other hand, several periodic X-ray sources identified in previous studies, as outlined in Section~\ref{sec:intro}, are not recovered in this work, for various reasons.

XMMU J004252.5+411540 has been identified as the brightest persistent SSS in M31. The {\it XMM-Newton} observations revealed a periodic modulation in it with a period of 217.7 sec. Our analysis using the {\it Chandra} observations indicates that only one HRC observation displayed similar periodic variation, but with a low GL probability ($P_{\rm GL}\sim 0.5$, ObsID=9829). This discrepancy might be attributed to the lower effective area of {\it Chandra} compared to {\it XMM-Newton}, particularly in the energy range of 0.2--0.5 keV over which the SSS is primarily detected. 
XMMU J004319.4+411758 is also an SSS with a period of $P \sim$ 865 sec detected by {\it XMM-Newton}. 
No point source is detected at the same position by the {\it Chandra} observations, probably due to its super-soft spectrum and the transient nature.
M31N 2006-04a and M31N 2007-12b are both novae, with their period and outburst detected by \cite{2010A&A...523A..89H,2011A&A...531A..22P}. We did not detect their periodicity as they were not in an outburst state during all {\it Chandra} observations.
XMMU J004314.1+410724 is identified as an LMXB within the globular cluster Bo158, with a period of 2.78 hr detected by \citet{2002ApJ...581L..27T}. However, it falls near the edge of the field-of-view of the {\it Chandra} observations with a poor PSF, which is not included in our period research.
3XMM J004232.1+411314 has been detected with periodic dipping signals of 4.02 hr only during a low-luminosity state ($L_{\rm 0.2-12~keV}< 10^{38}\rm~erg~s^{-1}$) by \citet{2017ApJ...851L..27M}. We detected the same period from {\it Chandra} data only in one long observation during a low-luminosity state, but with a low GL probability ($P_{\rm GL}\sim 0.72$, ObsID=14196).
3XMM J004301.4+413017 has been designated as an NS-XRB owing to the presence of a pulsation of 1.2 sec, a timescale shorter than the ACIS frame time. Furthermore, its orbital period of 1.2 days lies beyond our detection limit. 

To summarize, in the bulge of M31, there have been eight periodic X-ray signals reported in the literature, which are not recovered by our systematic search with the {\it Chandra} data, due primarily to a mismatch of the data capability. These include 2 SSSs, 2 novae and 3 LMXBs (XMMU J004314.1+410724, 3XMM J004232.1+411314, 3XMM J004301.4+413017).

\begin{figure}
\hspace{-0.5cm}
\includegraphics[width=0.5\textwidth,]{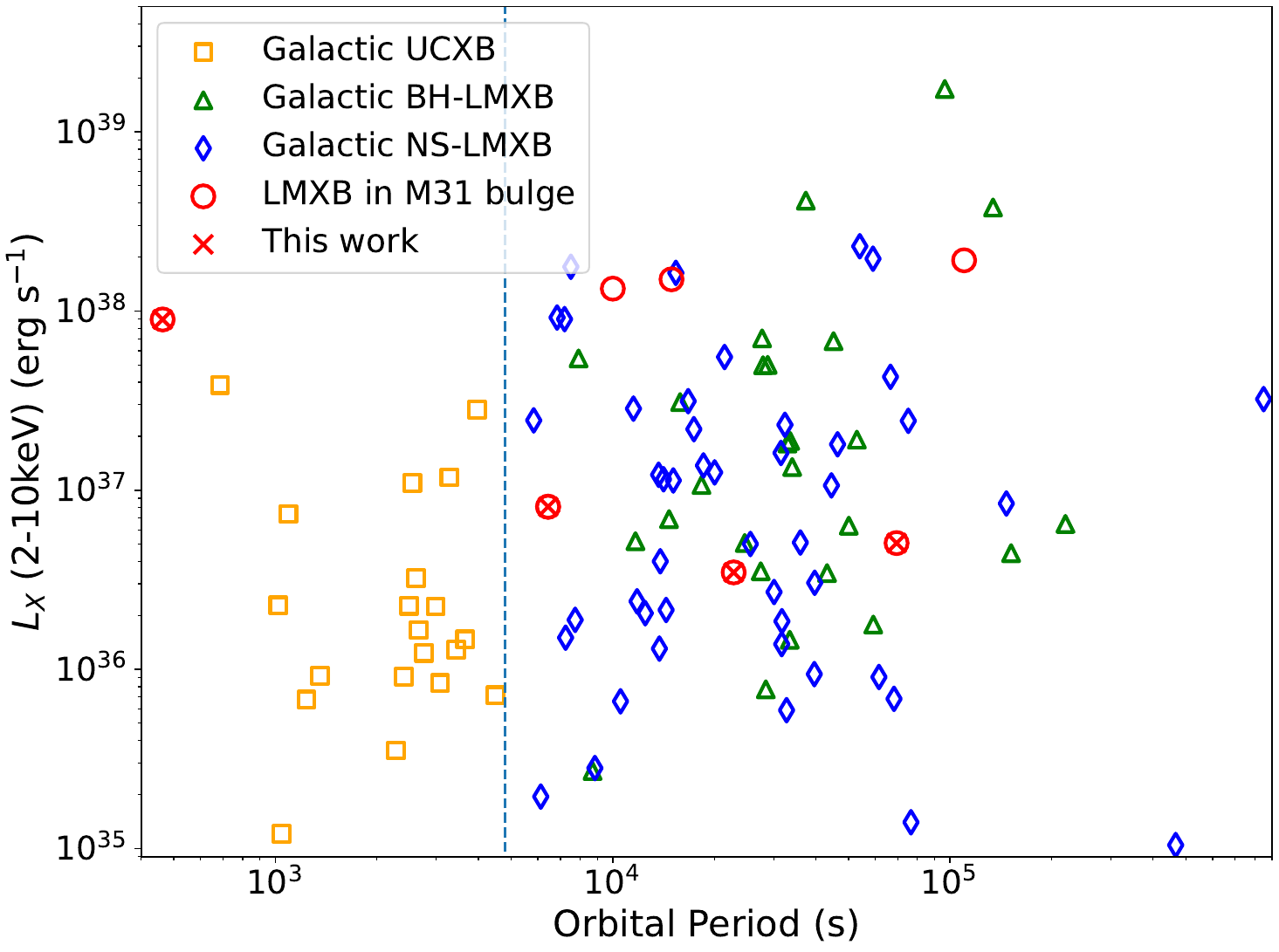}
\caption{The 2-10 keV X-ray luminosity versus orbital period for the seven M31 bulge LMXBs (red circles, with an additional `x' marker for the four sources discovered in this work). Galactic UCXBs, BH-LMXBs, NS-LMXBs from \citet{2023A&A...675A.199A} and \citet{2023A&A...677A.186A} are shown by orange squares, green triangles and blue diamonds, respectively. The vertical dashed line marks the upper bound of the orbital period of UCXBs.}
\label{fig:pl}
\end{figure}

\subsection{Implications on the LMXB population of the M31 bulge}

As summarized in the previous sections, a total of seven LMXBs with a periodic X-ray signal are found in the M31 bulge. Most, if not all, of these periodic signals can be associated with the binary orbital motion. 
Together, these sources may provide a useful diagnosis to the underlying close binary populations in the M31 bulge,  which in turn allows for a meaningful comparison with the Galactic populations.

In Figure~\ref{fig:pl}, we plot the X-ray (2--10 keV) luminosity versus the putative orbital period of the seven LMXBs of the M31 bulge.
The 2--10 keV luminosity of the four {\it Chandra}-detected sources are derived from the best-fit spectral model in Table~\ref{tab:spec}, whereas the values of the {\it XMM-Newton}-detected sources are taken from the literature \citep{2002ApJ...581L..27T,2017ApJ...851L..27M,2017ApJ...839..125Z}.
For comparison, Figure~\ref{fig:pl} also plots Galactic LMXBs with a known orbital period. In particular, we extract from the Galactic LMXB catalog compiled by \cite{2023A&A...675A.199A} a total of 64 NS-LMXB candidates and 31 BH-LMXB candidtaes with an orbital period $\gtrsim$ 80 minutes. 
This sample is supplied by \citet{2023A&A...677A.186A}, which have classified 21 Galactic UCXBs characterized by their orbital periods shorter than 80 minutes.
These sources are shown by different colored symbols in Figure~\ref{fig:pl}.

Certainly, not all intrinsic X-ray periodic signals in the M31 bulge have been identified by the GL algorithm.
Following \citet{2020MNRAS.498.3513B},
we evaluate the detection rate using a large set of synthetic time series.
Specifically, we adopt a sinusoidal light curve with a 30\% variation amplitude as the fiducial input of the periodic variation. 
Taking a stepwise light curve, which might better mimic an eclipsing behavior but requires an additional free parameter, gives a similar result according to \citet{2020MNRAS.498.3513B}. 
We try three sets of representative periods (793 s, 7093 s, 17093 s) and six sets of total counts (250, 500, 1000, 2000, 5000, 10000). A constant light curve with six sets of total counts is also simulated as the no-signal control group.
The synthetic light curves, taking into account the Poisson fluctuation and the observing cadence and length same as the true {\it Chandra} observations, are then fed to the GL algorithm.  For each parameter set, 5000 light curves are generated in order to minimize random fluctuations.
A reported period with $P_{\rm GL} > 0.9$ and within few percents of the input period is considered a valid detection.

\begin{figure}
\hspace{-0.5 cm}
    \centering
    \includegraphics[width=0.5\textwidth]{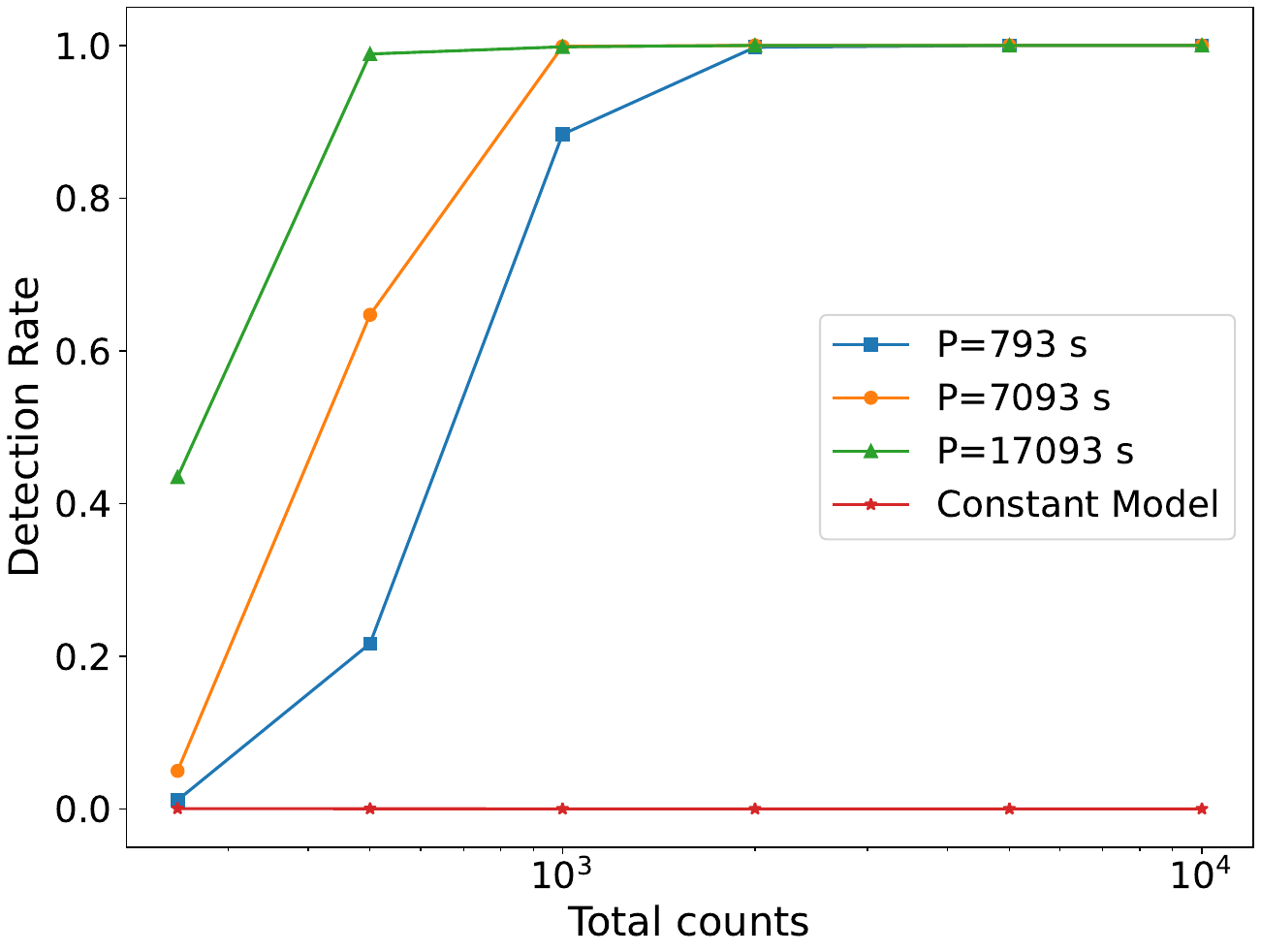}
    \caption{Detection rate as a function of total counts, based on simulated light curves with a sinusoidal variation fed to the GL algorithm. Six values of total counts (250, 500, 1000, 2000, 5000, 10000) and three values of periods (793 sec, 7093 sec, 17093 sec) are simulated. The model with a constant count rate (i.e., no periodic signal) is also plotted.
    }
    \label{fig:rate}
\end{figure}

The detection rate for a given combination of parameters is taken to be the fraction of the synthetic light curves having a valid detection, as shown in Figure~\ref{fig:rate}. The detection rate of the no-signal control group is nearly zero, showing that false alarm from a constant light curve is negligible, regardless of the source counts.
For the periodic light curves, a general trend revealed is that the detection rate increases as the total counts increase, which is naturally understood. 
Moreover, for the same number of total counts, a higher detection rate is found for a larger period, which is due to the nature of the GL algorithm \citep{2020MNRAS.498.3513B}.
The detection rate for source counts of a few $10^3$ (i.e., the case of the M31 sources, Table~\ref{tab:result}), is close to 100\% for all three periods. 
We point out that for long-period signals, despite the simulation results indicating a detection rate up to 100\%, 
the true signal might be accompanied by numerous false signals due to noise, as illustrated by the simulations presented in Figure~\ref{fig:glhist}, a practical effect that would significantly hamper the identification of the true period. 

Based on the predicted detection rate and the distribution of source counts among the X-ray sources in the M31 bulge, we can conclude that essentially all sources with a total count greater than $10^3$ and an intrinsic periodic variation at a period lower than about $10^4$ sec have been identified, whereas a significant fraction of periodic signals from fainter sources might have been hidden in the noise.  
Combining the estimated detection rate, the observed distribution of source counts,  
and further assuming 30\% of LMXBs having an orbital period below $10^4$ sec (i.e., same as the Milky Way population; Figure~\ref{fig:pl}),  
we expect to have $\sim$ 45 short-period ($< 10^4$ sec) LMXBs in the M31 bulge, which can be identified by the GL algorithm. 
A further factor to consider is the intrinsic fraction of orbital modulation, which should occur only when the inclination angle of the orbital plane is high. According to \cite{1987A&A...178..137F}, an inclination angle ranging between $60^{\circ} <i< 80 ^{\circ}$ would lead to eclipsing or dipping signals, which translates to a fraction of $\sim$ 20\% of all LMXBs for a randomly distributed inclination angle.
Hence we expect $\sim$ 9 detectable periodic LMXBs with a period shorter than 10$^4$ sec, which is several times higher than our actual detections (2 sources). Such a difference implies that the fraction of short-period LMXBs ($P\lesssim$ $10^4$ sec) in the M31 bulge may be significantly lower than that of the Milky Way LMXB populations, although we caution that selection bias could exist in both the M31 bulge and Galactic LMXB samples.

\section{Summary}
\label{sec:sum}
We have conducted a systematic search for periodic X-ray signals from several hundreds of X-ray sources in the M31 bulge observed by extensive {\it Chandra}  observations in a total of 2 Ms spanning a temporal baseline of 16 yrs. 
Thanks to the GL algorithm that works well for phased-folded X-ray light curves, seven periodic signals are confidently identified, among which four are newly discovered. Three of these sources are novae, the identified periods of which range between 1.3--2.0 h and is most likely the orbital period. The other four sources are LMXBs, the identified period of which is also likely orbital due to a clear eclipsing/dipping behavior in the light curve. 
One of them might be classified as an UCXB due to the very short (463 sec) period. 

 This is the first systematic search for periodic X-ray sources in an external galaxy. 
Our study demonstrates the potential of using archival X-ray observations to systematically identify periodic X-ray sources in external galaxies, especially LMXBs with short orbital periods, which would provide valuable information about the underlying exotic stellar populations.  

\section*{Data Availability}
The data underlying this article will be shared on reasonable request to the corresponding author. 
The Chandra data used in this article are available in the Chandra Data Archive (\href{https://cxc.harvard.edu/cda/}{https://cxc.harvard.edu/cda/}) by searching the Observation ID listed in Table \ref{tabacis}, \ref{tabhrc} and in the Search and Retrieval interface, ChaSeR (\href{https://cda.harvard.edu/chaser/}{https://cda.harvard.edu/chaser/}).

\section*{Acknowledgements}

This work is supported by the National Natural Science Foundation of China (grant 12225302) and the CNSA program (grant D050102). The authors wish to thank Zhongqun Cheng, Zhao Su and Jianguo He for helpful conversations. The authors wish to thank the anonymous referee for helpful comments that improve our work.

\bibliography{M31XRB.bib}

\begin{thebibliography}{}
\makeatletter
\relax
\def\mn@urlcharsother{\let\do\@makeother \do\$\do\&\do\#\do\^\do\_\do\%\do\~}
\def\mn@doi{\begingroup\mn@urlcharsother \@ifnextchar [ {\mn@doi@} {\mn@doi@[]}}
\def\mn@doi@[#1]#2{\def\@tempa{#1}\ifx\@tempa\@empty \href {http://dx.doi.org/#2} {doi:#2}\else \href {http://dx.doi.org/#2} {#1}\fi \endgroup}
\def\mn@eprint#1#2{\mn@eprint@#1:#2::\@nil}
\def\mn@eprint@arXiv#1{\href {http://arxiv.org/abs/#1} {{\tt arXiv:#1}}}
\def\mn@eprint@dblp#1{\href {http://dblp.uni-trier.de/rec/bibtex/#1.xml} {dblp:#1}}
\def\mn@eprint@#1:#2:#3:#4\@nil{\def\@tempa {#1}\def\@tempb {#2}\def\@tempc {#3}\ifx \@tempc \@empty \let \@tempc \@tempb \let \@tempb \@tempa \fi \ifx \@tempb \@empty \def\@tempb {arXiv}\fi \@ifundefined {mn@eprint@\@tempb}{\@tempb:\@tempc}{\expandafter \expandafter \csname mn@eprint@\@tempb\endcsname \expandafter{\@tempc}}}

\bibitem[\protect\citeauthoryear{{Armas Padilla}, {Corral-Santana}, {Borghese}, {C{\'u}neo}, {Mu{\~n}oz-Darias}, {Casares}  \& {Torres}}{{Armas Padilla} et~al.}{2023}]{2023A&A...677A.186A}
{Armas Padilla} M.,  {Corral-Santana} J.~M.,  {Borghese} A.,  {C{\'u}neo} V.~A.,  {Mu{\~n}oz-Darias} T.,  {Casares} J.,   {Torres} M.~A.~P.,  2023, \mn@doi [\aap] {10.1051/0004-6361/202346797}, \href {https://ui.adsabs.harvard.edu/abs/2023A&A...677A.186A} {677, A186}

\bibitem[\protect\citeauthoryear{{Avakyan}, {Neumann}, {Zainab}, {Doroshenko}, {Wilms}  \& {Santangelo}}{{Avakyan} et~al.}{2023}]{2023A&A...675A.199A}
{Avakyan} A.,  {Neumann} M.,  {Zainab} A.,  {Doroshenko} V.,  {Wilms} J.,   {Santangelo} A.,  2023, \mn@doi [\aap] {10.1051/0004-6361/202346522}, \href {https://ui.adsabs.harvard.edu/abs/2023A&A...675A.199A} {675, A199}

\bibitem[\protect\citeauthoryear{{Babak}, {Hewitson}  \& {Petiteau}}{{Babak} et~al.}{2021}]{2021arXiv210801167B}
{Babak} S.,  {Hewitson} M.,   {Petiteau} A.,  2021, \mn@doi [arXiv e-prints] {10.48550/arXiv.2108.01167}, \href {https://ui.adsabs.harvard.edu/abs/2021arXiv210801167B} {p. arXiv:2108.01167}

\bibitem[\protect\citeauthoryear{{Baluev}}{{Baluev}}{2008}]{2008MNRAS.385.1279B}
{Baluev} R.~V.,  2008, \mn@doi [\mnras] {10.1111/j.1365-2966.2008.12689.x}, \href {https://ui.adsabs.harvard.edu/abs/2008MNRAS.385.1279B} {385, 1279}

\bibitem[\protect\citeauthoryear{{Bao} \& {Li}}{{Bao} \& {Li}}{2020}]{2020MNRAS.498.3513B}
{Bao} T.,  {Li} Z.,  2020, \mn@doi [\mnras] {10.1093/mnras/staa2603}, \href {https://ui.adsabs.harvard.edu/abs/2020MNRAS.498.3513B} {498, 3513}

\bibitem[\protect\citeauthoryear{{Bao} \& {Li}}{{Bao} \& {Li}}{2022}]{2022MNRAS.509.3504B}
{Bao} T.,  {Li} Z.,  2022, \mn@doi [\mnras] {10.1093/mnras/stab3259}, \href {https://ui.adsabs.harvard.edu/abs/2022MNRAS.509.3504B} {509, 3504}

\bibitem[\protect\citeauthoryear{{Bao}, {Li}  \& {Cheng}}{{Bao} et~al.}{2023}]{2023MNRAS.521.4257B}
{Bao} T.,  {Li} Z.,   {Cheng} Z.,  2023, \mn@doi [\mnras] {10.1093/mnras/stad836}, \href {https://ui.adsabs.harvard.edu/abs/2023MNRAS.521.4257B} {521, 4257}

\bibitem[\protect\citeauthoryear{{Bao}, {Li}, {Cheng}  \& {Belloni}}{{Bao} et~al.}{2024}]{2024MNRAS.527.7173B}
{Bao} T.,  {Li} Z.,  {Cheng} Z.,   {Belloni} D.,  2024, \mn@doi [\mnras] {10.1093/mnras/stad3665}, \href {https://ui.adsabs.harvard.edu/abs/2024MNRAS.527.7173B} {527, 7173}

\bibitem[\protect\citeauthoryear{{Barnard}, {Garcia}  \& {Murray}}{{Barnard} et~al.}{2013}]{2013ApJ...770..148B}
{Barnard} R.,  {Garcia} M.~R.,   {Murray} S.~S.,  2013, \mn@doi [\apj] {10.1088/0004-637X/770/2/148}, \href {https://ui.adsabs.harvard.edu/abs/2013ApJ...770..148B} {770, 148}

\bibitem[\protect\citeauthoryear{{Barnard}, {Garcia}, {Primini}, {Li}, {Baganoff}  \& {Murray}}{{Barnard} et~al.}{2014}]{2014ApJ...780...83B}
{Barnard} R.,  {Garcia} M.~R.,  {Primini} F.,  {Li} Z.,  {Baganoff} F.~K.,   {Murray} S.~S.,  2014, \mn@doi [\apj] {10.1088/0004-637X/780/1/83}, \href {https://ui.adsabs.harvard.edu/abs/2014ApJ...780...83B} {780, 83}

\bibitem[\protect\citeauthoryear{{Barnard}, {Garcia}  \& {Murray}}{{Barnard} et~al.}{2015}]{2015ApJ...801...65B}
{Barnard} R.,  {Garcia} M.~R.,   {Murray} S.~S.,  2015, \mn@doi [\apj] {10.1088/0004-637X/801/1/65}, \href {https://ui.adsabs.harvard.edu/abs/2015ApJ...801...65B} {801, 65}

\bibitem[\protect\citeauthoryear{{Chou}}{{Chou}}{2014}]{2014RAA....14.1367C}
{Chou} Y.,  2014, \mn@doi [Research in Astronomy and Astrophysics] {10.1088/1674-4527/14/11/001}, \href {https://ui.adsabs.harvard.edu/abs/2014RAA....14.1367C} {14, 1367}

\bibitem[\protect\citeauthoryear{{De Luca} et~al.,}{{De Luca} et~al.}{2021}]{2021A&A...650A.167D}
{De Luca} A.,  et~al., 2021, \mn@doi [\aap] {10.1051/0004-6361/202039783}, \href {https://ui.adsabs.harvard.edu/abs/2021A&A...650A.167D} {650, A167}

\bibitem[\protect\citeauthoryear{{Esposito} et~al.,}{{Esposito} et~al.}{2016}]{2016MNRAS.457L...5E}
{Esposito} P.,  et~al., 2016, \mn@doi [\mnras] {10.1093/mnrasl/slv194}, \href {https://ui.adsabs.harvard.edu/abs/2016MNRAS.457L...5E} {457, L5}

\bibitem[\protect\citeauthoryear{{Foreman-Mackey}, {Hogg}, {Lang}  \& {Goodman}}{{Foreman-Mackey} et~al.}{2013}]{2013PASP..125..306F}
{Foreman-Mackey} D.,  {Hogg} D.~W.,  {Lang} D.,   {Goodman} J.,  2013, \mn@doi [\pasp] {10.1086/670067}, \href {https://ui.adsabs.harvard.edu/abs/2013PASP..125..306F} {125, 306}

\bibitem[\protect\citeauthoryear{{Frank}, {King}  \& {Lasota}}{{Frank} et~al.}{1987}]{1987A&A...178..137F}
{Frank} J.,  {King} A.~R.,   {Lasota} J.~P.,  1987, \aap, \href {https://ui.adsabs.harvard.edu/abs/1987A&A...178..137F} {178, 137}

\bibitem[\protect\citeauthoryear{{Gregory} \& {Loredo}}{{Gregory} \& {Loredo}}{1992}]{1992ApJ...398..146G}
{Gregory} P.~C.,  {Loredo} T.~J.,  1992, \mn@doi [\apj] {10.1086/171844}, \href {https://ui.adsabs.harvard.edu/abs/1992ApJ...398..146G} {398, 146}

\bibitem[\protect\citeauthoryear{{He}, {Shao}, {Gao}  \& {Li}}{{He} et~al.}{2023}]{2023ApJ...953..153H}
{He} J.-G.,  {Shao} Y.,  {Gao} S.-J.,   {Li} X.-D.,  2023, \mn@doi [\apj] {10.3847/1538-4357/ace348}, \href {https://ui.adsabs.harvard.edu/abs/2023ApJ...953..153H} {953, 153}

\bibitem[\protect\citeauthoryear{{Henze} et~al.,}{{Henze} et~al.}{2010}]{2010A&A...523A..89H}
{Henze} M.,  et~al., 2010, \mn@doi [\aap] {10.1051/0004-6361/201014710}, \href {https://ui.adsabs.harvard.edu/abs/2010A&A...523A..89H} {523, A89}

\bibitem[\protect\citeauthoryear{{Henze} et~al.,}{{Henze} et~al.}{2011}]{2011A&A...533A..52H}
{Henze} M.,  et~al., 2011, \mn@doi [\aap] {10.1051/0004-6361/201015887}, \href {https://ui.adsabs.harvard.edu/abs/2011A&A...533A..52H} {533, A52}

\bibitem[\protect\citeauthoryear{{Henze} et~al.,}{{Henze} et~al.}{2014}]{2014A&A...563A...2H}
{Henze} M.,  et~al., 2014, \mn@doi [\aap] {10.1051/0004-6361/201322426}, \href {https://ui.adsabs.harvard.edu/abs/2014A&A...563A...2H} {563, A2}

\bibitem[\protect\citeauthoryear{{Hofmann}, {Pietsch}, {Henze}, {Haberl}, {Sturm}, {Della Valle}, {Hartmann}  \& {Hatzidimitriou}}{{Hofmann} et~al.}{2013}]{2013A&A...555A..65H}
{Hofmann} F.,  {Pietsch} W.,  {Henze} M.,  {Haberl} F.,  {Sturm} R.,  {Della Valle} M.,  {Hartmann} D.~H.,   {Hatzidimitriou} D.,  2013, \mn@doi [\aap] {10.1051/0004-6361/201321165}, \href {https://ui.adsabs.harvard.edu/abs/2013A&A...555A..65H} {555, A65}

\bibitem[\protect\citeauthoryear{{Kaaret}}{{Kaaret}}{2002}]{2002ApJ...578..114K}
{Kaaret} P.,  2002, \mn@doi [\apj] {10.1086/342475}, \href {https://ui.adsabs.harvard.edu/abs/2002ApJ...578..114K} {578, 114}

\bibitem[\protect\citeauthoryear{{Kahabka} \& {van den Heuvel}}{{Kahabka} \& {van den Heuvel}}{1997}]{1997ARA&A..35...69K}
{Kahabka} P.,  {van den Heuvel} E.~P.~J.,  1997, \mn@doi [\araa] {10.1146/annurev.astro.35.1.69}, \href {https://ui.adsabs.harvard.edu/abs/1997ARA&A..35...69K} {35, 69}

\bibitem[\protect\citeauthoryear{{Knigge}, {Baraffe}  \& {Patterson}}{{Knigge} et~al.}{2011}]{2011ApJS..194...28K}
{Knigge} C.,  {Baraffe} I.,   {Patterson} J.,  2011, \mn@doi [\apjs] {10.1088/0067-0049/194/2/28}, \href {https://ui.adsabs.harvard.edu/abs/2011ApJS..194...28K} {194, 28}

\bibitem[\protect\citeauthoryear{{Kong}, {Garcia}, {Primini}, {Murray}, {Di Stefano}  \& {McClintock}}{{Kong} et~al.}{2002}]{2002ApJ...577..738K}
{Kong} A. K.~H.,  {Garcia} M.~R.,  {Primini} F.~A.,  {Murray} S.~S.,  {Di Stefano} R.,   {McClintock} J.~E.,  2002, \mn@doi [\apj] {10.1086/342116}, \href {https://ui.adsabs.harvard.edu/abs/2002ApJ...577..738K} {577, 738}

\bibitem[\protect\citeauthoryear{{Lazzarini} et~al.,}{{Lazzarini} et~al.}{2018}]{2018ApJ...862...28L}
{Lazzarini} M.,  et~al., 2018, \mn@doi [\apj] {10.3847/1538-4357/aacb2a}, \href {https://ui.adsabs.harvard.edu/abs/2018ApJ...862...28L} {862, 28}

\bibitem[\protect\citeauthoryear{{Lomb}}{{Lomb}}{1976}]{1976Ap&SS..39..447L}
{Lomb} N.~R.,  1976, \mn@doi [\apss] {10.1007/BF00648343}, \href {https://ui.adsabs.harvard.edu/abs/1976Ap&SS..39..447L} {39, 447}

\bibitem[\protect\citeauthoryear{{Mangano}, {Israel}  \& {Stella}}{{Mangano} et~al.}{2004}]{2004A&A...419.1045M}
{Mangano} V.,  {Israel} G.~L.,   {Stella} L.,  2004, \mn@doi [\aap] {10.1051/0004-6361:20040099}, \href {https://ui.adsabs.harvard.edu/abs/2004A&A...419.1045M} {419, 1045}

\bibitem[\protect\citeauthoryear{{Marelli} et~al.,}{{Marelli} et~al.}{2017}]{2017ApJ...851L..27M}
{Marelli} M.,  et~al., 2017, \mn@doi [\apjl] {10.3847/2041-8213/aa9b2e}, \href {https://ui.adsabs.harvard.edu/abs/2017ApJ...851L..27M} {851, L27}

\bibitem[\protect\citeauthoryear{{Marelli}, {De Martino}, {Mereghetti}, {De Luca}, {Salvaterra}, {Sidoli}, {Israel}  \& {Rodriguez}}{{Marelli} et~al.}{2018}]{2018ApJ...866..125M}
{Marelli} M.,  {De Martino} D.,  {Mereghetti} S.,  {De Luca} A.,  {Salvaterra} R.,  {Sidoli} L.,  {Israel} G.,   {Rodriguez} G.,  2018, \mn@doi [\apj] {10.3847/1538-4357/aadc67}, \href {https://ui.adsabs.harvard.edu/abs/2018ApJ...866..125M} {866, 125}

\bibitem[\protect\citeauthoryear{{McConnachie}, {Irwin}, {Ferguson}, {Ibata}, {Lewis}  \& {Tanvir}}{{McConnachie} et~al.}{2005}]{2005MNRAS.356..979M}
{McConnachie} A.~W.,  {Irwin} M.~J.,  {Ferguson} A.~M.~N.,  {Ibata} R.~A.,  {Lewis} G.~F.,   {Tanvir} N.,  2005, \mn@doi [\mnras] {10.1111/j.1365-2966.2004.08514.x}, \href {https://ui.adsabs.harvard.edu/abs/2005MNRAS.356..979M} {356, 979}

\bibitem[\protect\citeauthoryear{{Neumann}, {Avakyan}, {Doroshenko}  \& {Santangelo}}{{Neumann} et~al.}{2023}]{2023A&A...677A.134N}
{Neumann} M.,  {Avakyan} A.,  {Doroshenko} V.,   {Santangelo} A.,  2023, \mn@doi [\aap] {10.1051/0004-6361/202245728}, \href {https://ui.adsabs.harvard.edu/abs/2023A&A...677A.134N} {677, A134}

\bibitem[\protect\citeauthoryear{{Osborne} et~al.,}{{Osborne} et~al.}{2001}]{2001A&A...378..800O}
{Osborne} J.~P.,  et~al., 2001, \mn@doi [\aap] {10.1051/0004-6361:20011228}, \href {https://ui.adsabs.harvard.edu/abs/2001A&A...378..800O} {378, 800}

\bibitem[\protect\citeauthoryear{{Painter} et~al.,}{{Painter} et~al.}{2024}]{2024MNRAS.529..245P}
{Painter} C.,  et~al., 2024, \mn@doi [\mnras] {10.1093/mnras/stae164}, \href {https://ui.adsabs.harvard.edu/abs/2024MNRAS.529..245P} {529, 245}

\bibitem[\protect\citeauthoryear{{Pietsch}, {Freyberg}  \& {Haberl}}{{Pietsch} et~al.}{2005}]{2005A&A...434..483P}
{Pietsch} W.,  {Freyberg} M.,   {Haberl} F.,  2005, \mn@doi [\aap] {10.1051/0004-6361:20041990}, \href {https://ui.adsabs.harvard.edu/abs/2005A&A...434..483P} {434, 483}

\bibitem[\protect\citeauthoryear{{Pietsch}, {Henze}, {Haberl}, {Hernanz}, {Sala}, {Hartmann}  \& {Della Valle}}{{Pietsch} et~al.}{2011}]{2011A&A...531A..22P}
{Pietsch} W.,  {Henze} M.,  {Haberl} F.,  {Hernanz} M.,  {Sala} G.,  {Hartmann} D.~H.,   {Della Valle} M.,  2011, \mn@doi [\aap] {10.1051/0004-6361/201116756}, \href {https://ui.adsabs.harvard.edu/abs/2011A&A...531A..22P} {531, A22}

\bibitem[\protect\citeauthoryear{{Robinson}}{{Robinson}}{1976}]{1976ARA&A..14..119R}
{Robinson} E.~L.,  1976, \mn@doi [\araa] {10.1146/annurev.aa.14.090176.001003}, \href {https://ui.adsabs.harvard.edu/abs/1976ARA&A..14..119R} {14, 119}

\bibitem[\protect\citeauthoryear{{Rodr{\'\i}guez Castillo}, {Israel}, {Esposito}, {Papitto}, {Stella}, {Tiengo}, {De Luca}  \& {Marelli}}{{Rodr{\'\i}guez Castillo} et~al.}{2018}]{2018ApJ...861L..26R}
{Rodr{\'\i}guez Castillo} G.~A.,  {Israel} G.~L.,  {Esposito} P.,  {Papitto} A.,  {Stella} L.,  {Tiengo} A.,  {De Luca} A.,   {Marelli} M.,  2018, \mn@doi [\apjl] {10.3847/2041-8213/aacf40}, \href {https://ui.adsabs.harvard.edu/abs/2018ApJ...861L..26R} {861, L26}

\bibitem[\protect\citeauthoryear{{Shirey} et~al.,}{{Shirey} et~al.}{2001}]{2001A&A...365L.195S}
{Shirey} R.,  et~al., 2001, \mn@doi [\aap] {10.1051/0004-6361:20000243}, \href {https://ui.adsabs.harvard.edu/abs/2001A&A...365L.195S} {365, L195}

\bibitem[\protect\citeauthoryear{{Timmer} \& {Koenig}}{{Timmer} \& {Koenig}}{1995}]{1995A&A...300..707T}
{Timmer} J.,  {Koenig} M.,  1995, \aap, \href {https://ui.adsabs.harvard.edu/abs/1995A&A...300..707T} {300, 707}

\bibitem[\protect\citeauthoryear{{Trudolyubov} \& {Priedhorsky}}{{Trudolyubov} \& {Priedhorsky}}{2008}]{2008ApJ...676.1218T}
{Trudolyubov} S.~P.,  {Priedhorsky} W.~C.,  2008, \mn@doi [\apj] {10.1086/526397}, \href {https://ui.adsabs.harvard.edu/abs/2008ApJ...676.1218T} {676, 1218}

\bibitem[\protect\citeauthoryear{{Trudolyubov}, {Borozdin}, {Priedhorsky}, {Osborne}, {Watson}, {Mason}  \& {Cordova}}{{Trudolyubov} et~al.}{2002}]{2002ApJ...581L..27T}
{Trudolyubov} S.~P.,  {Borozdin} K.~N.,  {Priedhorsky} W.~C.,  {Osborne} J.~P.,  {Watson} M.~G.,  {Mason} K.~O.,   {Cordova} F.~A.,  2002, \mn@doi [\apjl] {10.1086/345786}, \href {https://ui.adsabs.harvard.edu/abs/2002ApJ...581L..27T} {581, L27}

\bibitem[\protect\citeauthoryear{{Voss} \& {Gilfanov}}{{Voss} \& {Gilfanov}}{2006}]{2006A&A...447...71V}
{Voss} R.,  {Gilfanov} M.,  2006, \mn@doi [\aap] {10.1051/0004-6361:20053420}, \href {https://ui.adsabs.harvard.edu/abs/2006A&A...447...71V} {447, 71}

\bibitem[\protect\citeauthoryear{{Voss} \& {Gilfanov}}{{Voss} \& {Gilfanov}}{2007a}]{2007MNRAS.380.1685V}
{Voss} R.,  {Gilfanov} M.,  2007a, \mn@doi [\mnras] {10.1111/j.1365-2966.2007.12223.x}, \href {https://ui.adsabs.harvard.edu/abs/2007MNRAS.380.1685V} {380, 1685}

\bibitem[\protect\citeauthoryear{{Voss} \& {Gilfanov}}{{Voss} \& {Gilfanov}}{2007b}]{2007A&A...468...49V}
{Voss} R.,  {Gilfanov} M.,  2007b, \mn@doi [\aap] {10.1051/0004-6361:20066614}, \href {https://ui.adsabs.harvard.edu/abs/2007A&A...468...49V} {468, 49}

\bibitem[\protect\citeauthoryear{{Warner}}{{Warner}}{1989}]{1989IBVS.3383....1W}
{Warner} B.,  1989, Information Bulletin on Variable Stars, \href {https://ui.adsabs.harvard.edu/abs/1989IBVS.3383....1W} {3383, 1}

\bibitem[\protect\citeauthoryear{{White} \& {Holt}}{{White} \& {Holt}}{1982}]{1982ApJ...257..318W}
{White} N.~E.,  {Holt} S.~S.,  1982, \mn@doi [\apj] {10.1086/159991}, \href {https://ui.adsabs.harvard.edu/abs/1982ApJ...257..318W} {257, 318}

\bibitem[\protect\citeauthoryear{{Zhu}, {Li}  \& {Morris}}{{Zhu} et~al.}{2018}]{2018ApJS..235...26Z}
{Zhu} Z.,  {Li} Z.,   {Morris} M.~R.,  2018, \mn@doi [\apjs] {10.3847/1538-4365/aab14f}, \href {https://ui.adsabs.harvard.edu/abs/2018ApJS..235...26Z} {235, 26}

\bibitem[\protect\citeauthoryear{{Zolotukhin}, {Bachetti}, {Sartore}, {Chilingarian}  \& {Webb}}{{Zolotukhin} et~al.}{2017}]{2017ApJ...839..125Z}
{Zolotukhin} I.~Y.,  {Bachetti} M.,  {Sartore} N.,  {Chilingarian} I.~V.,   {Webb} N.~A.,  2017, \mn@doi [\apj] {10.3847/1538-4357/aa689d}, \href {https://ui.adsabs.harvard.edu/abs/2017ApJ...839..125Z} {839, 125}

\bibitem[\protect\citeauthoryear{{van Speybroeck} et~al.,}{{van Speybroeck} et~al.}{1979}]{1979BAAS...11..609V}
{van Speybroeck} L.,  et~al., 1979, in Bulletin of the American Astronomical Society. p.~609

\makeatother
\end{thebibliography}
\bibliographystyle{mnras}

\appendix
\section{A brief overview of the Gregory-Loredo algorithm}
\label{appendix}
To help understand the results of the Gregory-Loredo Algorithm, we provide here a brief overview of its working principle (more mathematical details can be found in \citealp{1992ApJ...398..146G} and the appendix in \citealp{2020MNRAS.498.3513B}). The key of this algorithm is the multiplicity $W$ of the phase distribution of events (i.e. detected counts). For a specific time series of photons with total counts $N$, the algorithm searches over a frequency ($\omega$) range with a given frequency resolution. For each $\omega$, the algorithm folds the time series into the corresponding period and divide the phase into $m$ bins. The number of photons in the $i_{\rm th}$ bin is $n_i$ ($i$=1, 2, ···, m). The multiplicity is defined as
\begin{equation}
\centering
    W_m(\omega,\phi)=\frac{N!}{n_1!n_2!...n_m!}.
\end{equation}

The odds ratio ($O_{m1}$) is defined as the following integral of $1/\omega$ over $\omega_{\rm lo}$ to $\omega_{\rm hi}$ and $\phi$ over 0 to $2\pi$,
\begin{equation}
\centering
    O_{m1}=\frac{1}{2\pi\nu\ln{\frac{\omega_{\rm hi}}{\omega_{\rm lo}}}}\left(\begin{matrix} N+m-1 \\ N\end{matrix}\right)^{-1} \times \int_{\omega_{\rm lo}}^{\omega_{\rm hi}}\frac{d\omega}{\omega}\times\int_0^{2\pi}d\phi \frac{S(\omega,\phi)m^N}{W_m(\omega,\phi)}.
\end{equation}

Here $S(\omega,\phi)$ is a weighting function that accounts for dead time in the otherwise continuous light curve.
The probability of a periodic signal, $P_{\rm GL}$, is calculated from the sum of the odds ratios,
\begin{equation}
\centering
    P_{\rm GL}=\frac{\sum_{m=2}^{m_{max}}O_{m1}}{1+\sum_{m=2}^{m_{max}}O_{m1}}.
\label{eq:Pgl}
\end{equation}

The posterior probability as a function of the frequency can be solved to find the most probable period $P = 2\pi/\omega$, when $O_{m1}(\omega)$ takes the maximum value,
\begin{equation}
\centering
    O_{m1}(\omega)=\frac{1}{2\pi\nu}\left(\begin{matrix} N+m-1 \\ N\end{matrix}\right)^{-1} \times\int_0^{2\pi}d\phi \frac{S(\omega,\phi)m^N}{W_m(\omega,\phi)}.
\label{eq:om1w}
\end{equation}

In \cite{1992ApJ...398..146G}, $P_{\rm GL} > 0.5$ is taken to indicate the tendency of a periodic signal.
From Eqn.~\ref{eq:Pgl}, we can see that this requires $O_{\rm m1}$ to be greater than 1. The higher the odds ratio, the greater the probability that a periodicity exists. 
However, we stress that the ``odds ratio'' cannot be directly interpreted as the significance of the signal. Rather, one relies on synthetic light curves (which reflect the real observations as much as possible) to evaluate the false alarm probability, as we have done and presented in Section~\ref{subsec:rednoise} (see also \citealp{2023MNRAS.521.4257B}). 

Diagrams of the GL ``odds ratio'' (Eqn.~\ref{eq:om1w}) as a function of $\omega$ for the seven periodic sources are shown in Figure~\ref{fig:gl1}. 
In each diagram, the identified period corresponds to the most prominent peak of $O_{m1}(\omega)$ across the searched frequency range. 
Although the Lomb-Scargle (LS) periodogram \citep{1976Ap&SS..39..447L}, a popular period searching method, is not utilized in our study, we have also computed the LS periodograms for the seven sources and shown them in Figure~\ref{fig:gl1} for a close comparison, along with their corresponding confidence levels (confidence = 1 $-$ FAP) calculated following the method in \citet{2008MNRAS.385.1279B}. For Seq.1, 2, 3, 5, 6, the HRC data is used for the diagram in Figure~\ref{fig:gl1}. For Seq.4 and Seq.7, the ACIS data is used. We use the same ObsIDs as in Table~\ref{tab:result}. 

In most cases, the periodic signal revealed by the LS periodogram is roughly consistent with the period identified by the GL algorithm, with Seq.7 an exception, in which the LS periodogram fails to identify the true signal. However, it is evident that the GL algorithm performs significantly better against red noise. This comparison further supports the advantage and robustness of the GL algorithm.

\onecolumn
\begin{figure*}
    \centering
    \includegraphics[width=0.51\linewidth]{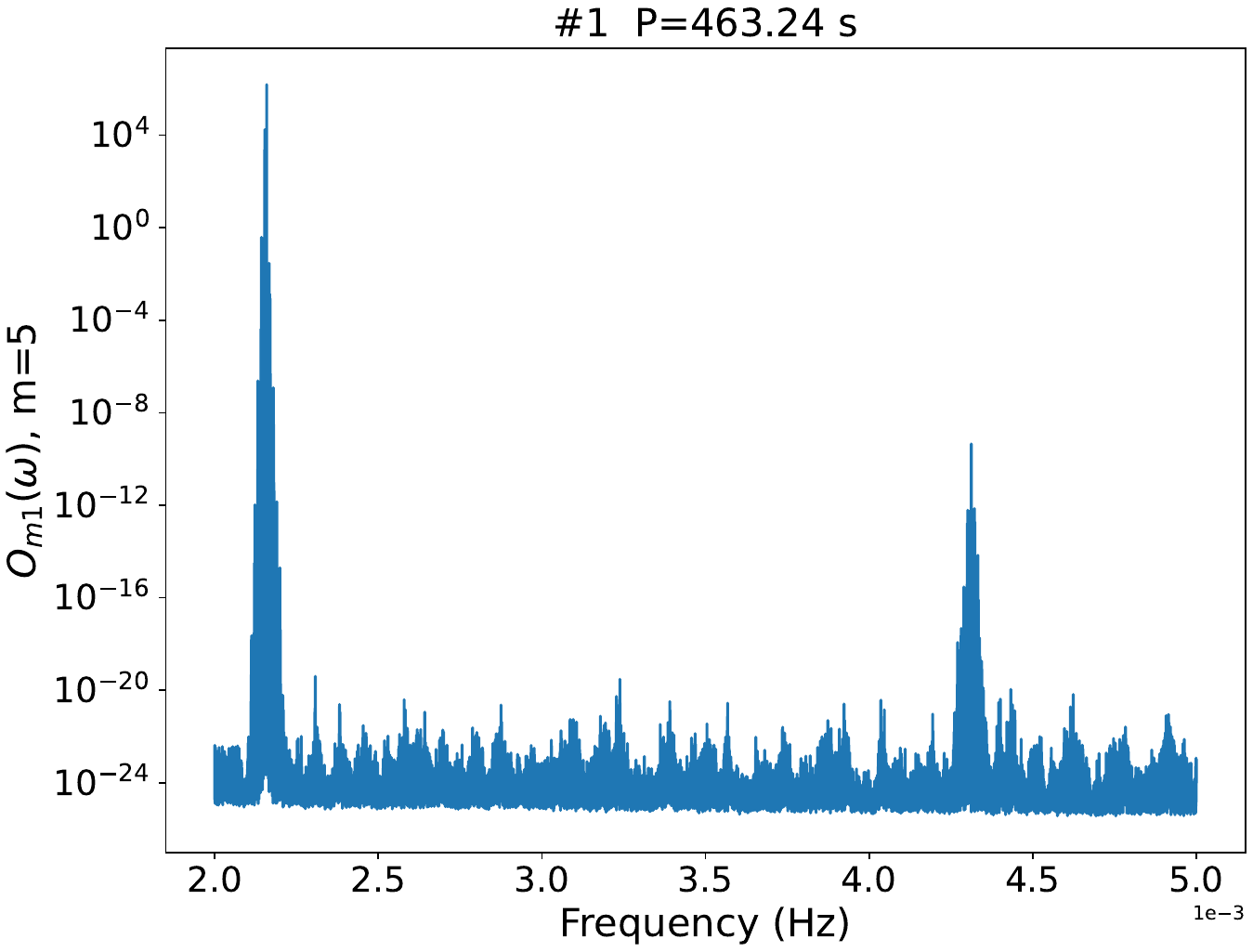}\includegraphics[width=0.5\linewidth]{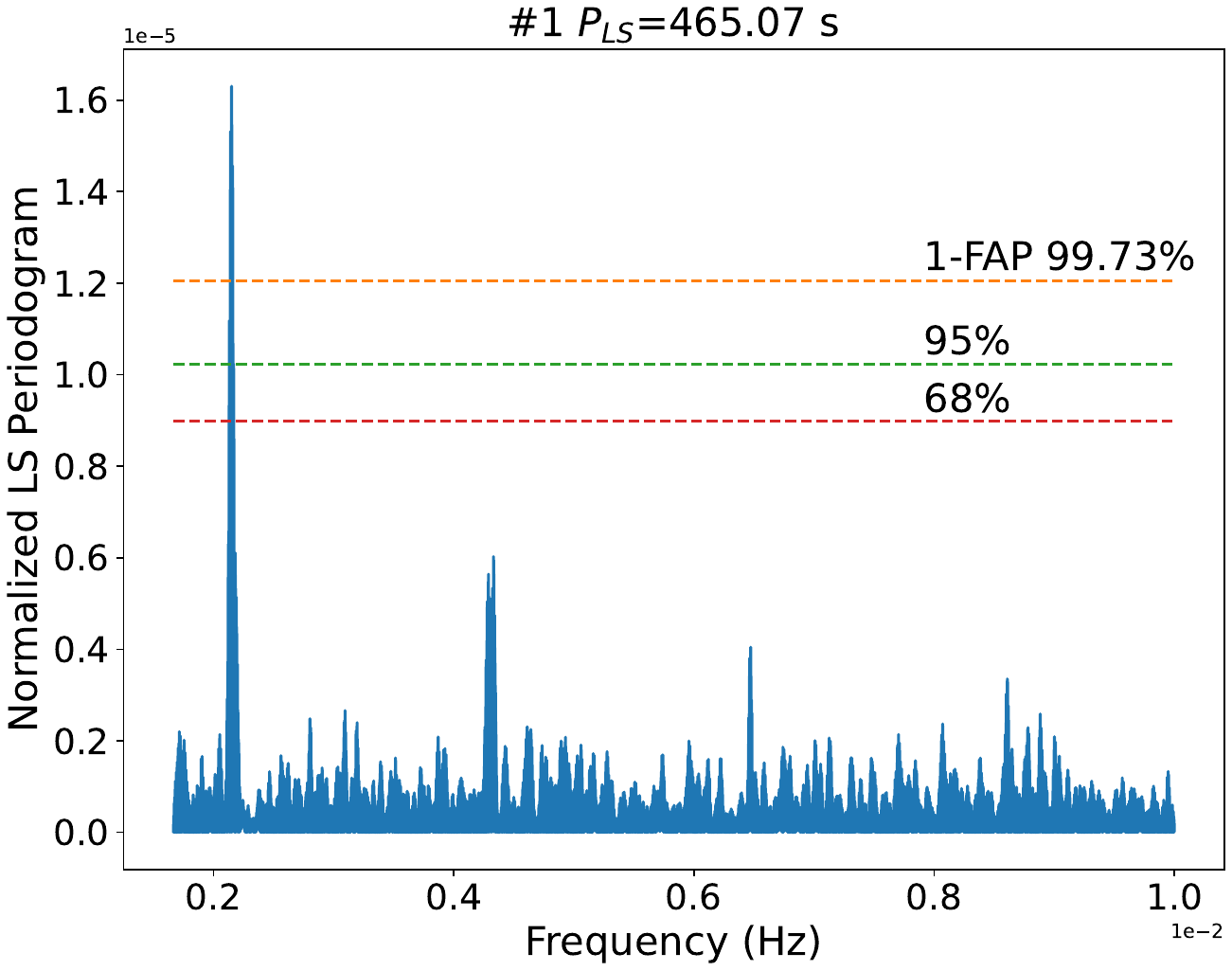}\\
    \includegraphics[width=0.51\linewidth]{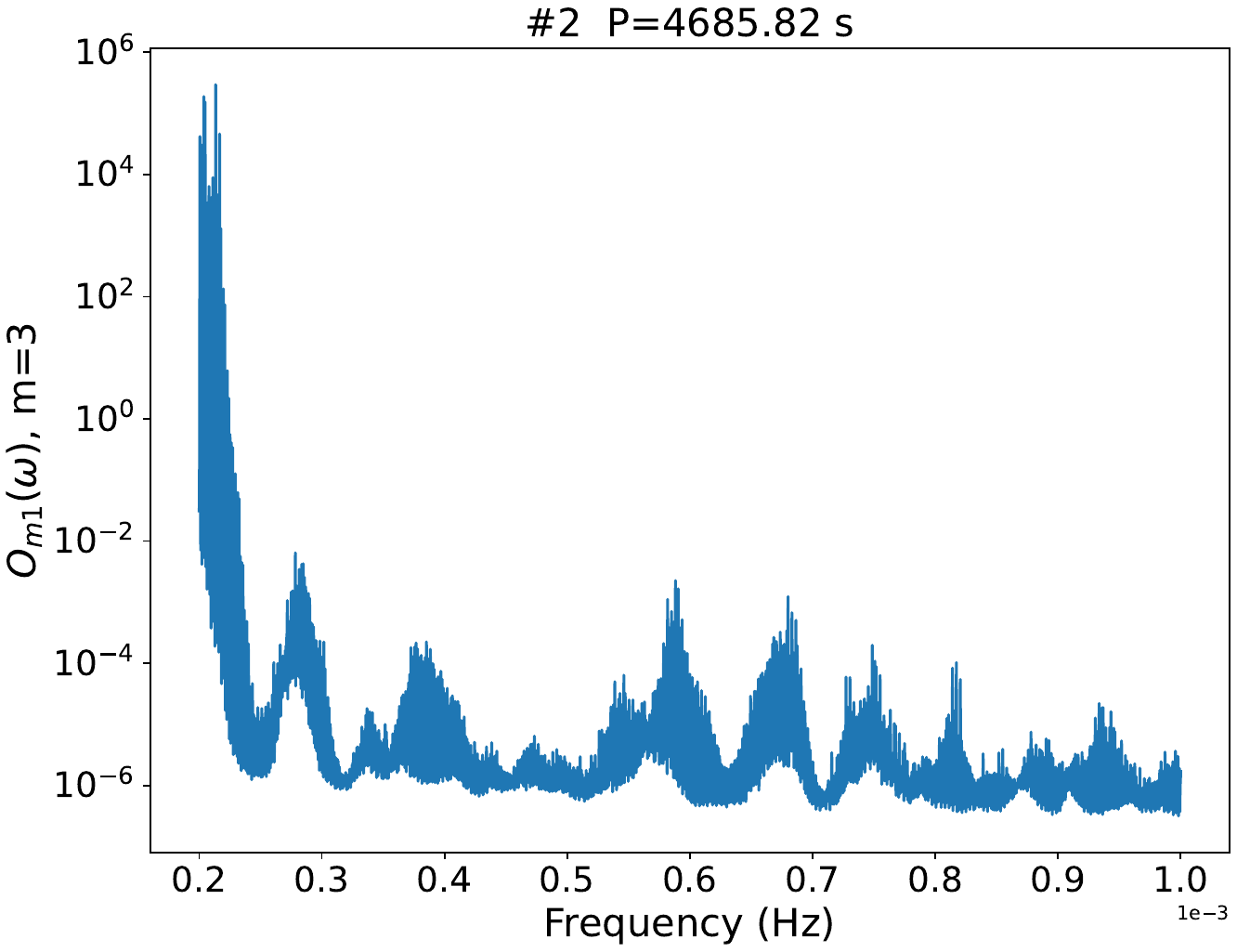}\includegraphics[width=0.5\linewidth]{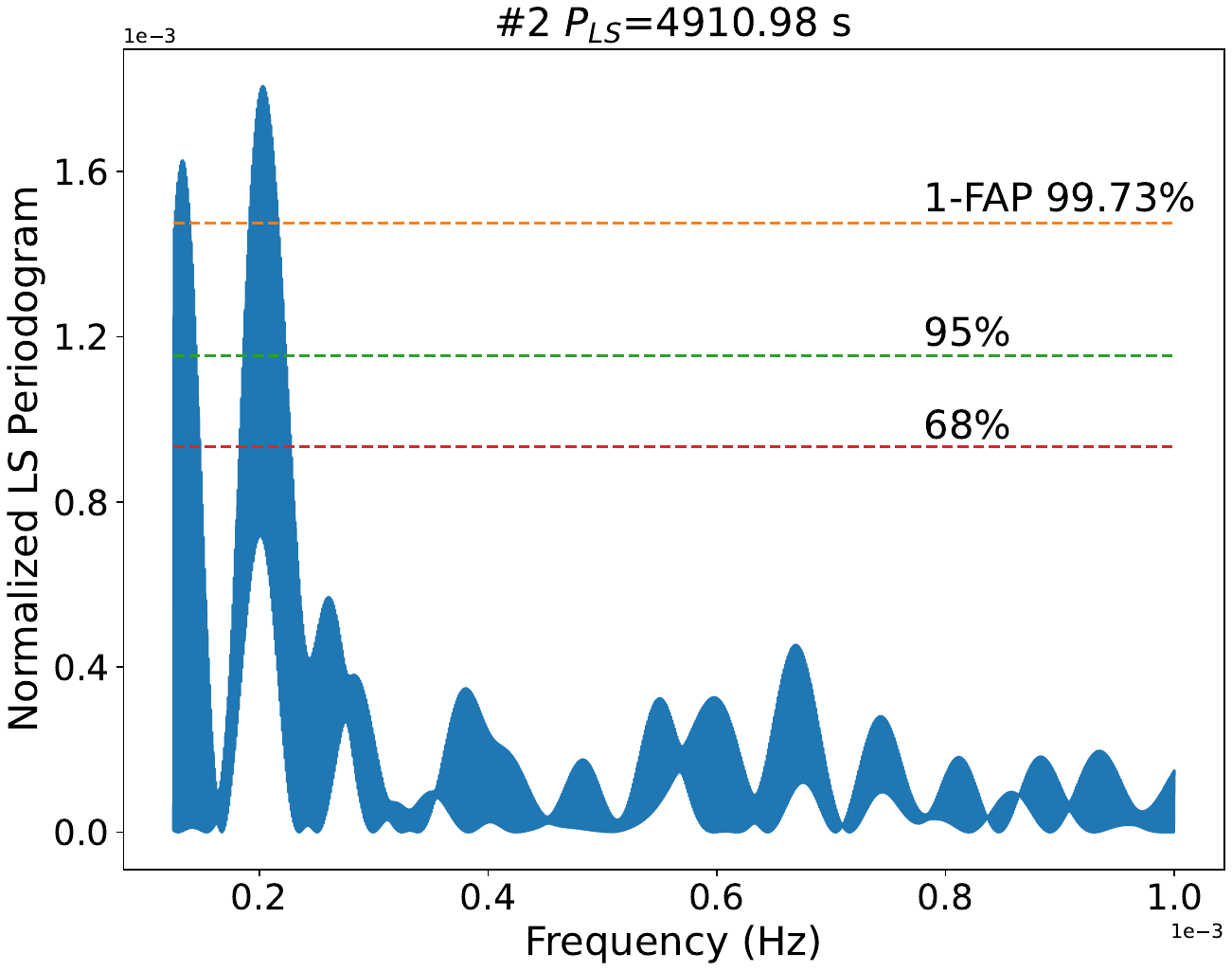}\\
    \includegraphics[width=0.51\linewidth]{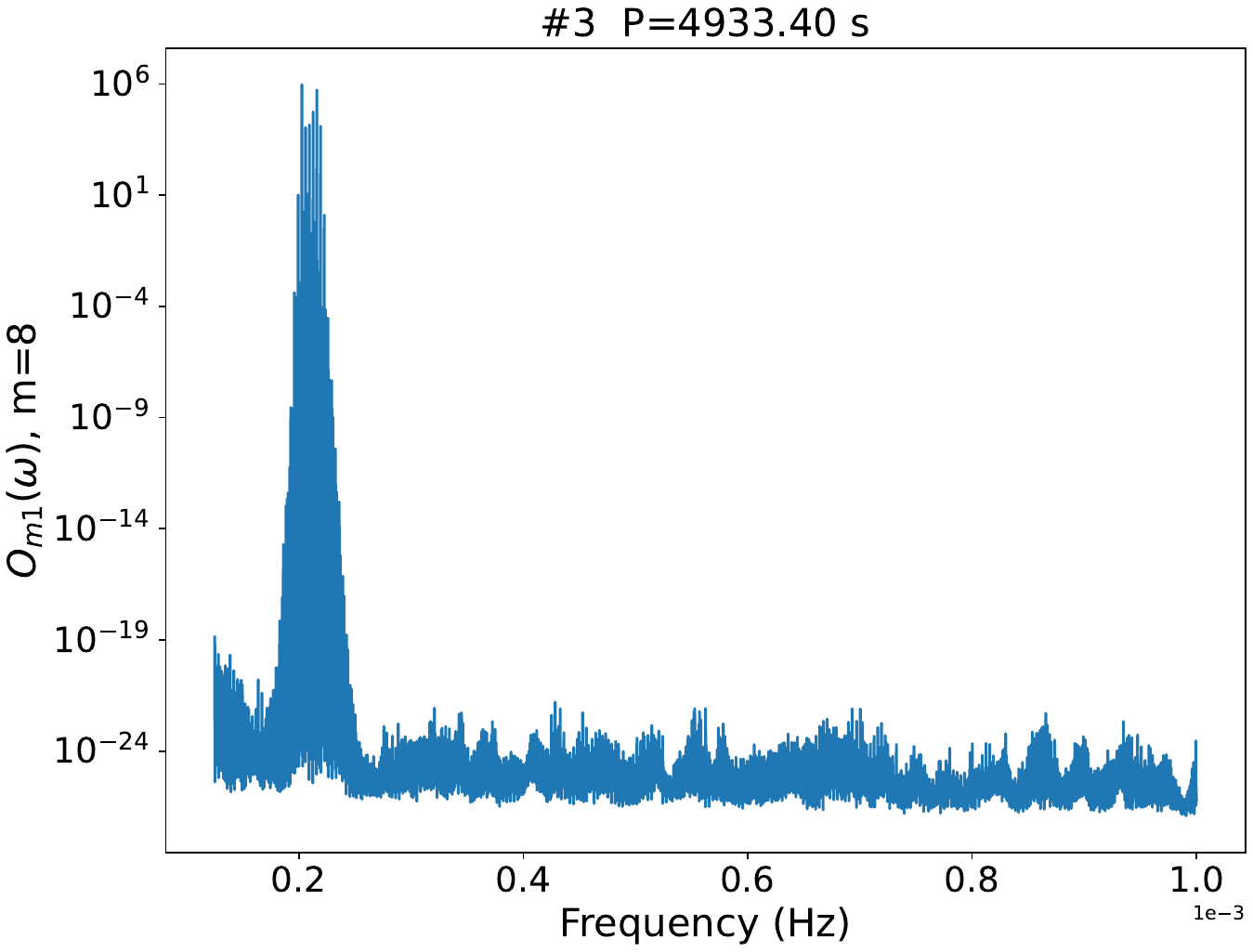}\includegraphics[width=0.5\linewidth]{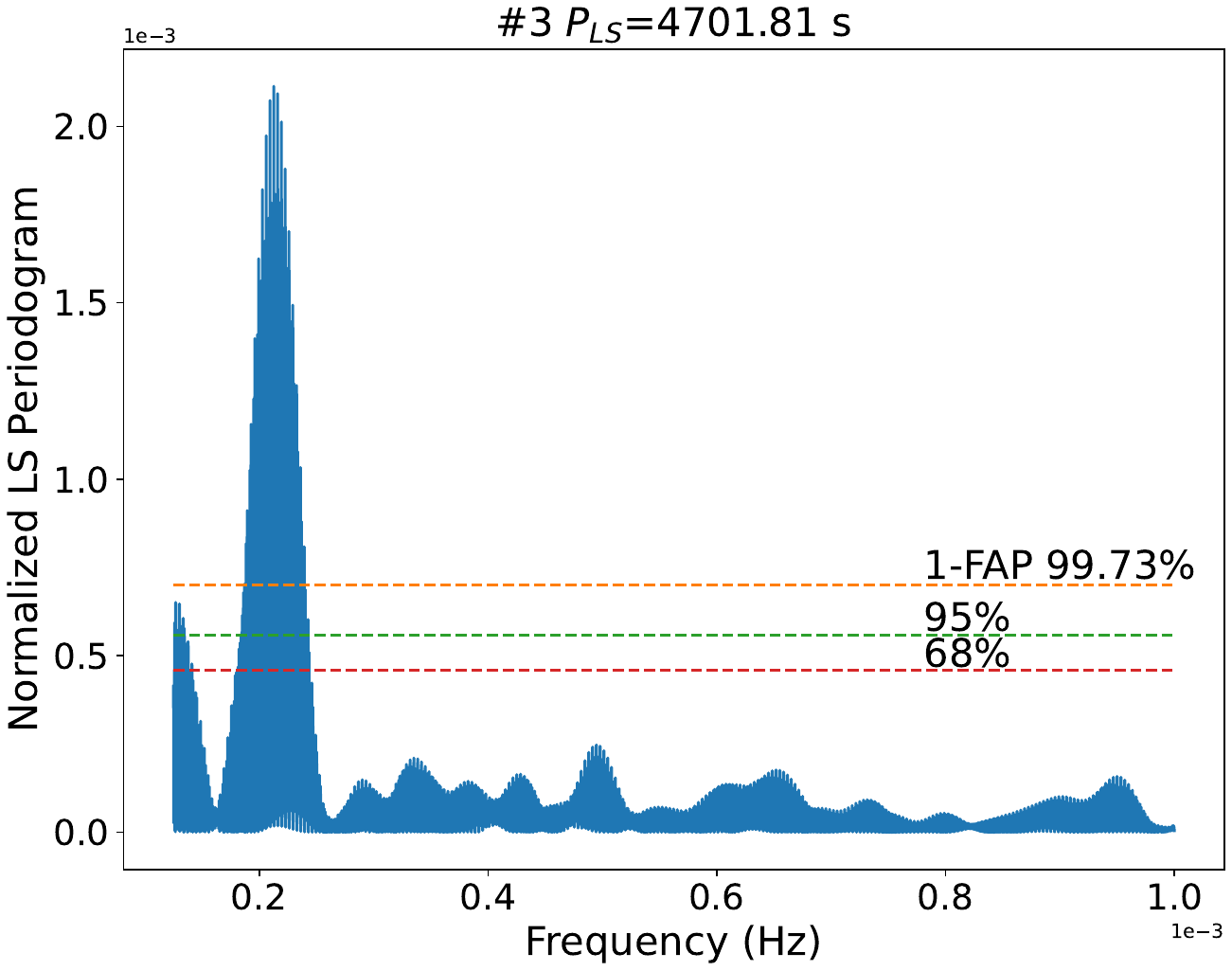}
    \caption{}
\end{figure*}
\begin{figure*}
        \centering
        \ContinuedFloat
    \includegraphics[width=0.51\linewidth]{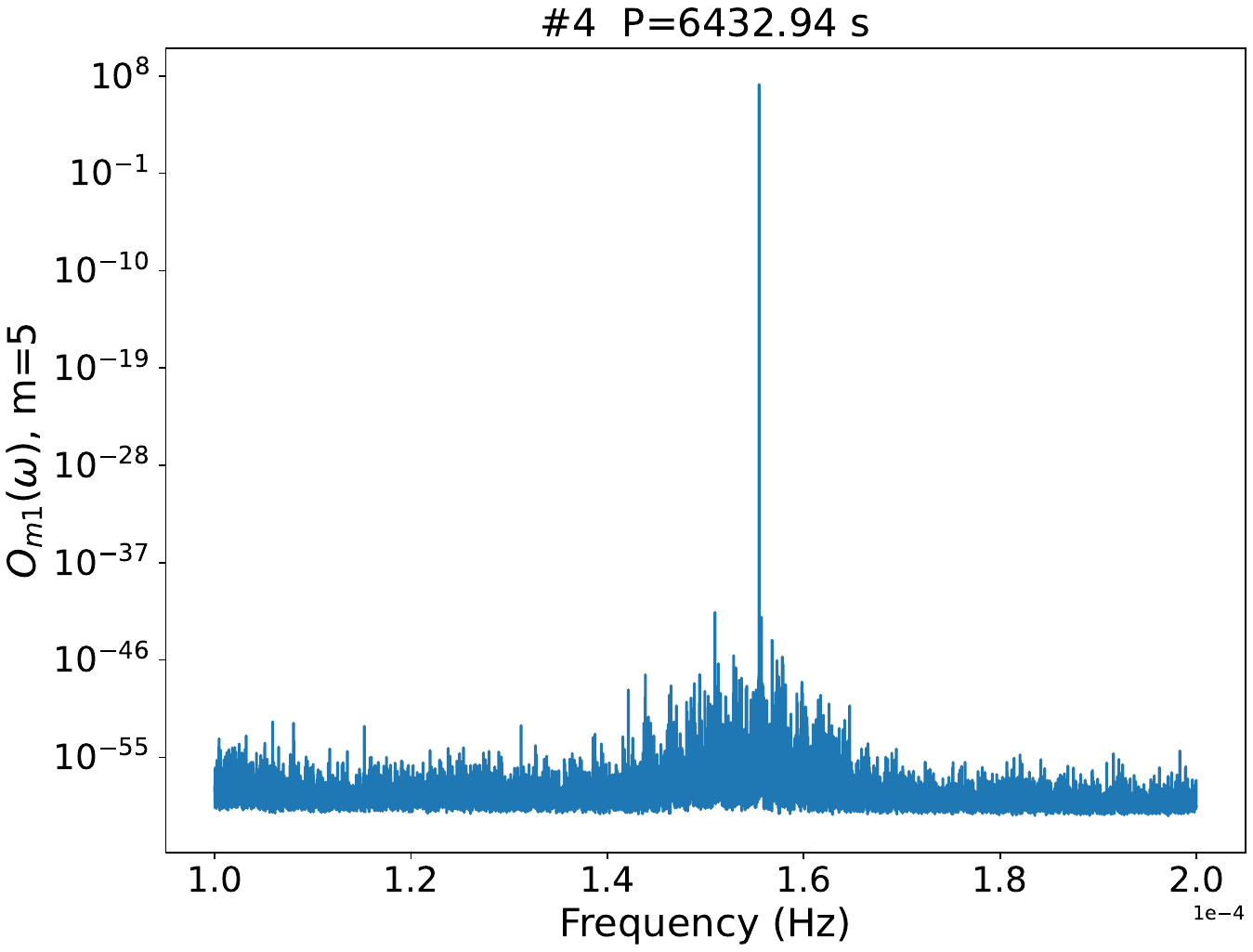}\includegraphics[width=0.5\linewidth]{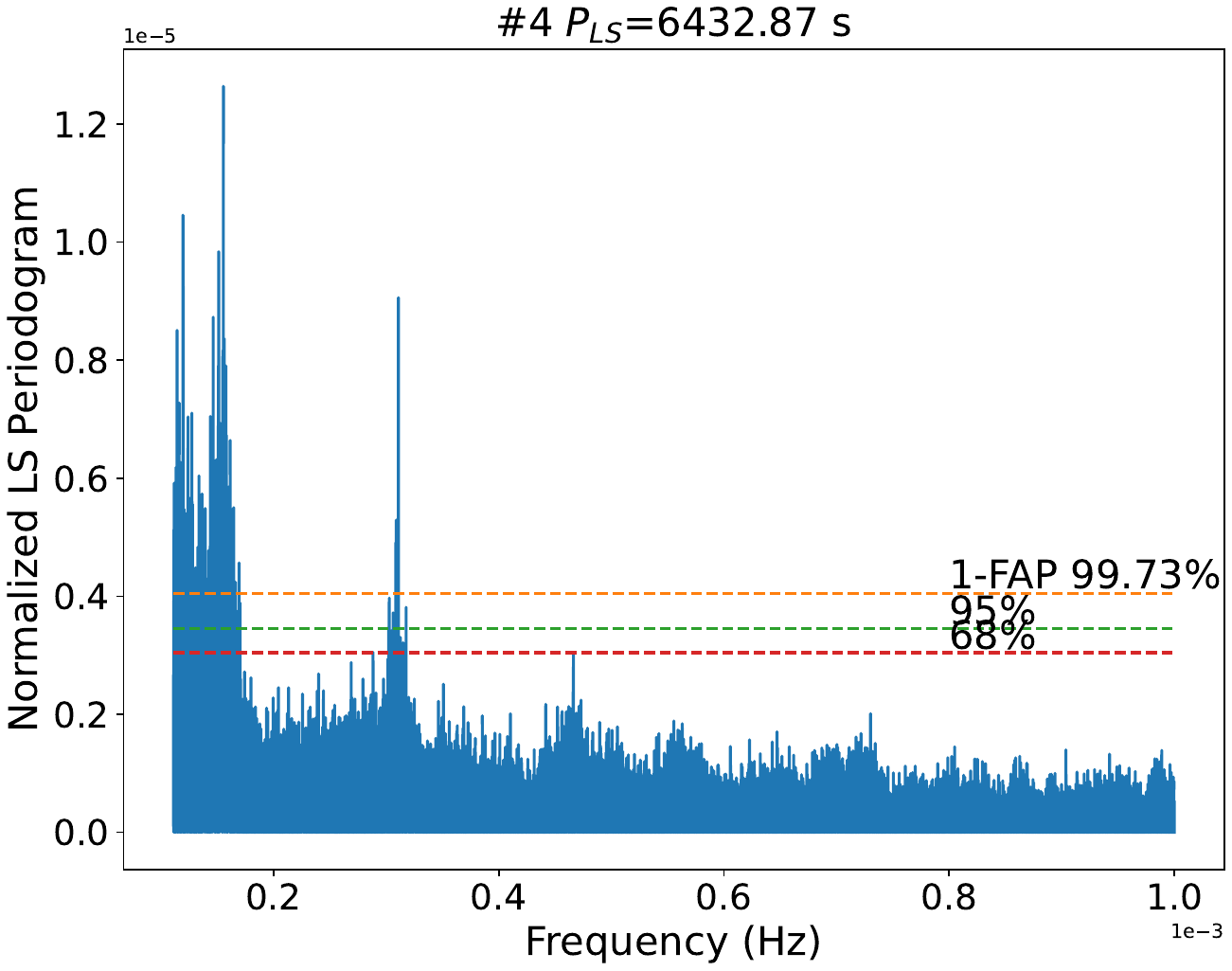}
    \includegraphics[width=0.51\linewidth]{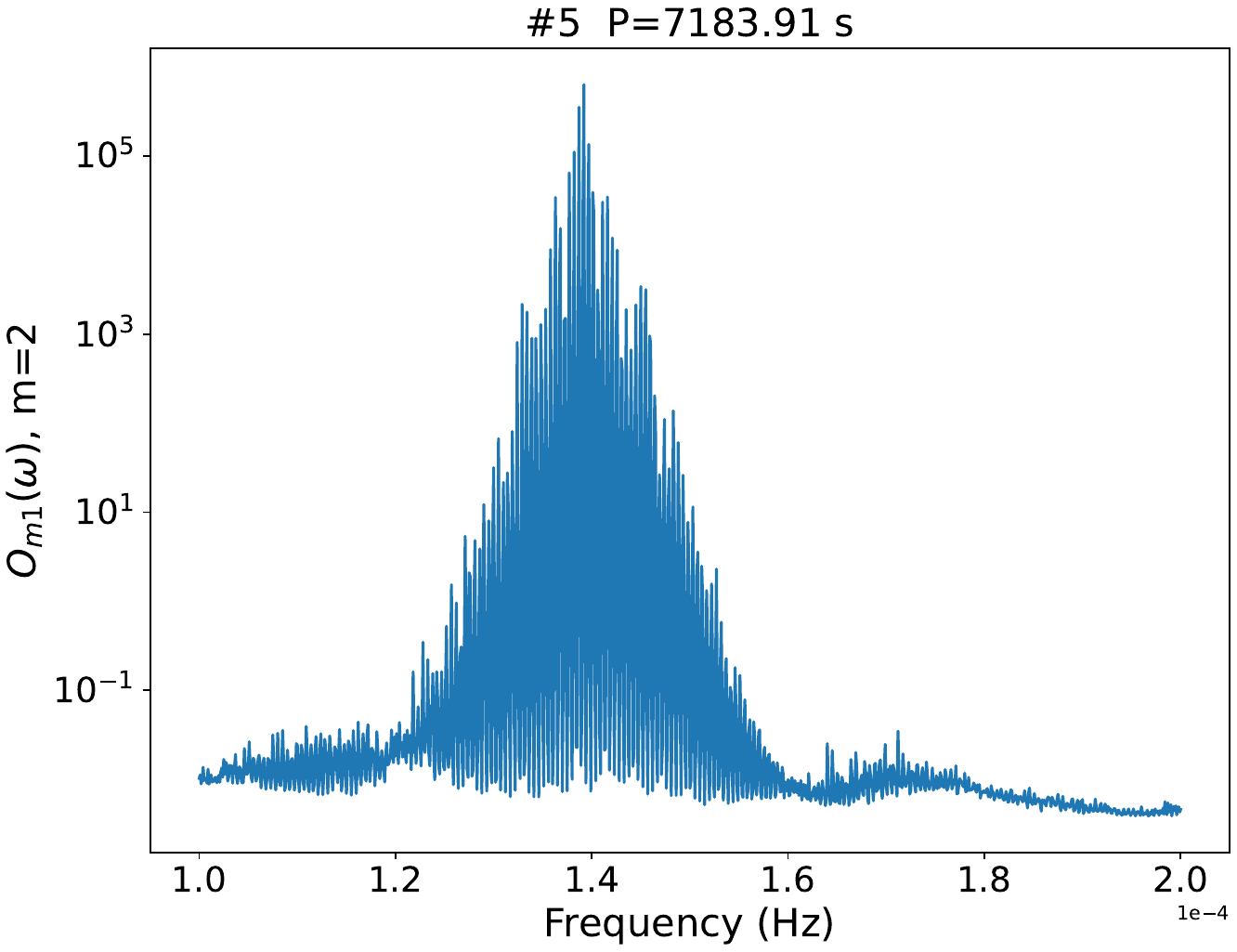}\includegraphics[width=0.5\linewidth]{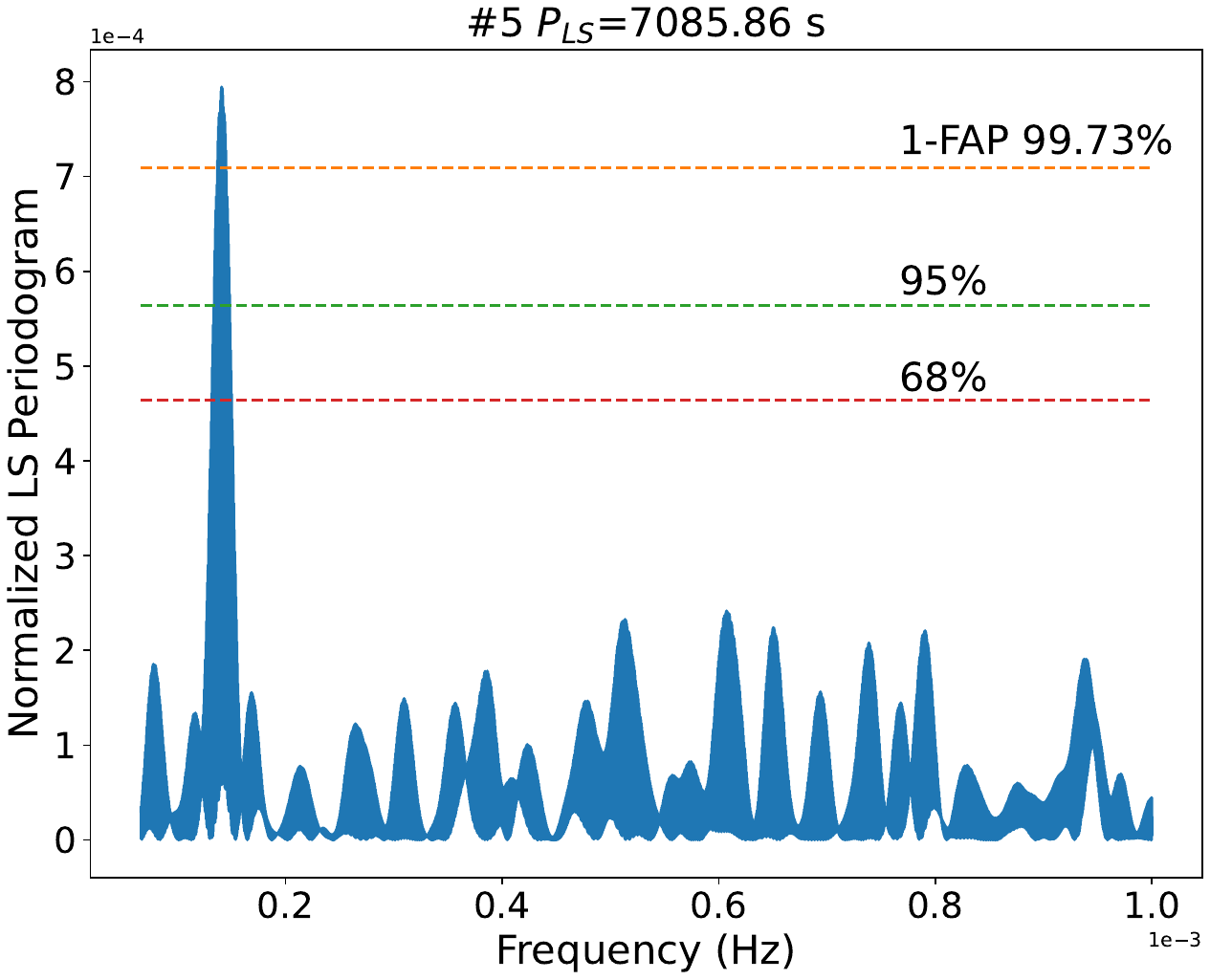}
    \includegraphics[width=0.51\linewidth]{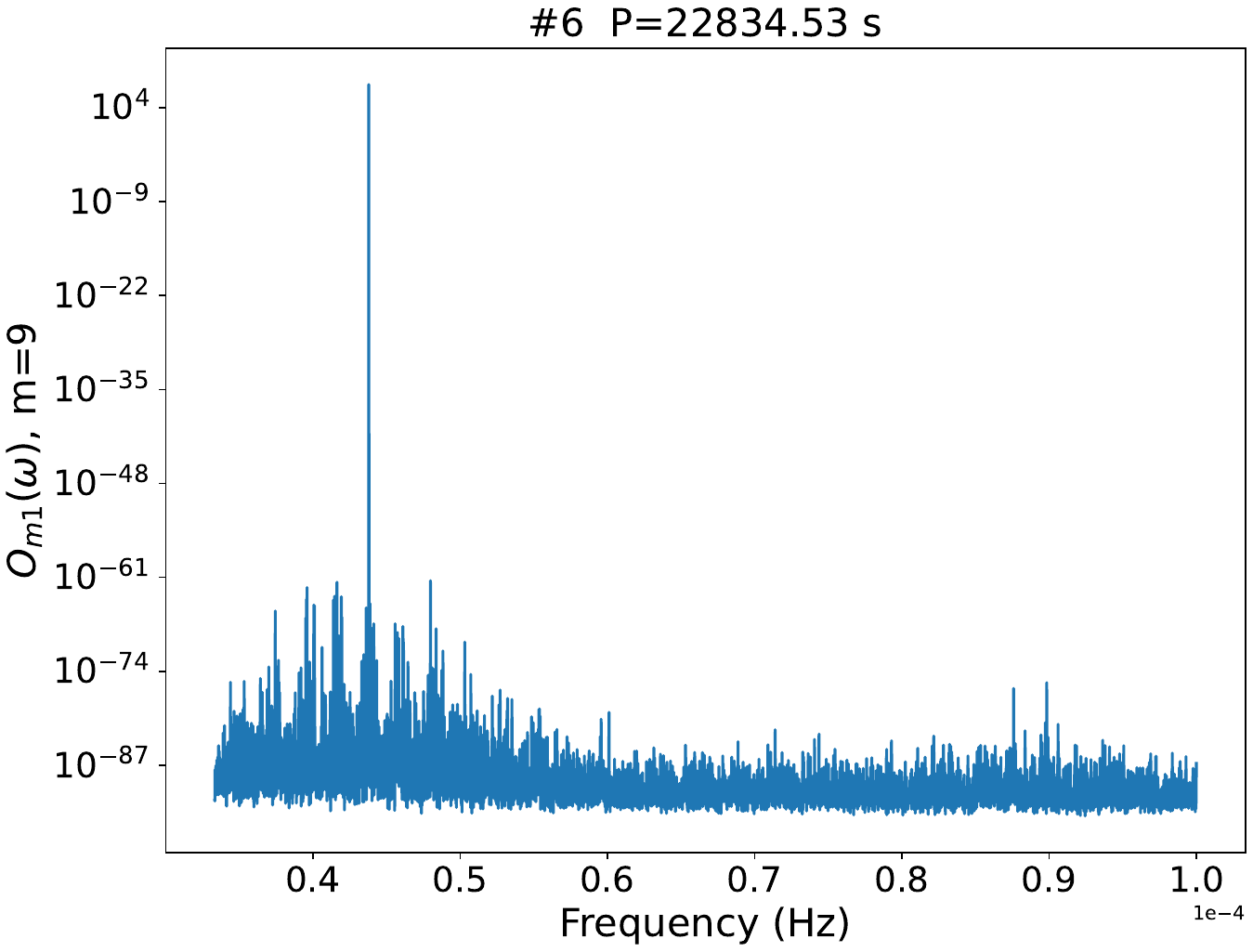}\includegraphics[width=0.5\linewidth]{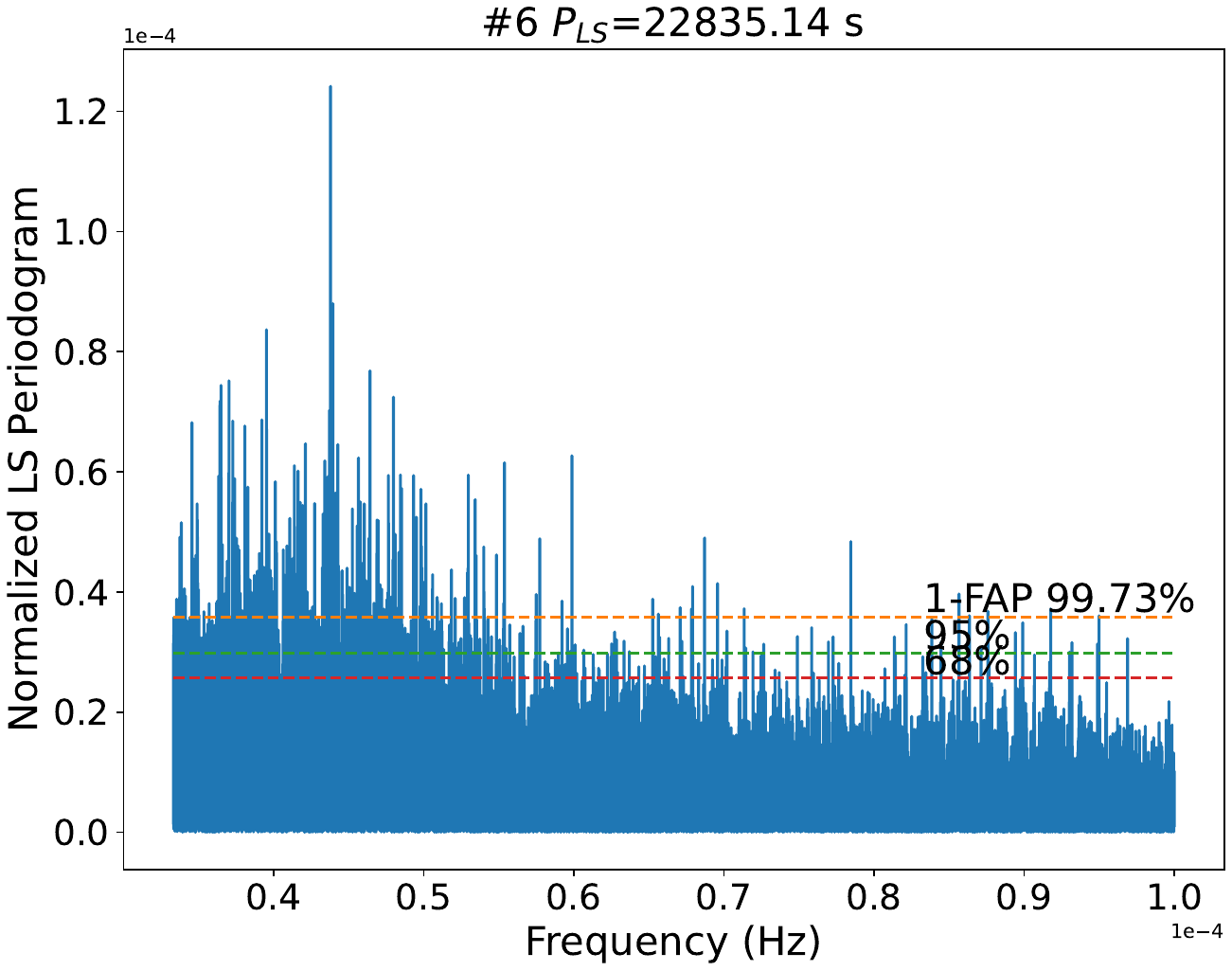}
    \caption{}
\end{figure*}
\begin{figure*}
        \centering
        \ContinuedFloat
            \includegraphics[width=0.51\linewidth]{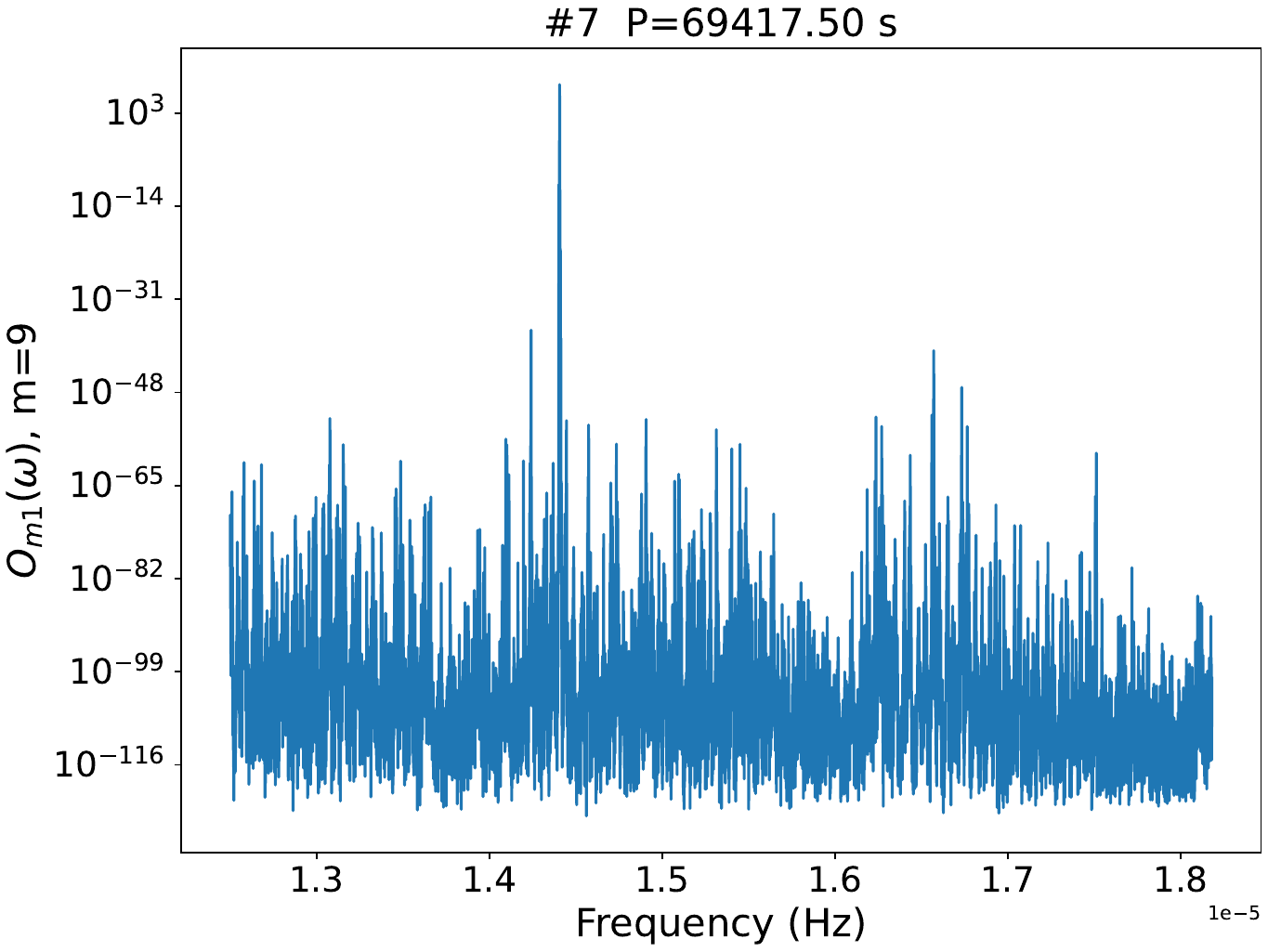}\includegraphics[width=0.5\linewidth]{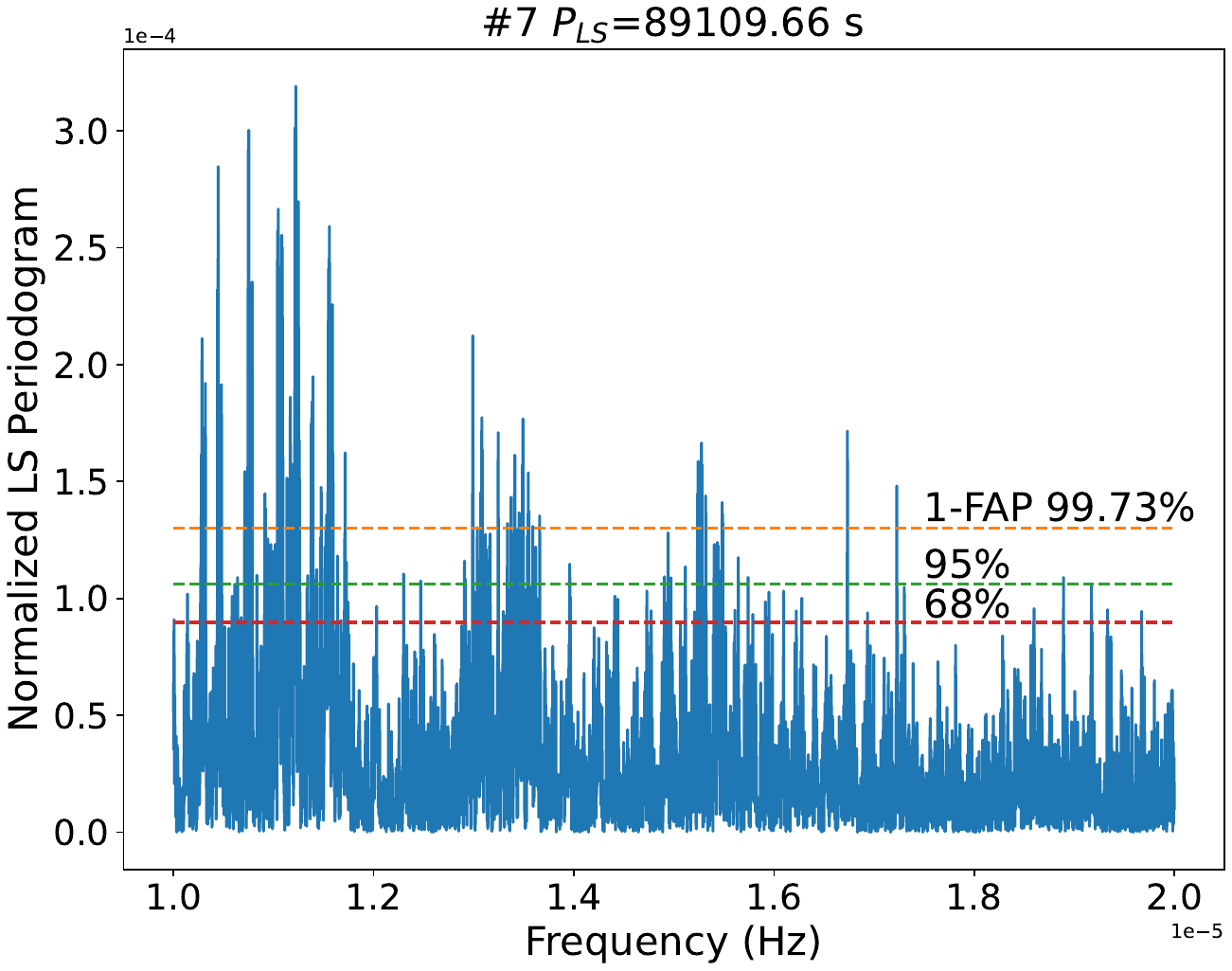}
    \caption{{\it Left}: the Gregory-Loredo diagram of odds ratio for the seven periodic sources. The $x$-axis is the frequency ($\omega$/2$\pi$). The $y$-axis is the odds ratio for an optimal number of $m$ bins. 
    {\it Right}: the normalized Lomb-Scargle periodogram for the seven periodic sources. Three false alarm levels are denoted by the colored dashed lines.}
    \label{fig:gl1}
\end{figure*}

\section{{\it Chandra} observations}
\centering
\begin{longtable}{cccccc}
\caption{{\it Chandra} ACIS Observations of M31}\\
\label{tabacis}\\
\hline
	Obs ID	&	Start Time	&	Target RA	&	Target Dec	&	Data Mode	&	Exposure\\
	&	UT	&		&		&		&	ks	\\
\hline
\hline
\endhead
\hline
\endfoot
	303	&	1999/10/13 05:15	&	00 42 44.40	&	+41 16 08.30	&	FAINT	&	14.04 	\\
	305	&	1999/12/11 05:34	&	00 42 44.40	&	+41 16 08.30	&	FAINT	&	4.18 	\\
	306	&	1999/12/27 16:24	&	00 42 44.40	&	+41 16 08.30	&	FAINT	&	4.18 	\\
	307	&	2000/1/29 14:10	&	00 42 44.40	&	+41 16 08.30	&	FAINT	&	4.16 	\\
	308	&	2000/2/16 02:35	&	00 42 44.40	&	+41 16 08.30	&	FAINT	&	4.06 	\\
	309	&	2000/6/1 01:23	&	00 42 44.40	&	+41 16 08.30	&	FAINT	&	5.15 	\\
	310	&	2000/7/2 22:11	&	00 42 44.40	&	+41 16 08.30	&	FAINT	&	5.14 	\\
	311	&	2000/7/29 00:26	&	00 42 44.40	&	+41 16 08.30	&	FAINT	&	4.96 	\\
	312	&	2000/8/27 20:09	&	00 42 44.40	&	+41 16 08.30	&	FAINT	&	4.72 	\\
	1581	&	2000/12/13 01:36&	00 42 44.40	&	+41 16 08.30	&	FAINT	&	4.46 	\\
	1854	&	2001/1/13 10:28	&	00 42 40.80	&	+41 15 54.00	&	FAINT	&	4.75 	\\
	1582	&	2001/2/18 15:42	&	00 42 44.40	&	+41 16 08.30	&	FAINT	&	4.36 	\\
	1583	&	2001/6/10 19:35	&	00 42 44.40	&	+41 16 08.30	&	FAINT	&	4.97 	\\
	1575	&	2001/10/5 00:32	&	00 42 44.40	&	+41 16 08.30	&	FAINT	&	38.15 	\\
	4360	&	2002/8/11 17:56	&	00 42 44.40	&	+41 16 08.90	&	FAINT	&	4.96 	\\
	4678	&	2003/11/9 07:21	&	00 42 44.40	&	+41 16 08.30	&	FAINT	&	4.87 	\\
	4679	&	2003/11/26 22:21	&	00 42 44.40	&	+41 16 08.30	&	FAINT	&	4.78 	\\
	4680	&	2003/12/27 07:34	&	00 42 44.40	&	+41 16 08.30	&	FAINT	&	5.25 	\\
	4681	&	2004/1/31 19:18	&	00 42 44.40	&	+41 16 08.30	&	FAINT	&	5.12 	\\
	4682	&	2004/5/23 17:38	&	00 42 44.40	&	+41 16 08.30	&	FAINT	&	4.93 	\\
	4719	&	2004/7/17 22:22	&	00 42 44.30	&	+41 16 08.40	&	FAINT	&	5.15 	\\
	4720	&	2004/9/2 14:08	&	00 42 44.30	&	+41 16 08.40	&	FAINT	&	5.13 	\\
	4721	&	2004/10/4 20:38	&	00 42 44.30	&	+41 16 08.40	&	FAINT	&	5.17 	\\
	4722	&	2004/10/31 01:36&	00 42 44.30	&	+41 16 08.40	&	FAINT	&	4.87 	\\
	4723	&	2004/12/5 08:50	&	00 42 50.00	&	+41 17 15.00	&	FAINT	&	5.04 	\\
	7136	&	2006/1/6 20:08	&	00 42 44.40	&	+41 16 08.30	&	VFAINT	&	4.97 	\\
	7137	&	2006/5/26 04:11	&	00 42 44.40	&	+41 16 08.30	&	VFAINT	&	4.92 	\\
	7138	&	2006/6/9 16:15	&	00 42 44.40	&	+41 16 08.30	&	VFAINT	&	5.11 	\\
	7139	&	2006/7/31 00:22	&	00 42 44.40	&	+41 16 08.30	&	VFAINT	&	4.96 	\\
	7140	&	2006/9/24 18:15	&	00 42 44.40	&	+41 16 08.30	&	VFAINT	&	5.12 	\\
	7064	&	2006/12/4 21:09	&	00 42 44.40	&	+41 16 08.30	&	VFAINT	&	29.07 	\\
	8183	&	2007/1/14 20:33	&	00 42 44.40	&	+41 16 08.30	&	VFAINT	&	4.96 	\\
	8184	&	2007/2/14 01:49	&	00 42 44.40	&	+41 16 08.30	&	VFAINT	&	5.18 	\\
	8185	&	2007/3/10 05:56	&	00 42 44.40	&	+41 16 08.30	&	VFAINT	&	4.96 	\\
	7068	&	2007/6/2 22:22	&	00 42 44.40	&	+41 16 08.30	&	VFAINT	&	9.63 	\\
	8191	&	2007/6/18 06:47	&	00 42 44.40	&	+41 16 08.30	&	VFAINT	&	4.96 	\\
	8192	&	2007/7/5 12:59	&	00 42 44.40	&	+41 16 08.30	&	VFAINT	&	5.09 	\\
	8193	&	2007/7/31 03:27	&	00 42 44.40	&	+41 16 08.30	&	VFAINT	&	5.16 	\\
	8194	&	2007/8/28 12:36	&	00 42 44.40	&	+41 16 08.30	&	VFAINT	&	5.04 	\\
	8195	&	2007/9/26 15:04	&	00 42 44.40	&	+41 16 08.30	&	VFAINT	&	4.96 	\\
	8186	&	2007/11/3 04:22	&	00 42 44.40	&	+41 16 08.30	&	VFAINT	&	5.17 	\\
	8187	&	2007/11/27 03:42&	00 42 44.40	&	+41 16 08.30	&	VFAINT	&	4.80 	\\
	9520	&	2007/12/29 16:54&	00 42 44.40	&	+41 16 08.30	&	VFAINT	&	4.96 	\\
	9529	&	2008/5/31 11:20	&	00 42 44.40	&	+41 16 08.30	&	VFAINT	&	5.14 	\\
	9522	&	2008/7/15 16:41	&	00 42 44.40	&	+41 16 08.30	&	VFAINT	&	5.06 	\\
	9523	&	2008/9/1 07:45	&	00 42 44.40	&	+41 16 08.30	&	VFAINT	&	5.18 	\\
	9524	&	2008/10/13 03:30	&	00 42 44.40	&	+41 16 08.30	&	VFAINT	&	5.15 	\\
	9521	&	2008/11/27 17:17	&	00 42 44.40	&	+41 16 08.30	&	VFAINT	&	4.96 	\\
	10551	&	2009/1/9 23:31	&	00 42 44.40	&	+41 16 08.30	&	VFAINT	&	4.96 	\\
	10552	&	2009/2/7 10:36	&	00 42 44.40	&	+41 16 08.30	&	VFAINT	&	4.96 	\\
	10553	&	2009/3/11 14:07	&	00 42 44.40	&	+41 16 08.30	&	VFAINT	&	5.13 	\\
	10554	&	2009/5/29 03:06	&	00 42 44.40	&	+41 16 08.30	&	VFAINT	&	5.05 	\\
	10555	&	2009/7/3 08:15	&	00 42 44.40	&	+41 16 08.30	&	VFAINT	&	5.08 	\\
	10715	&	2009/9/18 02:52	&	00 42 44.40	&	+41 16 08.30	&	VFAINT	&	4.96 	\\
	10716	&	2009/9/25 04:21	&	00 42 44.40	&	+41 16 08.30	&	VFAINT	&	5.13 	\\
	10717	&	2009/10/22 04:35	&	00 42 44.40	&	+41 16 08.30	&	VFAINT	&	5.18 	\\
	11275	&	2009/11/11 05:23	&	00 42 44.40	&	+41 16 08.30	&	VFAINT	&	5.17 	\\
	11276	&	2009/12/8 20:50	&	00 42 44.40	&	+41 16 08.30	&	VFAINT	&	4.95 	\\
	10719	&	2009/12/27 19:53&	00 42 44.40	&	+41 16 08.30	&	VFAINT	&	5.13 	\\
	11277	&	2010/1/1 19:50	&	00 42 44.40	&	+41 16 08.30	&	VFAINT	&	5.17 	\\
	11278	&	2010/2/4 06:21	&	00 42 44.40	&	+41 16 08.30	&	VFAINT	&	4.96 	\\
	11279	&	2010/3/5 14:10	&	00 42 44.40	&	+41 16 08.30	&	VFAINT	&	5.17 	\\
	11838	&	2010/5/27 14:05	&	00 42 44.40	&	+41 16 08.30	&	VFAINT	&	4.96 	\\
	11839	&	2010/6/23 14:28	&	00 42 44.40	&	+41 16 08.30	&	VFAINT	&	4.97 	\\
	11840	&	2010/7/20 20:40	&	00 42 44.40	&	+41 16 08.30	&	VFAINT	&	4.96 	\\
	11841	&	2010/8/24 10:40	&	00 42 44.40	&	+41 16 08.30	&	VFAINT	&	4.96 	\\
	11842	&	2010/9/25 00:43	&	00 42 44.40	&	+41 16 08.30	&	VFAINT	&	4.97 	\\
	12160	&	2010/10/19 11:12	&	00 42 44.40	&	+41 16 08.30	&	VFAINT	&	4.96 	\\
	12161	&	2010/11/16 13:10	&	00 42 44.40	&	+41 16 08.30	&	VFAINT	&	4.97 	\\
	12162	&	2010/12/12 11:30	&	00 42 44.40	&	+41 16 08.30	&	VFAINT	&	4.96 	\\
	12163	&	2011/1/13 03:49	&	00 42 44.40	&	+41 16 08.30	&	VFAINT	&	4.97 	\\
	12164	&	2011/2/16 14:26	&	00 42 44.40	&	+41 16 08.30	&	VFAINT	&	4.97 	\\
	12970	&	2011/5/27 09:21	&	00 42 44.40	&	+41 16 08.30	&	VFAINT	&	4.97 	\\
	12971	&	2011/6/30 11:38	&	00 42 44.40	&	+41 16 08.30	&	VFAINT	&	4.97 	\\
	12972	&	2011/7/25 05:19	&	00 42 44.40	&	+41 16 08.30	&	VFAINT	&	4.97 	\\
	12973	&	2011/8/25 09:31	&	00 42 44.40	&	+41 16 08.30	&	VFAINT	&	4.93 	\\
	14197	&	2011/9/1 20:12	&	00 42 44.40	&	+41 16 08.30	&	FAINT	&	46.86 	\\
	14198	&	2011/9/6 23:10	&	00 42 44.40	&	+41 16 08.30	&	FAINT	&	49.88 	\\
	12974	&	2011/9/28 18:33	&	00 42 44.40	&	+41 16 08.30	&	FAINT	&	4.97 	\\
	13833	&	2011/10/31 04:57	&	00 42 44.40	&	+41 16 08.30	&	FAINT	&	4.97 	\\
	13834	&	2011/11/24 17:46	&	00 42 44.40	&	+41 16 08.30	&	FAINT	&	4.89 	\\
	13835	&	2011/12/19 19:38	&	00 42 44.40	&	+41 16 08.30	&	FAINT	&	4.88 	\\
	13836	&	2012/1/16 23:02	&	00 42 44.40	&	+41 16 08.30	&	FAINT	&	4.88 	\\
	13837	&	2012/2/19 15:16	&	00 42 44.40	&	+41 16 08.30	&	FAINT	&	4.88 	\\
	13298	&	2012/5/26 23:46	&	00 42 44.40	&	+41 16 08.30	&	FAINT	&	4.88 	\\
	13825	&	2012/6/1 07:00	&	00 42 44.40	&	+41 16 08.30	&	FAINT	&	49.99 	\\
	13826	&	2012/6/6 13:27	&	00 42 44.40	&	+41 16 08.30	&	FAINT	&	45.88 	\\
	13827	&	2012/6/12 11:59	&	00 42 44.40	&	+41 16 08.30	&	FAINT	&	51.89 	\\
	13299	&	2012/6/21 22:01	&	00 42 44.40	&	+41 16 08.30	&	VFAINT	&	4.89 	\\
	13828	&	2012/7/1 10:55	&	00 42 44.40	&	+41 16 08.30	&	FAINT	&	49.89 	\\
	13300	&	2012/7/20 08:33	&	00 42 44.40	&	+41 16 08.30	&	FAINT	&	4.89 	\\
	14195	&	2012/8/14 15:14	&	00 42 44.40	&	+41 16 08.30	&	FAINT	&	34.89 	\\
	15267	&	2012/8/16 23:47	&	00 42 44.40	&	+41 16 08.30	&	FAINT	&	13.88 	\\
	13301	&	2012/8/19 01:41	&	00 42 44.40	&	+41 16 08.30	&	FAINT	&	4.70 	\\
	13302	&	2012/9/12 16:25	&	00 42 44.40	&	+41 16 08.30	&	FAINT	&	4.88 	\\
	14196	&	2012/10/28 15:45&	00 42 44.40	&	+41 16 08.30	&   FAINT	&	53.83 	\\
	14927	&	2012/12/9 05:38	&	00 42 44.40	&	+41 16 08.30	&	FAINT	&	4.89 	\\
	14928	&	2012/12/31 08:24&	00 42 44.40	&	+41 16 08.30	&	FAINT	&	4.88 	\\
	14929	&	2013/1/21 17:15	&	00 42 44.40	&	+41 16 08.30	&	FAINT	&	4.88 	\\
	14930	&	2013/2/18 08:02	&	00 42 44.40	&	+41 16 08.30	&	FAINT	&	4.89 	\\
	14931	&	2013/3/12 00:43	&	00 42 44.40	&	+41 16 08.30	&	FAINT	&	4.89 	\\
	15324	&	2013/6/2 14:23	&	00 42 44.40	&	+41 16 08.30	&	FAINT	&	4.89 	\\
        15325   &   2013/6/26 09:23	&  00 42 44.40  &+41 16 08.30       &   FAINT   &   4.89 \\
	15326	&	2013/7/24 02:10	&	00 42 44.40	&	+41 16 08.30	&	FAINT	&	4.89 	\\
	15327	&	2013/8/26 09:43	&	00 42 44.40	&	+41 16 08.30	&	FAINT	&	4.88 	\\
	15328	&	2013/9/27 13:46	&	00 42 44.40	&	+41 16 08.30	&	FAINT	&	4.99 	\\
	16294	&	2014/5/26 08:51	&	00 42 44.40	&	+41 16 08.30	&	FAINT	&	4.88 	\\
	16295	&	2014/6/24 04:52	&	00 42 44.40	&	+41 16 08.30	&	FAINT	&	4.88 	\\
	16296	&	2014/7/22 15:16	&	00 42 44.40	&	+41 16 08.30	&	FAINT	&	4.89 	\\
	16297	&	2014/8/18 02:13	&	00 42 44.40	&	+41 16 08.30	&	FAINT	&	4.89 	\\
	16298	&	2014/9/16 21:03	&	00 42 44.40	&	+41 16 08.30	&	FAINT	&	4.89 	\\
	17443	&	2015/5/26 07:20	&	00 42 44.40	&	+41 16 08.50	&	FAINT	&	5.05 	\\
	17444	&	2015/7/20 10:26	&	00 42 44.40	&	+41 16 08.50	&	FAINT	&	5.04 	\\
	17445	&	2015/8/30 07:00	&	00 42 44.40	&	+41 16 08.50	&	FAINT	&	5.06 	\\
	17446	&	2015/10/14 09:58	&	00 42 44.40	&	+41 16 08.50	&	FAINT	&	5.06 	\\
	17447	&	2015/11/28 18:41	&	00 42 44.40	&	+41 16 08.50	&	FAINT	&	5.06 	\\
	\hline
\end{longtable}

\centering
\begin{longtable}{ccccc}
	\caption{\it{Chandra} HRC Observations of M31}\\
	\label{tabhrc}\\
	\hline
	Obs ID	&	Start Date	&	Target RA	&	Target Dec	&	Exposure	\\						
	&	UT	&		&		&	ks	\\						
	\hline
	\hline
	\endhead
	\hline
	\endfoot
		267	&	1999/11/30 18:29	&	00 42 44.40	&	+41 16 08.30	&	1.28 	\\						
		268	&	1999/12/23 02:20	&	00 42 44.40	&	+41 16 08.30	&	5.21 	\\						
		269	&	2000/1/19 21:50	&	00 42 44.40	&	+41 16 08.30	&	1.21 	\\						
		270	&	2000/2/13 05:12	&	00 42 44.40	&	+41 16 08.30	&	1.48 	\\						
		271	&	2000/3/8 18:06	&	00 42 44.40	&	+41 16 08.30	&	2.48 	\\						
		272	&	2000/5/26 16:46	&	00 42 44.40	&	+41 16 08.30	&	1.21 	\\						
		273	&	2000/6/21 04:52	&	00 42 44.40	&	+41 16 08.30	&	1.20 	\\						
		275	&	2000/8/18 10:15	&	00 42 44.40	&	+41 16 08.30	&	1.19 	\\						
		276	&	2000/9/11 20:41	&	00 42 44.40	&	+41 16 08.30	&	1.19 	\\						
		277	&	2000/10/12 04:14	&	00 42 44.40	&	+41 16 08.30	&	1.19 	\\						
		278	&	2000/11/17 04:34	&	00 42 44.40	&	+41 16 08.30	&	1.19 	\\						
		1569	&	2001/2/1 00:42	&	00 42 44.40	&	+41 16 08.30	&	1.19 	\\						
		1570	&	2001/6/10 22:11	&	00 42 44.40	&	+41 16 08.30	&	1.19 	\\						
		1912	&	2001/10/31 23:48	&	00 42 42.30	&	+41 16 08.40	&	46.98 	\\						
		2904	&	2001/11/19 21:01	&	00 42 44.40	&	+41 16 08.30	&	1.19 	\\						
		2905	&	2002/1/16 04:46	&	00 42 44.40	&	+41 16 08.30	&	1.10 	\\						
		2906	&	2002/6/2 21:29	&	00 42 44.40	&	+41 16 08.30	&	1.19 	\\						
		5925	&	2004/12/6 18:07	&	00 42 44.40	&	+41 16 08.30	&	46.68 	\\						
		5926	&	2004/12/27 20:13	&	00 42 44.40	&	+41 16 08.30&	28.47 	\\						
		5927	&	2005/1/28 19:53	&	00 42 44.40	&	+41 16 08.30	&	27.19 	\\						
		5928	&	2005/2/21 16:16	&	00 42 44.40	&	+41 16 08.30	&	45.15 	\\						
		6177	&	2004/12/27 07:18&	00 42 44.40	&	+41 16 08.30	&	20.19 	\\						
		6202	&	2005/1/28 01:31	&	00 42 44.40	&	+41 16 08.30	&	18.17 	\\						
		7283	&	2006/6/5 06:52	&	00 42 44.30	&	+41 16 09.40	&	20.12 	\\						
		7284	&	2006/9/30 21:16	&	00 42 44.30	&	+41 16 09.40	&	20.19 	\\						
		7285	&	2006/11/13 07:16	&	00 42 44.30	&	+41 16 09.40&	18.68 	\\						
		7286	&	2007/3/11 14:35	&	00 42 44.30	&	+41 16 09.40	&	18.89 	\\						
		8526	&	2007/11/7 15:16	&	00 42 44.30	&	+41 16 09.40	&	20.16 	\\						
		8527	&	2007/11/17 18:08	&	00 42 44.30	&	+41 16 09.40&	20.19 	\\						
		8528	&	2007/11/28 18:58	&	00 42 44.30	&	+41 16 09.40&	20.18 	\\						
		8529	&	2007/12/7 13:44	&	00 42 44.30	&	+41 16 09.40	&	19.13 	\\						
		8530	&	2007/12/17 11:48	&	00 42 44.30	&	+41 16 09.40&	20.16 	\\						
		9825	&	2008/11/8 08:02	&	00 42 44.30	&	+41 16 09.40	&	20.44 	\\						
		9826	&	2008/11/17 03:20	&	00 42 44.30	&	+41 16 09.40&	20.14 	\\						
		9827	&	2008/11/28 05:44	&	00 42 44.30	&	+41 16 09.40&	20.19 	\\						
		9828	&	2008/12/7 09:48	&	00 42 44.30	&	+41 16 09.40	&	20.19 	\\						
		9829	&	2008/12/18 00:24	&	00 42 44.30	&	+41 16 09.40&	10.19 	\\						
		10683	&	2009/2/16 21:28	&	00 42 44.30	&	+41 16 09.40	&	20.13 	\\						
		10684	&	2009/2/26 04:02	&	00 42 44.30	&	+41 16 09.40	&	18.92 	\\						
		10838	&	2008/12/18 11:43	&	00 42 44.30	&	+41 16 09.40&	10.15 	\\						
		10882	&	2009/11/7 05:26	&	00 42 44.30	&	+41 16 09.40	&	19.08 	\\						
		10883	&	2009/11/16 05:51	&	00 42 44.30	&	+41 16 09.40&	18.55 	\\						
		10884	&	2009/11/27 15:01	&	00 42 44.30	&	+41 16 09.40&	18.59 	\\						
		10885	&	2009/12/8 22:38	&	00 42 44.30	&	+41 16 09.40	&	18.50 	\\						
		10886	&	2009/12/17 21:34	&	00 42 44.30	&	+41 16 09.40&	18.58 	\\						
		11808	&	2010/2/15 20:43	&	00 42 44.30	&	+41 16 09.40	&	17.33 	\\						
		11809	&	2010/2/26 06:31	&	00 42 44.30	&	+41 16 09.40	&	18.64 	\\						
		12110	&	2010/11/14 03:57	&	00 42 44.30	&	+41 16 09.40&	20.17 	\\						
		12111	&	2010/11/23 04:18	&	00 42 44.30	&	+41 16 09.40&	20.13 	\\						
		12112	&	2010/12/3 15:53	&	00 42 44.30	&	+41 16 09.40	&	20.16 	\\						
		12113	&	2010/12/12 13:23	&	00 42 44.30	&	+41 16 09.40&	19.18 	\\						
		12114	&	2010/12/22 04:12	&	00 42 44.30	&	+41 16 09.40&	20.19 	\\						
		13178	&	2011/2/17 03:31	&	00 42 44.30	&	+41 16 09.40	&	17.65 	\\						
		13179	&	2011/2/27 05:53	&	00 42 44.30	&	+41 16 09.40	&	17.68 	\\						
		13180	&	2011/3/10 02:52	&	00 42 44.30	&	+41 16 09.40	&	17.68 	\\						
		13227	&	2011/11/12 02:29	&	00 42 44.30	&	+41 16 09.40&	20.18 	\\						
		13228	&	2011/11/21 05:18	&	00 42 44.30	&	+41 16 09.40&	19.19 	\\						
		13229	&	2011/11/30 23:29	&	00 42 44.30	&	+41 16 09.40&	19.77 	\\						
		13230	&	2011/12/11 13:26	&	00 42 44.30	&	+41 16 09.40&	19.12 	\\						
		13231	&	2011/12/20 07:53	&	00 42 44.30	&	+41 16 09.40&	19.68 	\\						
		13278	&	2012/2/17 18:11	&	00 42 44.30	&	+41 16 09.40	&	19.15 	\\						
		13279	&	2012/2/28 06:09	&	00 42 44.30	&	+41 16 09.40	&	19.15 	\\						
		13280	&	2012/3/13 04:57	&	00 42 44.30	&	+41 16 09.40	&	19.45 	\\						
		13281	&	2012/6/1 21:29	&	00 42 44.30	&	+41 16 09.40	&	19.08 	\\	
		\hline					
\end{longtable}

\end{document}